\documentclass{aa}

\usepackage{natbib}
\usepackage[varg]{txfonts}
\usepackage{booktabs}
\usepackage{bm}
\usepackage{comment}
\usepackage{color}
\usepackage{graphicx}  
\usepackage{subfig}
\usepackage{upgreek}
\usepackage{multirow}
\bibpunct{(}{)}{;}{a}{}{,} 

\begin{document}

\title{Evolution of the atomic component in protostellar outflows}
 
\author{T.\,Sperling\inst{1}
\and J.\,Eislöffel\inst{1}
\and C.\,Fischer \inst{3}
\and B.\,Nisini\inst{2} 
\and T.\,Giannini\inst{2}
\and A.\,Krabbe \inst{3}}
\institute{Thüringer Landessternwarte, Sternwarte 5, D-07778, Tautenburg, Germany
\and INAF—Osservatorio Astronomico di Roma, via Frascati 33, I-00040 Monte Porzio, Italy
\and Deutsches SOFIA Institut University of Stuttgart, D-70569 Stuttgart, Germany}
\date{Received: 03 December 2020 / Accepted: 15 April 2021}

\abstract {We present SOFIA/FIFI-LS observations of three Class 0 and one Class I outflows (Cep\,E, HH\,1, HH\,212, and L1551\,IRS5) in the far-infrared [O\,I]$_{63\upmu\text{m}}$ and [O\,I]$_{145\upmu\text{m}}$ transitions. Spectroscopic [O\,I]$_{63\upmu\text{m}}$ maps enabled us to infer the spatial extent of warm ($T \sim 500-1200\,\text{K}$), low-excitation atomic gas within these protostellar outflows.} {Our main goal is to determine mass-loss rates from the obtained [OI]$_{63\upmu\text{m}}$ maps and compare these with accretion rates from other studies.  } {The far-infrared [O\,I]$_{63\upmu\text{m}}$ emission line is predicted to be the main coolant of dense, dissociative J-shocks caused by decelerated wind or jet shocks. If proper shock conditions prevail, the instantaneous mass-ejection rate is directly connected to the [O\,I]$_{63\upmu\text{m}}$ luminosity.  In order to unravel evolutionary trends, we analysed a set of 14 Class 0/I outflow sources that were spatially resolved in the [O\,I]$_{63}$ emission. We compared these data with a sample of 72 Class 0/I/II outflow sources that have been observed with Herschel (WISH, DIGIT, WILL, GASPS surveys) without spatially resolving the [O\,I]$_{63\upmu\text{m}}$ line.} {All our newly observed targets feature prominent [O\,I]$_{63\upmu\text{m}}$ emission either close to the driving source (L1551\,IRS5, HH\,1, HH\,212) or as extended jet-like  or  knotty  emission region away from it (Cep\,E). The detected [O\,I]$_{63\upmu\text{m}}$ emission can  mostly be attributed to dissociative shocks and photodissociation regions (PDRs).  Flux values at $63\,\upmu\text{m}$ and $145\,\upmu\text{m}$ of all four associated continuum sources are presented. We calculated mass-loss rates connected to the low-excitation, atomic outflow component in the range of $(5-50)\times 10^{-7}\,M_\odot\,\text{yr}^{-1}$.  
Estimated ratios between the mass loss in the outflow and the mass accretion onto the source (jet efficiency ratios)   are largely in the range of $\dot{M}_\text{out}/\dot{M}_\text{acc} \sim 0.05-0.5$  for the observed outflow sources, which are consistent with theoretical predictions and quoted Herschel data.  
 } {Our new observations and a comparison with the 72 outflow sources  observed with Herschel  indicate that the bulk ejected material in outflows from Class 0 sources resides in the molecular component, that is mass-loss rates derived from the [O\,I]$_{63}$ emission line significantly underestimate the total mass-loss rate during this and possibly also later phases of the star formation process.}     
\maketitle

{\renewcommand{\arraystretch}{1.2}
\begin{table*}[!htb]
\caption{\small{Target information.}}\label{table:objects}
\centering
\begin{tabular}{c c c c c c c c }
\hline\hline
Source & Class & Cloud   &  R.A.\,(J2000)\tablefootmark{a} & Decl.\,(J2000)\tablefootmark{a} & $D$\tablefootmark{b}  & $L_\text{bol}$\tablefootmark{c} & P.A.\tablefootmark{h}   \\[1.5pt]
 & & & (h m s) & ($\degr\,\arcmin\,\arcsec$) &   (pc)  & ($L_\odot$) & ($^\text{o}$) \\ 
\hline  
\small{IRAS 23011+6126 (Cep\,E)}  & \small{0}\tablefootmark{g} & \small{Cepheus, OB3}   & \small{23 03 12.8} & \small{$+$61 42 26} & $730$  &  $\sim 75$\tablefootmark{d} &  \small{25} \\ [1.5pt]
 \hline  
\small{VLA 1 (HH\,1)}   & \small{0}\tablefootmark{f}  & \small{Orion A, L1641}   & \small{05 36 22.8} & \small{$-$06 46 06} & $430$ & $\sim 46$\tablefootmark{d} & \small{145}\\ [1.5pt]
\hline  
\small{IRAS 05413-0104 (HH\,212)}   &\small{0}\tablefootmark{g} & \small{Orion B, L1630} & \small{05 43 51.4} & \small{$-$01 02 53} & $420$ &   $\sim 13$\tablefootmark{d} & \small{30} \\ [1.5pt]
\hline  
\small{L1551 IRS 5} & \small{I}\tablefootmark{g} & \small{Taurus, L1551}   & \small{04 31 34.2} & \small{$+$18 08 05} & $140$ & $\sim 23$\tablefootmark{e} & \small{75} \\ [1.5pt]
 \hline  
\end{tabular}
\tablefoot{ 
\tablefoottext{a}{\small{Taken from the Two Micron All Sky Survey (2MASS).}}
\tablefoottext{b}{\small{Rounded from \citet{zucker_2019} based on GAIA DR2, except  Cep\,E taken from \citet{kun_2008}.}}
\tablefoottext{c}{\small{We corrected the luminosities taken from the cited papers for our assumed distances: $L_\text{bol} = (D_{\text{adopted}}/D_\text{paper})^2 L_\text{bol}^\text{paper}$.}}
\tablefoottext{d}{\small{\citet{dishoeck_2011}.}}
\tablefoottext{e}{\small{\citet{karska_2018}.}}
\tablefoottext{f}{\small{\citet{ami_consortium_2012, chini_2001}.}}
\tablefoottext{g}{\small{\citet{froebrich_2005}.}}
\tablefoottext{h}{\small{Position angle of the outflow.}}
}
\end{table*}

\section{Introduction}

Protostellar jets and outflows are spectacular astrophysical phenomena associated with star formation \citep{frank_2014}. Outflows from young stellar objects (YSOs) are nowadays commonly observed throughout the spectral range. 
Depending on the evolutionary status of the driving source, jets are mostly seen in distinctive molecular (e.g. SiO, CO, H$_2$), atomic (e.g. [O\,I]), or ionic (e.g. [Fe\,II], [S\,II], [N\,II]) transitions \citep{bally_2016}, and until now only in a few cases in radio synchrotron emission \citep{vig_2018, anglada_2018, anton_2019}.\\
The broad categorisation of YSOs into Class 0, Class I, and Class II sources is well established \citep{lada_1987, andre_1993, greene_1994}, and in this context it may seem reasonable that a comparable evolutionary scheme for their connected outflows  can be sought out as well.\\
This idea has been tackled by a handful of studies \citep[e.g.][]{ellerbroek_2013, watson_2016, mottram_2017} mainly focusing on evolutionary indicators such as the efficiency ratio $f:=\dot{M}_\text{out}/\dot{M}_\text{acc}$ of the protostar-outflow system. Here, $\dot{M}_\text{out}$ and $\dot{M}_\text{acc}$ represent the mass-ejection rate and the mass-accretion rate, respectively. \citet{watson_2016} claimed to have identified a trend of decreasing mass loss among Class 0, Class I, and Class II sources. Reasonable doubts for this hypothesis are articulated in \citet{mottram_2017}, who identified different observing strategies as responsible for that finding. However, the determination of mass-ejection rates in all mentioned studies have to be reviewed since the outflow regions were observed with a poor spatial resolution. Furthermore, outflow rates, which are usually calculated from the far-infrared [O\,I]$_{63}$ line luminosity, can be meaningless if proper shock conditions of the underlying \citet{hollenbach_1989} shock model do not prevail \citep{hartigan_2019, sperling_2020}. \\
In their survey, \citet{alonso_martinez_2017} detected the [O\,I]$_{63}$ emission line towards all young outflow sources (Classes I, II, and III), supporting the notion that the [O\,I]$_{63}$ emission line traces the warm ($T\sim 500-1500\,\text{K}$), shock-excited gas component in protostellar outflows. In addition, the [O\,I]$_{63}$ emission line is expected to be comparably bright among other emission lines in low-excitation atomic jets, making it a reasonable basis for observational studies of protostellar outflows.\\
Only a handful protostellar outflows have been observed spatially resolving the far-infrared [O\,I]$_{63}$ emission line \citep{podio_2012, nisini_2015, dionatos_2017, sperling_2020}. A detailed analysis of these targets mostly supports the notion of the shock origin of the detected [O\,I]$_{63}$ emission. In addition, specific line ratios observed towards protostellar outflows can help to rule out other possible [O\,I]$_{63}$ origins such as photodissociation regions (PDRs)  or a disc \citep[e.g.][]{watson_2016}. \\
In this context, we present new SOFIA/FIFI-LS observations of four Class 0/I outflow sources (Cep\,E, L1551\,IRS5, HH\,212, HH\,1, see Table\,\ref{table:objects}) spatially resolving [O\,I]$_{63,145}$ along their outflow regions close to their respective driving sources. These maps enable us to  measure  the instantaneous mass-loss rates residing in the outflows.   
Together with accretion rates estimated from bolometric luminosities \citep{gullbring_1998, hartmann_1998, white_2004}, our main intention is to revisit the prominence of the atomic flow component at different evolutionary stages of the driving source. 

\section{Observations}\label{observations}
 
The observations were obtained with the FIFI-LS instrument \citep{fifi_looney_2000, fischer_2018, fifi_colditz_2018} aboard the flying  observatory SOFIA \citep{young_2012}. SOFIA is a modified Boeing 747SP aircraft equipped with a 2.5\,m effective diameter reflecting telescope. The FIFI-LS instrument is an integral field, far-infrared spectrometer consisting of two independent grating spectrometers simultaneously operating at 51-120\,$\upmu$m (blue channel) and 115-200\,$\upmu$m (red channel). Both channels are composed of a $5\times 5$ pixel array with a field of view of $30\arcsec \times 30\arcsec$ in the blue channel and $1\arcmin\times 1\arcmin$ in the red channel.  All four targets were mapped along their outflow axis close to their respective driving sources in the [O\,I]\,63.1837\,$\upmu$m  and [O\,I]\,145.5254\,$\upmu$m far-infrared fine structure lines\footnote{\tiny{\url{https://www.mpe.mpg.de/ir/ISO/linelists/FSlines.html}}}. The diffraction-limited   full width at half maximum  (FWHM) beam sizes are $5\farcs 4$ at 63\,$\upmu$m and $12\farcs 4$ at 145\,$\upmu$m. The spectral resolutions $R=\lambda/\Delta \lambda$ are specified as 1300 (63\,$\upmu$m) and 1000 (145\,$\upmu$m), which corresponds to a medium velocity resolution of 231\,$\text{km}\,\text{s}^{-1}$ and 300\,$\text{km}\,\text{s}^{-1}$. The data were acquired in observing Cycle 7 (programme IDs: 03\_0073, 07\_0069) in two-point symmetric chop mode.

\begin{table}[!htb]
\caption{\small{SOFIA flight information and chosen ATRAN parameters for the five observed targets (zenith angle $\theta$, flight altitude $H$, water vapour overburden $wvp$).}}\label{table:atranparameters}
\centering
\begin{tabular}{c c c c c c c }
\hline\hline
Object        & Obs.\,date & Flight & Time\tablefootmark{a} & $\theta$ & $H$ & $wvp$ \\
              &            &        & \tiny{(min)}                      & \tiny{($\degr$)} & \tiny{(kft)} & \tiny{($\upmu$m)} \\
\hline
\tiny{Cep\,E}  & \tiny{2019-10-30} & \tiny{631}      & \tiny{52}                &  \tiny{50}  & \tiny{44}                    &   \tiny{4.00}       \\
\tiny{VLA 1, HH\,1}  & \tiny{2018-11-09}     & \tiny{527} & \tiny{84}                   &   \tiny{45}   &   \tiny{43}  &     \tiny{4.00}    \\
\tiny{HH\,212}  & \tiny{2018-11-08}    & \tiny{526}   & \tiny{49}                  & \tiny{40} &  \tiny{43} &  \tiny{4.25}  \\
\tiny{L1551IRS5}  &   \tiny{2019-11-07}   & \tiny{637}     &\tiny{21}                 & \tiny{45}       &  \tiny{43}    & \tiny{4.50}           \\
\hline
\end{tabular}
\tablefoot{ 
\tablefoottext{a}{\small{Total effective on-source integration time.}}
}
\end{table}

\section{Data reduction}\label{data_reduction}
  
We applied the same data reduction technique as described in detail in \citet{sperling_2020}, which can be briefly summarised as follows.  
According to the flight parameters (i.e. the flight altitude $H$, the zenith angle $\theta$,
and water vapour overburden $wvp$), we specified a suitable synthetic transmission model $\tau (\lambda, H, \theta, wvp)$ for the atmosphere    \citep[ATRAN model,][]{atran_lord_1992}. The expected continuum emission together with a potentially present emission line is modelled as a four-parametric 1D-Gaussian function $\varphi(\lambda; A, \sigma, \mu, B)$. Thus, for each spatial pixel (also called spaxel) in our obtained data cubes, a non-linear least-squares fit to the function, 
\begin{equation}\label{equ:S}
y  =   \bigg[ \varphi(\lambda; A, \sigma, \mu, B) \cdot  \tau (\lambda, H, \theta, wvp)  \bigg]  *\text{SIF}\left(\lambda; R  \right) 
,\end{equation} 
is performed to extract the continuum as well as the continuum-subtracted line information in each spaxel (Figs.\,\ref{fig:all_minispectra_A}--\ref{fig:all_minispectra_D}). In Eq.\,\ref{equ:S},  the FIFI-LS spectral instrument function depending on the spectral resolution $R$ is denoted as $\text{SIF}\left(\lambda; R  \right)$.\\
Table\,\ref{table:atranparameters} lists the selected ATRAN parameters for our four targets. Water vapour values therein represent actual measurements during observations (Fischer et al. in prep.). We estimate the uncertainties introduced by the time-averaged atmospheric modelling to be not more than 5 \%. The total uncertainty in the absolute
flux calibration for the integrated line fluxes amounts to approximately 20 \%.\\
The observed [O\,I]$_{63}$ line widths ($\Delta v_\text{obs}^2 =  \Delta v_\text{line}^2 + \Delta v_\text{FIFI}^2 $) are of the order of $250-350\,\text{km}\,\text{s}^{-1 }$ , indicating that the line is spectrally unresolved. We therefore constrain the intrinsic line width in our  data reduction pipeline  to be in the range of $50-300\,\text{km}\,\text{s}^{-1}$.

\begin{figure*} 
\centering
\subfloat{\includegraphics[trim=0 0 0 0, clip, width=0.33\textwidth]{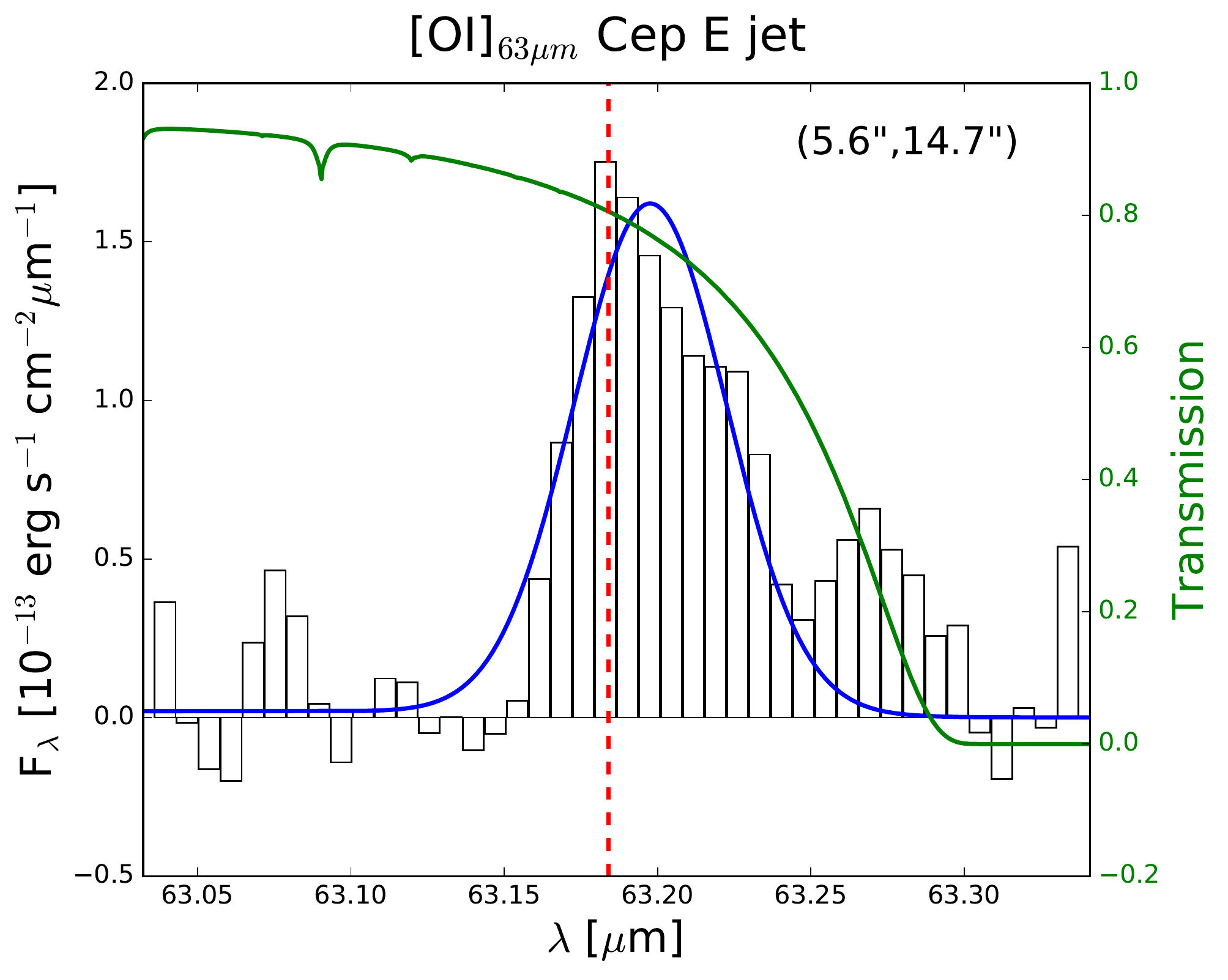}}
\hfill
\subfloat{\includegraphics[trim=0 0 0 0, clip, width=0.33\textwidth]{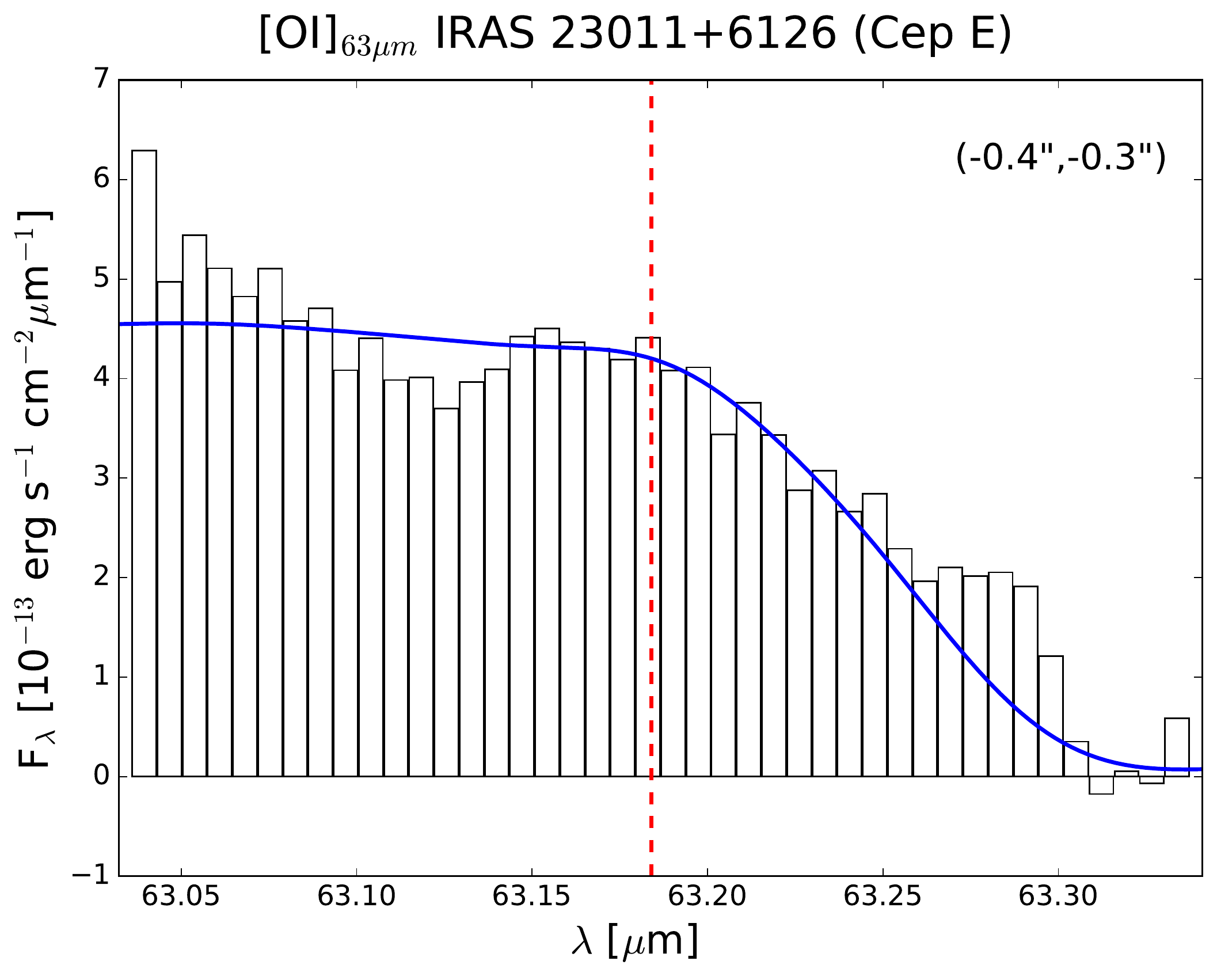}}
\hfill
\subfloat{\includegraphics[trim=0 0 0 0, clip, width=0.33\textwidth]{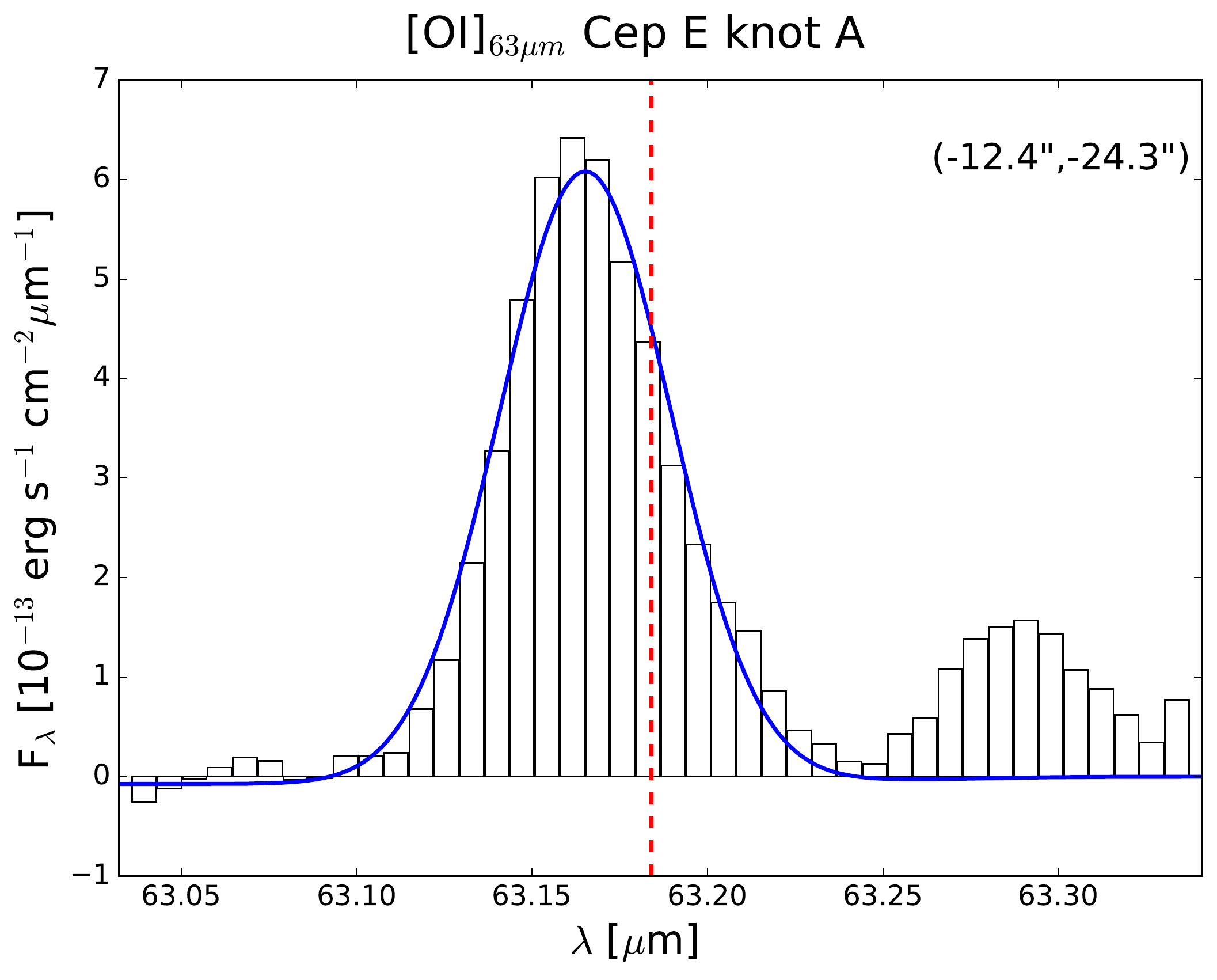}}
\hfill
\subfloat{\includegraphics[trim=0 0 0 0, clip, width=0.33\textwidth]{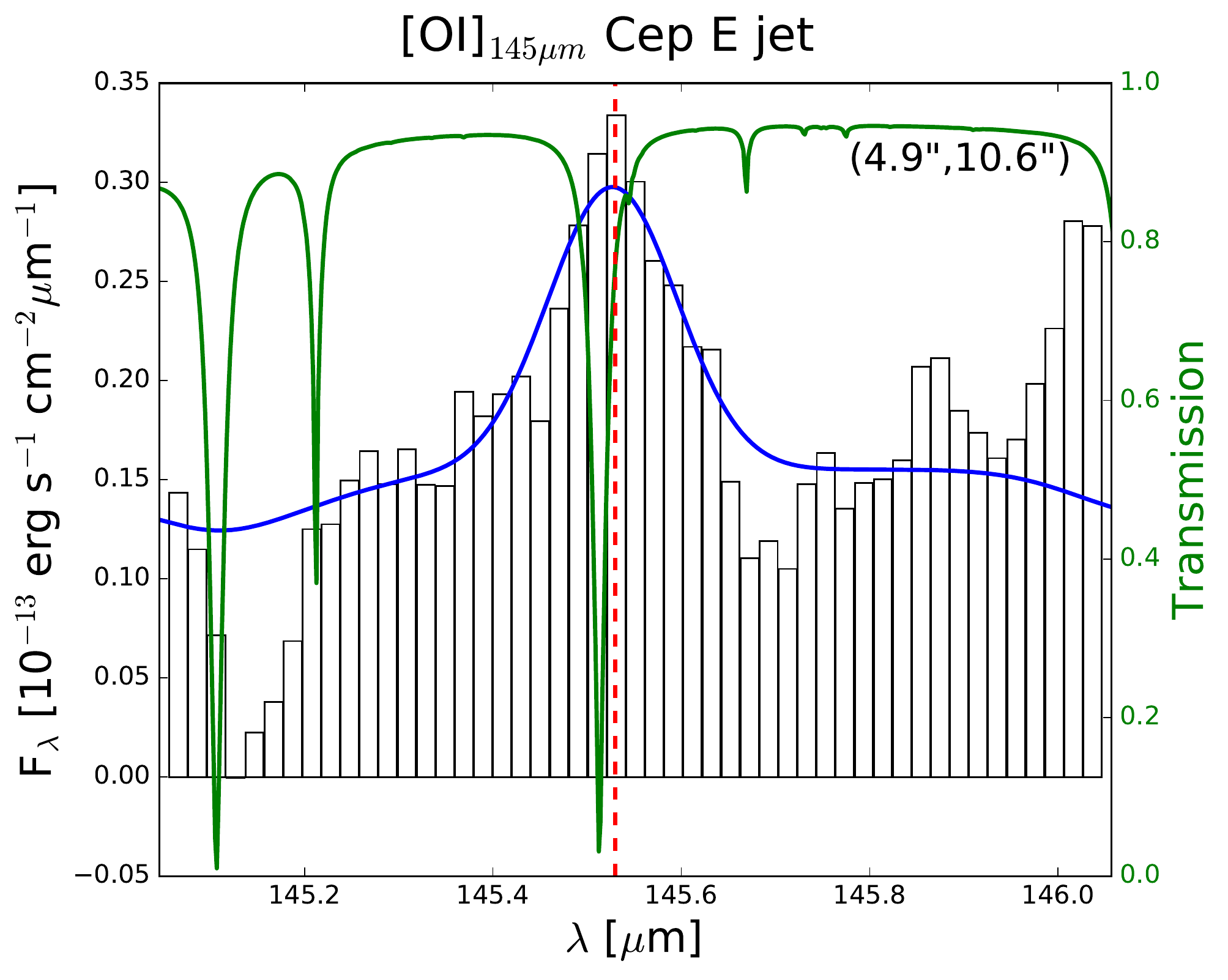}} 
\hfill
\subfloat{\includegraphics[trim=0 0 0 0, clip, width=0.33\textwidth]{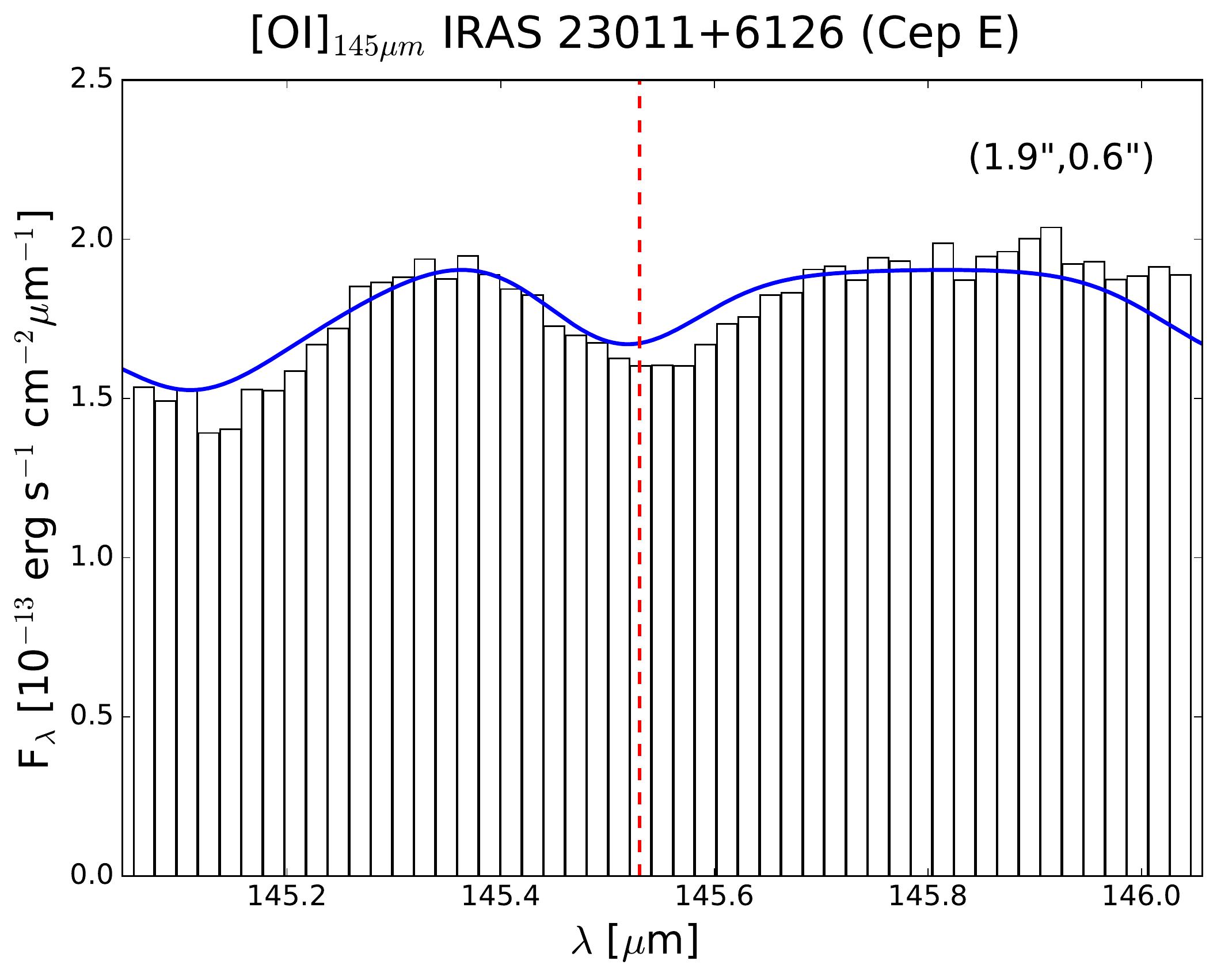}}
\hfill
\subfloat{\includegraphics[trim=0 0 0 0, clip, width=0.33\textwidth]{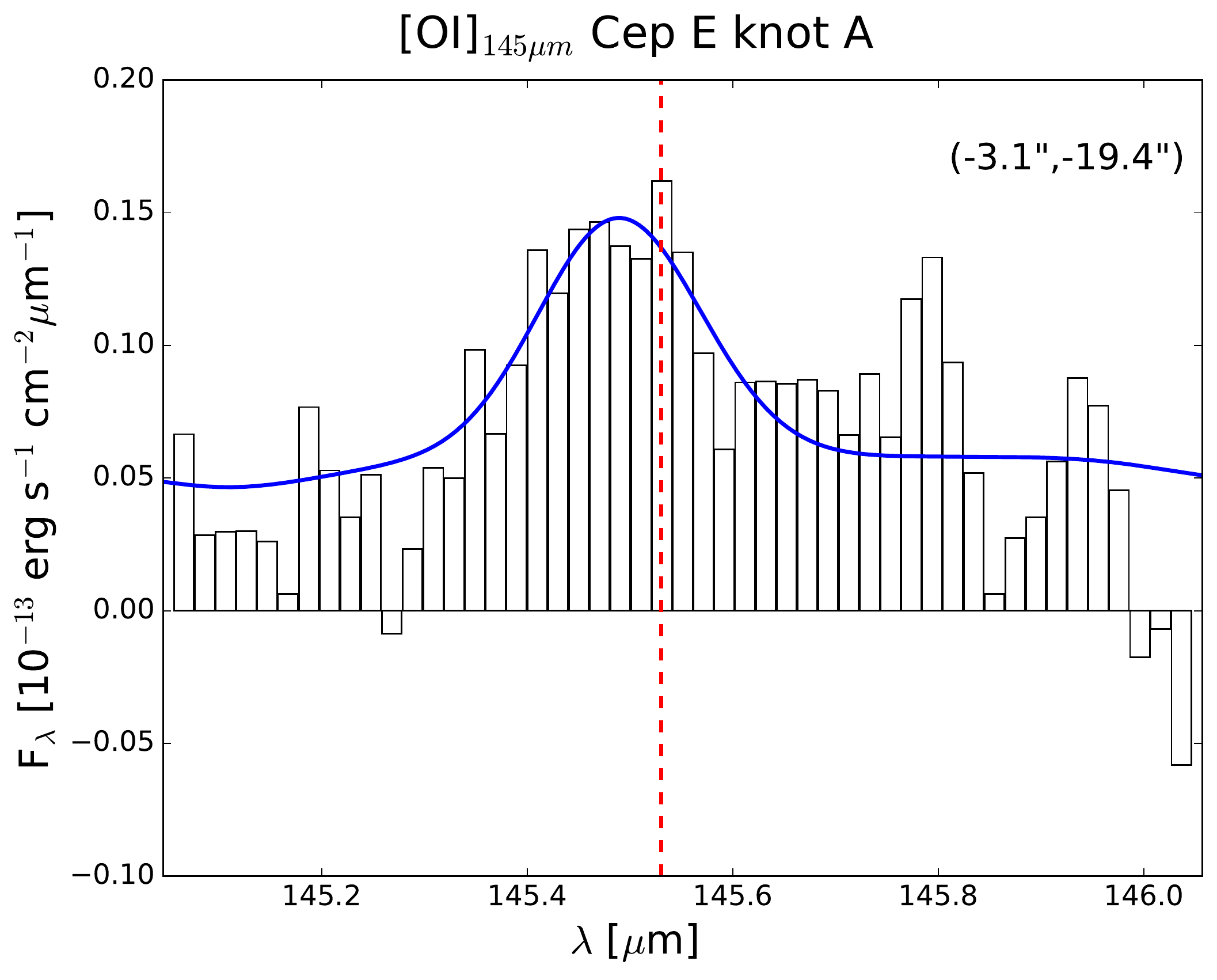}}
\caption{\small{Sample spectra of the detected [O\,I]$_{63,145}$ lines at different regions of Cep\,E. Green lines display the atmospheric transmission $\tau$ in the spectral regions around both relevant far-infrared [O\,I] transitions. The fitted model function $y$ (Eq.\,\ref{equ:S}) is plotted as a blue line in each spaxel. Vertical dashed red lines mark the position of the rest wavelengths of the [O\,I]$_{63, 145}$  emission lines. Numbers in parentheses in the upper right corner indicate the individual spaxel position with respect to the driving source  (see Fig.\,\ref{fig:emission_cepE_hh1}), that is the offset in right ascension and declination in units of arcseconds $(\Delta\text{RA}[\arcsec],\Delta\text{DEC}[\arcsec])$.}}\label{fig:all_minispectra_A}
\end{figure*} 

\begin{figure*} 
\centering
\subfloat{\includegraphics[trim=0 0 0 0, clip, width=0.49\textwidth]
{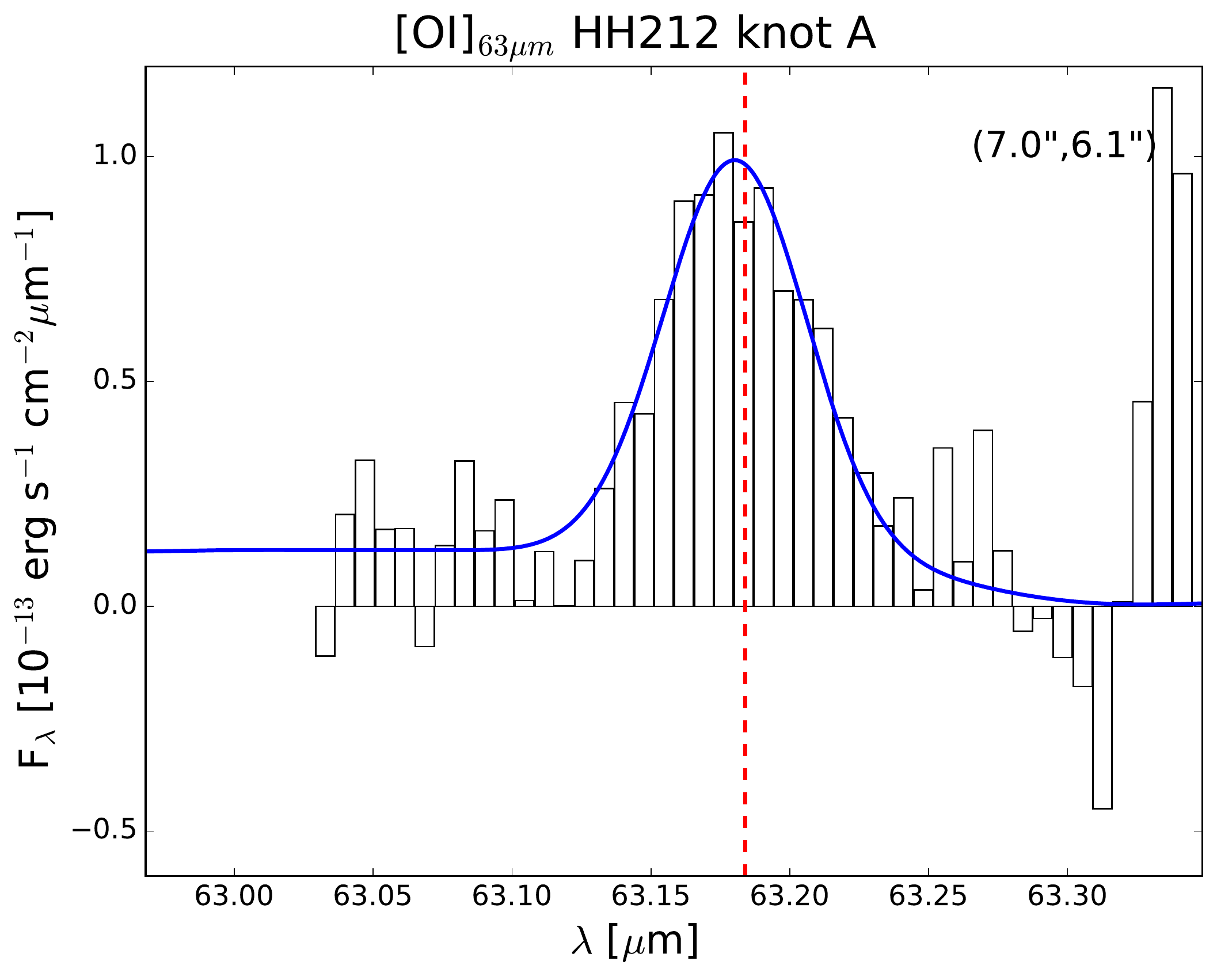}}
\hfill
\subfloat{\includegraphics[trim=0 0 0 0, clip, width=0.49\textwidth]{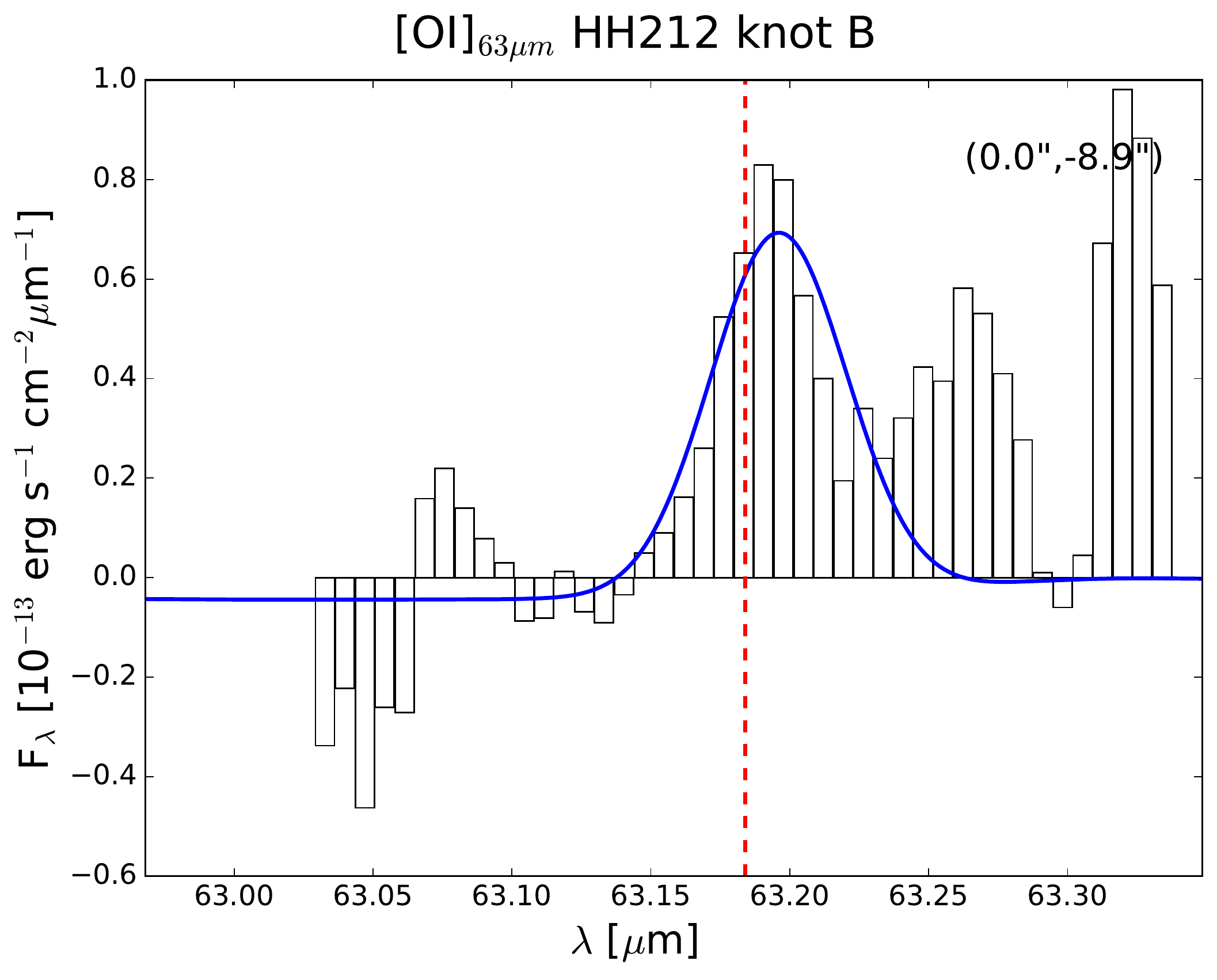}}
\caption{\small{Same as in Figure\,\ref{fig:all_minispectra_A}, but for HH\,212.}}\label{fig:all_minispectra_C}
\end{figure*} 

\begin{figure*} 
\centering
\subfloat{\includegraphics[trim=0 0 0 0, clip, width=0.49\textwidth]{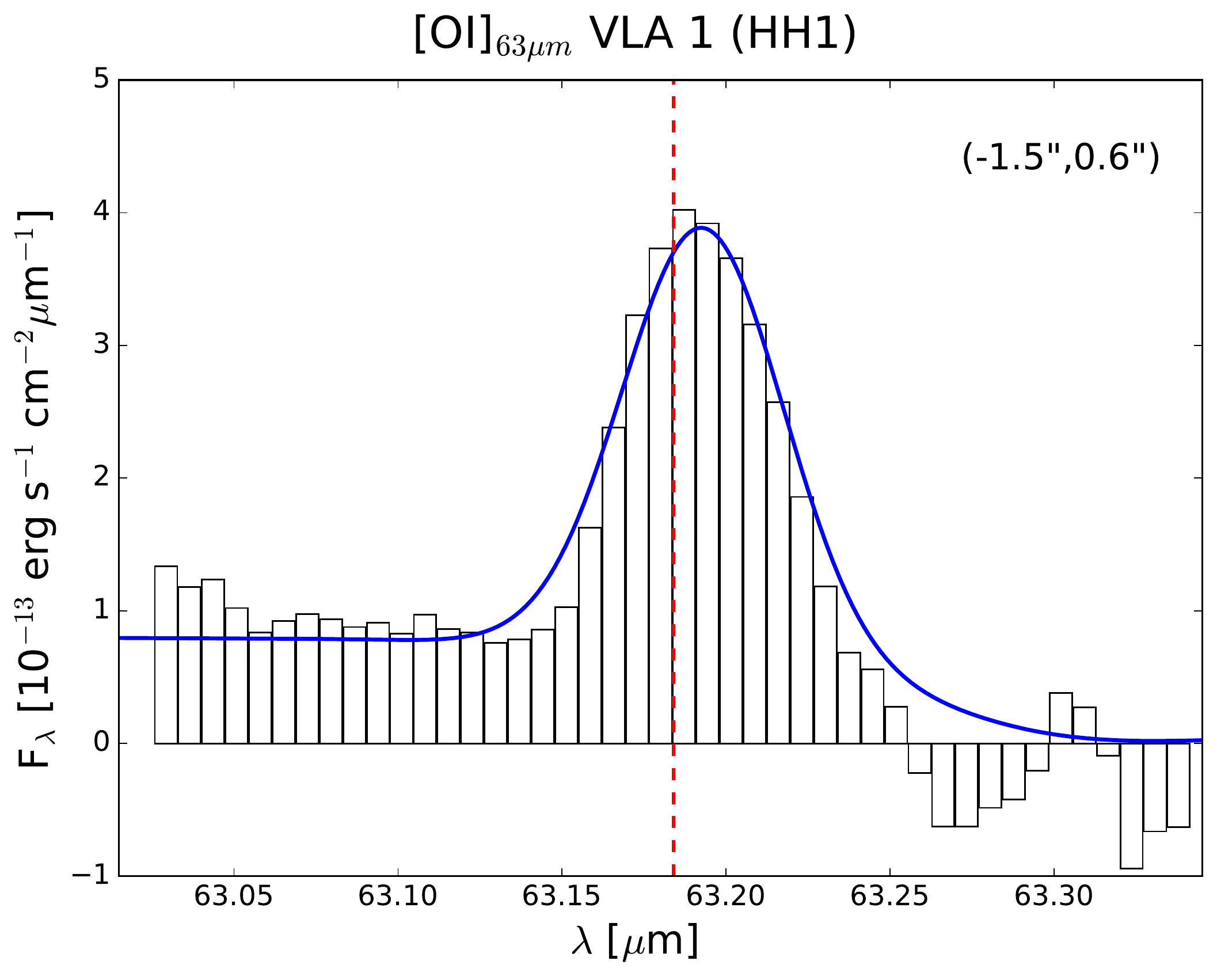}}
\hfill
\subfloat{\includegraphics[trim=0 0 0 0, clip, width=0.49\textwidth]{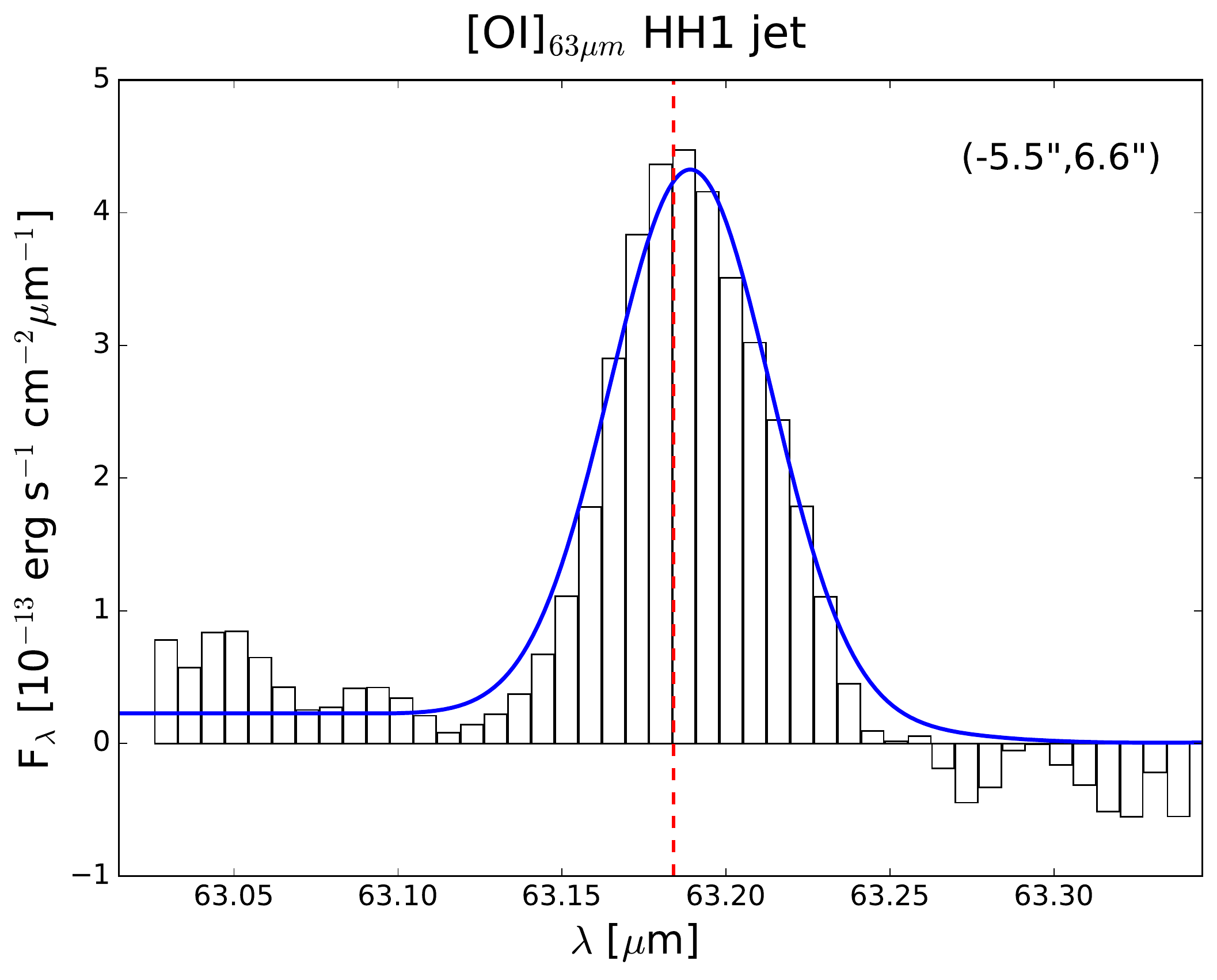}}
\hfill
\subfloat{\includegraphics[trim=0 0 0 0, clip, width=0.49\textwidth]{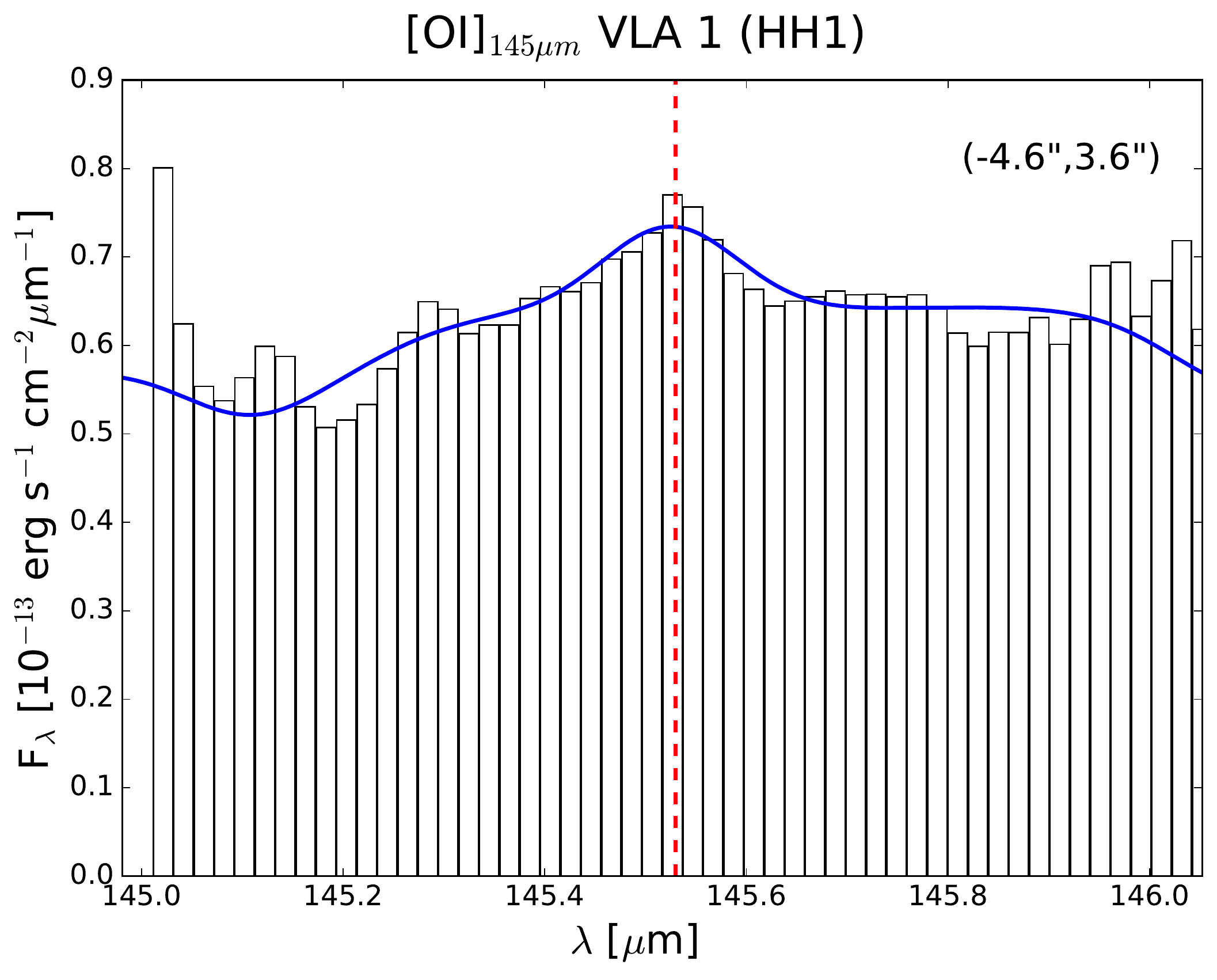}}
\hfill
\subfloat{\includegraphics[trim=0 0 0 0, clip, width=0.49\textwidth]
{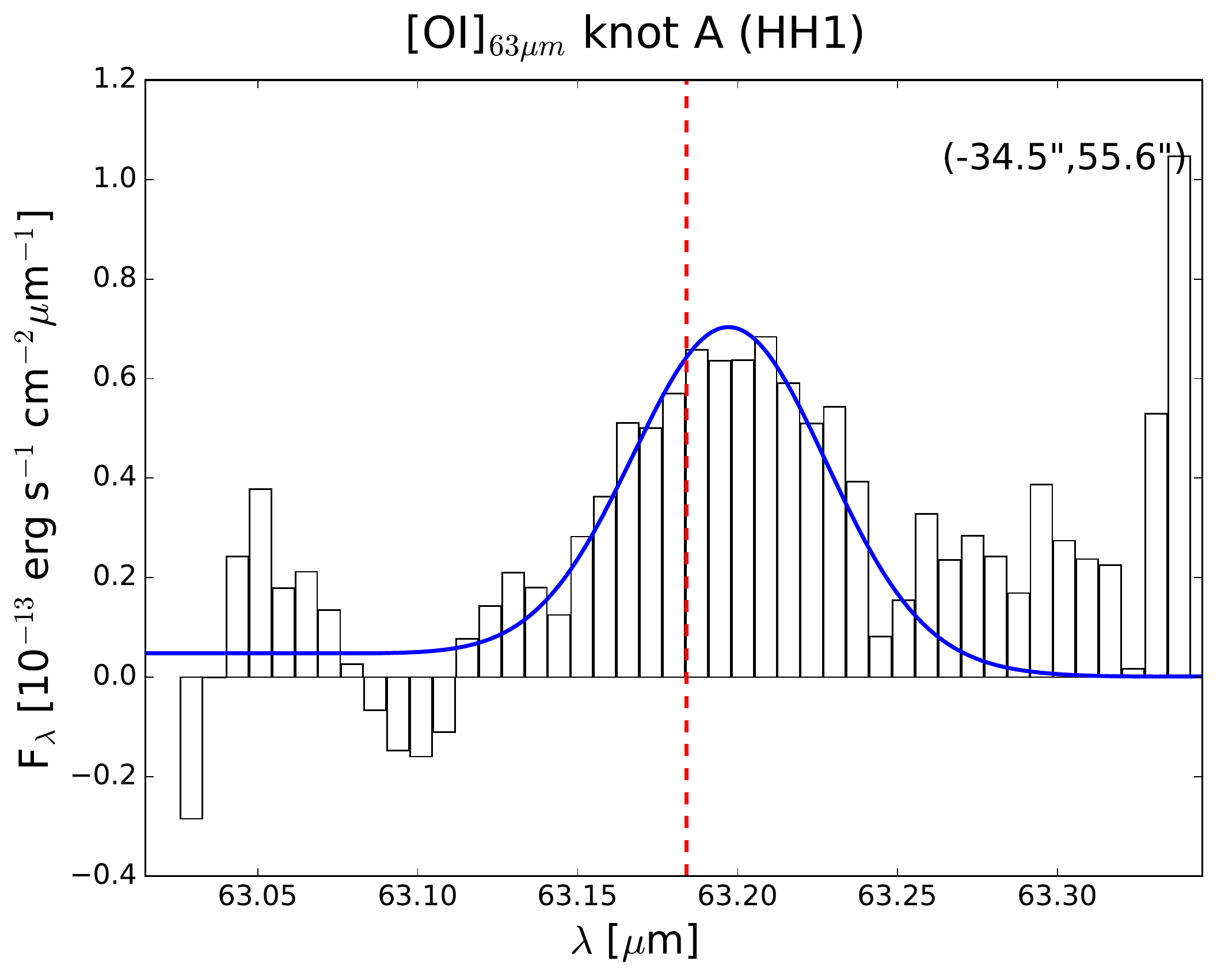}}
\caption{\small{Same as in Figure\,\ref{fig:all_minispectra_A}, but for HH\,1.}}\label{fig:all_minispectra_B}
\end{figure*} 

\begin{figure*} 
\centering

\subfloat{\includegraphics[trim=0 0 0 0, clip, width=0.49\textwidth]{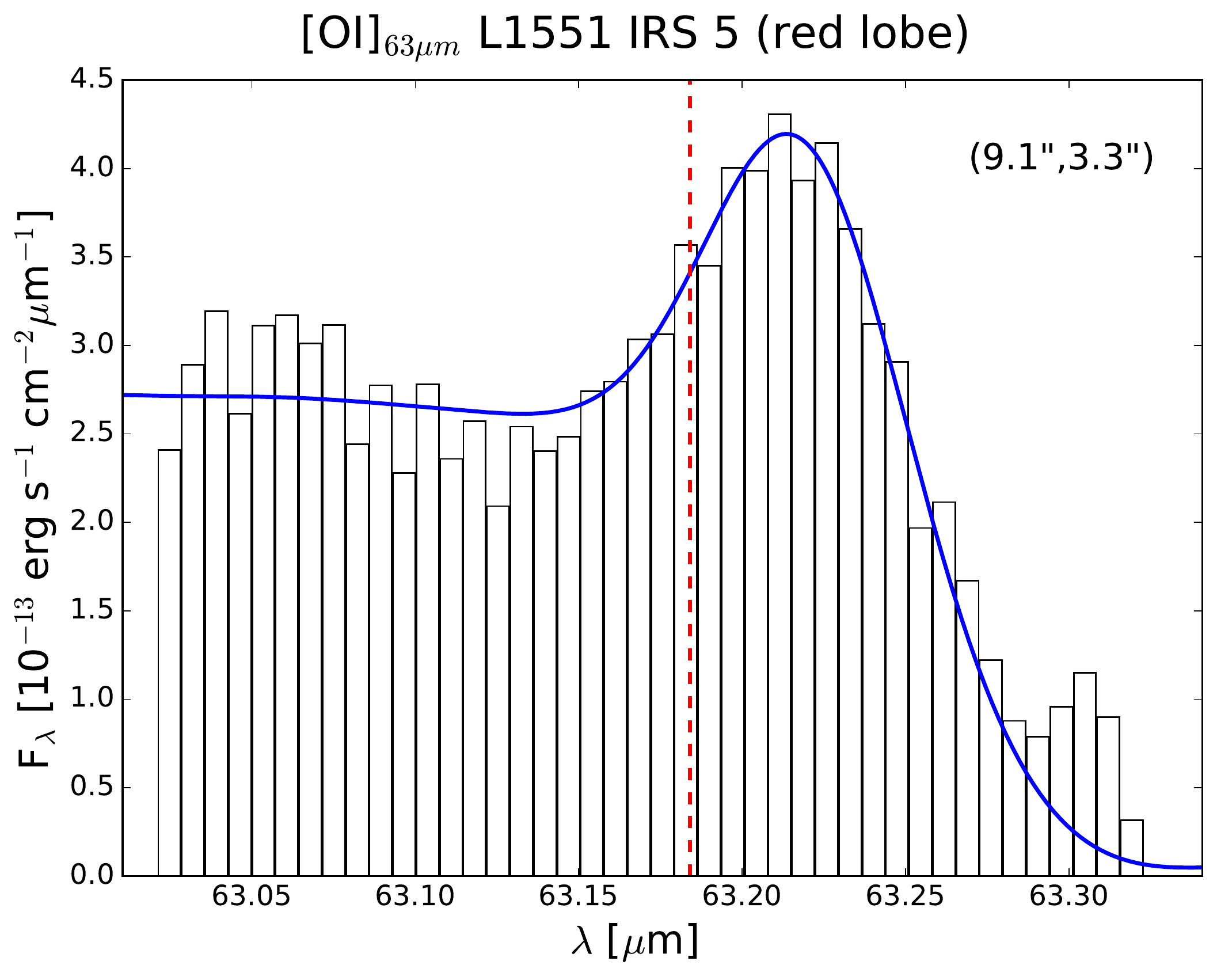}}
\hfill  
\subfloat{\includegraphics[trim=0 0 0 0, clip, width=0.49\textwidth]{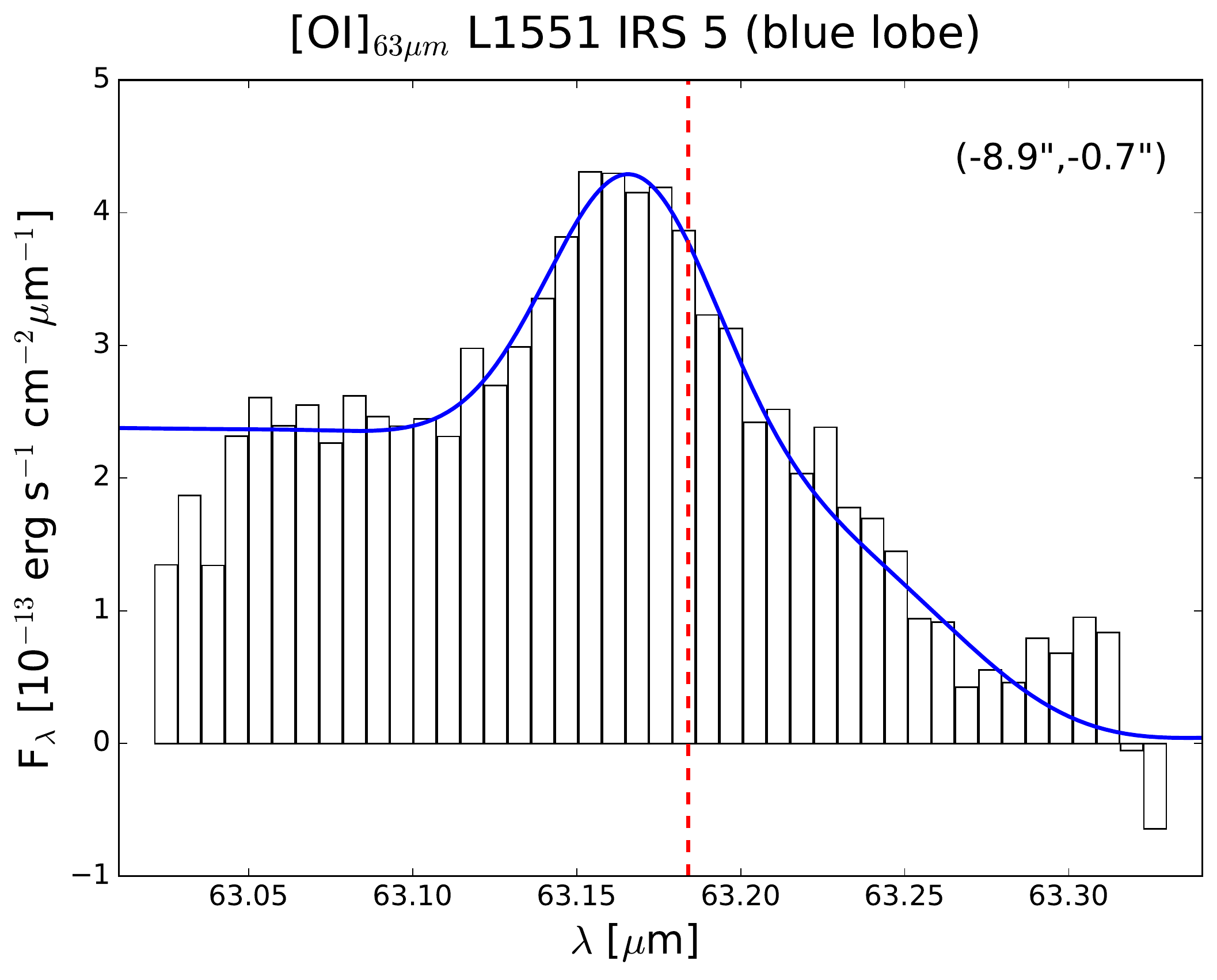}}
\hfill  
\caption{\small{Same as in Figure\,\ref{fig:all_minispectra_A}, but for L1551.}}\label{fig:all_minispectra_D}
\end{figure*} 
   
\section{Results}

{\renewcommand{\arraystretch}{1.2}
\begin{table}
\caption{\small{Measured background-corrected continuum fluxes of our objects (flux in units:  $10^{-13}\,\text{erg}\,\text{s}^{-1}\,\text{cm}^{-2}\,\upmu\text{m}^{-1}$).}}\label{table:continuum_fluxes}
\centering
\begin{tabular}{c  c c}
\hline\hline
Obj.      & $F_{63\upmu\text{m}} \pm\Delta F_{63\upmu\text{m}}$ & $F_{145\upmu\text{m}} \pm\Delta F_{145\upmu\text{m}}$ \\[1.5pt]
\hline  
\small{Cep\,E} & $ 361.4\pm 9.4$ & $ 170.3 \pm 1.7$ \\[1.5pt]
\small{HH\,1}   & $  127.7\pm 14.1 $ & $  117.8\pm 1.1 $ \\[1.5pt] 
\small{HH\,1 radio source}   & $   - $ & $ 223.2 \pm  3.0 $\tablefootmark{a} \\[1.5pt] 
\small{HH\,212} & $  95.3\pm  8.3$ & $ 81.3 \pm  1.5$ \\[1.5pt]
\small{L1551} & $  1857.7\pm  24.5$ & $ 572.4 \pm  7.0 $ \\[1.5pt]
\hline\hline
\end{tabular}
\tablefoot{ 
\tablefoottext{a}{\small{The detected radio source is located at the edge of the obtained continuum map and thus no aperture photometry could be done. The stated value is the integrated flux within the aperture of 1.5\,$\sigma_\text{s}$ of the fitted 2D Gaussian.}}
}
\end{table}

{\renewcommand{\arraystretch}{1.2}
\begin{table*} 
\caption{\small{Relevant information on the aperture boxes, which  are sketched in the continuum-subtracted [O\,I]$_{63}$ emission maps (Fig.\,\ref{fig:emission_cepE_hh1}). Fluxes measured within these boxes are listed in units of $10^{-13}\,\text{erg}\,\text{s}^{-1}\,\text{cm}^{-2}$.}}\label{table:main_results}
\centering
\begin{tabular}{c c c c c c |c}
\hline\hline
   &  &  & [O\,I]$_{63\upmu\text{m}}$ &     &    [O\,I]$_{145\upmu\text{m}}$  &    \\ \cmidrule{4-6}
Obj. & Region   & Size & $F_{63\upmu\text{m}}$  & $L\left(\text{[O\,I]}_{63}\right)/L_\odot$   &   $F_{145\upmu\text{m}}$ & [O\,I]$_{63}$ origin (see Appendix)\\[1.5pt]
\hline  
\small{Cep\,E} 
              & \tiny{knot A}   & \tiny{25\arcsec $\times$ 23\arcsec}  & $78.38 \pm 3.69$ & $ 13.06 \pm 0.62  \times 10^{-2}$  &   $< 7.1$\tablefootmark{a} &  \tiny{PDR + partly dissociative J-shocks}\\[1.5pt]
              & \tiny{jet}     & \tiny{35\arcsec $\times$ 23\arcsec}  & $55.99\pm 3.43$  & $ 9.33 \pm 0.57  \times 10^{-2}$  &   $2.39\pm 0.82$ & \tiny{PDR + a few  internal shocks} \\[1.5pt]
                               &   &    &   &    &     & \tiny{(mostly non-dissociative)} \\[1.5pt]
              \hline  
\small{HH\,1}   & \tiny{VLA\,1}    & \tiny{50\arcsec $\times$ 25\arcsec}  & $97.50\pm 2.59$  & $5.63 \pm 0.15 \times 10^{-2}$  &  $<56.3$\tablefootmark{a}  & \tiny{PDR + dissociative J-shocks} \\ [1.5pt]
              & \tiny{knot A}   & \tiny{20\arcsec $\times$ 20\arcsec}  & $10.74\pm 1.79$  & $6.21 \pm 1.04 \times 10^{-3}$  & $<3.0$\tablefootmark{a} & \tiny{PDR + cooling zone behind bow shocks}\\[1.5pt]
               \hline  
\small{HH\,212} & \tiny{knot A and B}   & \tiny{30\arcsec $\times$ 15\arcsec}  & $17.90\pm 3.47$  &  $9.87\pm 1.92 \times 10^{-3}$  &    $<6.6$\tablefootmark{a} & \tiny{dissociative J-shocks}\\ [1.5pt]
 \hline  
\small{L1551} & \tiny{IRS 5}   &  \tiny{25\arcsec $\times$ 18\arcsec} & $83.70\pm 4.52$   & $5.13\pm 0.28\times 10^{-3}$  & $< 18.7$\tablefootmark{a}   &  \tiny{dissociative J-shocks, disc winds, and deflections} \\ [1.5pt]
\hline\hline
\end{tabular}
\tablefoot{
\tablefoottext{a}{\small{The listed value corresponds to the $3\,\sigma$ upper limit.}}
}
\end{table*}

{\renewcommand{\arraystretch}{1.2}
\begin{table*}
\caption{\small{Mass-loss rates of the observed targets. Column 5: Derived from jet geometry (i.e. Eq.\,\ref{equ:sperling_2020_formula}). Column 6: Derived from the \citet{hollenbach_1989} shock model, that is Eq.\,\ref{equ:HM89_formula}. Column 7: Mass-loss rates from the literature that are connected to other outflow components, meaning they are traced by species other than [O\,I] as indicated. Mass accretion rates in Column 8 are calculated from Eq.\,\ref{equ:accretion_rates} unless stated otherwise.}}\label{table:main_results_II}
\centering
\begin{tabular}{c c c c c c || c |c}
\hline\hline
Obj. & Region  & $\theta$ & $v_\text{t}$ &  $\dot{M}_\text{out}^\text{lum}(\text{[O\,I]})$  & $\dot{M}_\text{out}^\text{shock}(\text{[O\,I]})$   & $\dot{M}_\text{out}^\text{other}$ \tiny{\& component} &   $\dot{M}_\text{acc}$\tablefootmark{e}\\[1.5pt]
            &     & ('')     & (km\,$\text{s}^{-1}$)       & $(10^{-7}\,M_\odot\,\text{yr}^{-1})$   & $(10^{-7}\,M_\odot\,\text{yr}^{-1})$             & $(10^{-7}\,M_\odot\,\text{yr}^{-1})$ & $(10^{-7}\,M_\odot\,\text{yr}^{-1})$      \\[1.5pt]
\hline  
\small{Cep E} & \tiny{knot A}      & 25  & 95\tablefootmark{a}  &  $\lesssim 22.4-45.5$ &  $\lesssim 124.4-136.7$ & $\sim 200$\tablefootmark{j} \tiny{\& CO}   & $170$\tablefootmark{f}   \\[3pt]
              & \tiny{jet}         & 35  &  60\tablefootmark{a} &  $\lesssim 7.2-14.7$ &  $\lesssim 87.5-99.0$     &   & \\[3pt]
\hline
\small{HH\,1} & \tiny{VLA\,1}     &  50 &  300\tablefootmark{b} &  $\lesssim 25.9- 52.6$   &  $\lesssim 54.9-57.8$  & $\sim 6$\tablefootmark{k} \tiny{\& [Fe\,II]} & $290$\tablefootmark{g}\\ [1.5pt]
            & \tiny{knot A}     &  20 & 400\tablefootmark{b}  & $\lesssim 9.5-19.4$  &  $\lesssim 5.2-7.2$    & $\sim 4$\tablefootmark{k} \tiny{\&  [S\,II]}  &  \\[3pt]
                            &   &    &    &   &     & $\sim 0.1$\tablefootmark{k} \tiny{\& H$_{2}$}&  \\[3pt]
                                     &  &    &   &   &   & $\sim 15$\tablefootmark{l} \tiny{\& CO} &  \\ [1.5pt]
            \hline
\small{HH\,212} & \tiny{knots A and B}       & 30  & $150$\tablefootmark{c}  &  $3.9-7.9$   & $8.0-11.8$ & $\sim 10$\tablefootmark{m} \tiny{\& CO, SO, SiO} & $80$\tablefootmark{h}\\ [1.5pt]
      &  &    &   &   &   & $\leq 3$\tablefootmark{n} \tiny{\& CO, SiO} &  \\ [1.5pt]
      &  &    &   &   &   & $\sim 1$\tablefootmark{o} \tiny{\& H$_2$} &  \\ [1.5pt]
\hline
\small{L1551} & \tiny{IRS 5}      & 25   & $120$\tablefootmark{d}  &  $5.8-11.8$ & $4.9-5.4$  & $\gtrsim  100$\tablefootmark{p}\tiny{\& CO, HCO$^+$, swept-up gas}   &  $30$\tablefootmark{i} \\ [1.5pt]
        &  &    &    &  &  & $\sim 8.6$\tablefootmark{q} \tiny{\& HI} &    \\ [1.5pt]
          &  &    &    &  &  & $\sim 1.7$\tablefootmark{r} \tiny{\& [Fe\,II]} &    \\ [1.5pt]
            &  &    &    &  &  & $\sim 0.4$\tablefootmark{r} \tiny{\& H$_2$} &    \\ [1.5pt]
\hline\hline
\end{tabular}
\tablefoot{  
\tablefoottext{a}{\small{Proper motions from \citet{noriega_crespo_2013}.}}
\tablefoottext{b}{\small{Proper motions from \citet{bally_2002_hh1_2}.}}
\tablefoottext{c}{\small{Proper motions from \citet{noriega_crespo_2020}.}}
\tablefoottext{d}{\small{Proper motions from \citet{fridlund_1994}.}}
\tablefoottext{e}{\small{Calculated from their bolometric luminosities (see Section\,\ref{sec:accretion_rates}).}}
\tablefoottext{f}{\small{Our estimate (see Section\,\ref{sec:accretion_rates}). From SED modeling, \citet{velusamy_2011} estimated  a surprisingly high  accretion rate for Cep\,E-mm of the order of $10^{-4}M_\odot\,\text{yr}^{-1}.$}}
\tablefoottext{g}{\small{The stated value is in good agreement with \citet{fischer_2010}.  Therein, the source of the HH\,1-2 outflow is labelled HOPS 203.}}
\tablefoottext{h}{\small{The stated value is in good agreement with \citet{lee_2014}.}}
\tablefoottext{i}{\small{The stated value is in good agreement with, for example, \citet{osorio_2003}, \citet{liseau_2005}, and \citet{gramajo_2007}.}}
\tablefoottext{j}{\small{\citet{lefloch_2015}: This material is directly connected to the  Cep\,E jet and not to swept-up gas.}}
\tablefoottext{k}{\small{\citet{nisini_2005}.}}
\tablefoottext{l}{\small{From CO J=1--0 observations undertaken by \citet{tanabe_2019}, it is unclear if it is entrained material or not.}}
\tablefoottext{m}{\small{\citet{lee_2007, podio_2015, lee_2015_hh212jet};  \citet{lee_2020_molecular_review}.}}
\tablefoottext{n}{\small{\citet{cabrit_2012}.}}
\tablefoottext{o}{\small{\citet{davis_2000}.}}
\tablefoottext{p}{\small{\citet{hogerheijde_1998, fridlund_2002, yildiz_2015}.}}
\tablefoottext{q}{\small{\citet{giovanardi_2000}.}}
\tablefoottext{r}{\small{\citet{davis_2003}.}}
}
\end{table*}

\subsection{Continuum sources}\label{sec:continuum_sources}

The obtained continuum maps of HH\,1, HH\,212, Cep\,E, and L1551 are presented in Appendix A (Fig.\,\ref{fig:continuum_maps}). With regard to Cep E, the continuum source IRAS 23 011+6126 \citep[also Cep\,E-mm,][]{lefloch_1996, chini_2001}, which drives the powerful   Cep\,E outflow \citep[e.g.][]{eisloeffel_1996, ayala_2000, lefloch_2015}, is detected in both FIFI-LS channels. Millimetre observations indicate that IRAS 23011+6126 is a system of two protostars (separation $\sim 2\arcsec$), namely Cep\,E-A and Cep\,E-B \citep{ospina_zamudio_2018}, which appear as one continuum source in our maps.  
After subtracting this continuum source, a possible second source becomes visible at location $(\alpha, \delta)_{\text{J}2000} =$(23$^\text{h}$03$^\text{m}$08$\fs$7, 61$\degr$42$\arcmin$  $38\farcs 0$) at 145\,$\upmu$m.

\noindent
The HH\,1/2 complex is among the brightest HH objects in the sky and has been studied  extensively \citep[e.g. see review in][]{raga_2011}. A central source at the location of VLA 1 \citep{pravdo_1985, rodriguez_2000} is detected prominently in the red channel but very faintly in the blue channel. Inbetween the HH\,1 jet and the bow-shock region HH\,1 lies the unrelated Cohen-Schwartz star \citep{cohen_1979}, which is not seen in our maps. In the red channel, the known radio source is detected at $(\alpha, \delta)_{\text{J}2000} =$(5$^\text{h}$36$^\text{m}$18$\fs$9, $-$06$\degr$45$\arcmin$26$\arcsec$), \citet{pravdo_1985}.
  
\noindent
The HH\,212 outflow in L1630 in Orion is driven by a Class 0 protostar IRAS 05413--0104 \citep{froebrich_2005}.  We detect one continuum source in the 63\,$\upmu$m and 145\,$\upmu$m maps located at $(\alpha, \delta)_{\text{J}2000} =$(5$^\text{h}$43$^\text{m}$51$\fs$2, $-$1$\degr$02$\arcmin$ $56\farcs 2$) and $(\alpha, \delta)_{\text{J}2000} =$(5$^\text{h}$43$^\text{m}$51$\fs$3, $-$1$\degr$02$\arcmin$ $53\farcs 8$), respectively; hence, their declinations differ by about $2\farcs 5$.     
 However, since the mentioned offset is within the accuracy of the SOFIA positions, we identify both detected sources as the HH\,212 driving source  IRAS 05413--0104. The refined coordinates of the associated VLA source \citep{galvan_madrid_2004, codella_2007} are more in line with the position of the detected 145\,$\upmu$m source. Additionally, \citet{chen_2013} detected three close by continuum sources (HH\,212 MMS1, MMS2, MMS3) at IRAS 05413--0104. This triple system is not resolved in our maps, since their mutual separation is smaller than $1\arcsec$.    About 7$\arcsec$ away from IRAS 05413--0104, a further source was detected in the near- and mid-infrared by  \citet{reipurth_2019}: IRS-B, which is located at $(\alpha, \delta)_\text{J2000} =$ (05$^\text{h}$43$^\text{m}$51$\fs 2$, $-$01$\degr$02$\arcmin$47$\arcsec$). This source is not detected in our continuum and [O\,I] maps.

\noindent
In the dark molecular cloud L1551,  the infrared source IRS5 \citep{strom_1976} drives powerful bipolar outflows that are associated with various Herbig-Haro objects \citep[e.g.][]{fridlund_1994, devine_1999, hayashi_2009}. We detect a continuum source in both channels. Its coordinates are in agreement with the location of IRS 5 \citep{bally_2003}. L1551 IRS5 is a protostellar binary system with a separation of only $\sim 0\farcs 3$ \citep[e.g.][]{bieging_1985, looney_1997, rodriguez_1986, rodriguez_1998, rodriguez_2003}. Tentatively, a third source was detected by \citet{lim_2006}. At the given spatial resolution in our maps, the sources appear as one continuum source.     

\noindent  
 We fitted a 2D Gaussian to the detected continuum sources to measure the continuum flux within a circular aperture of radius $1.5\,\sigma_{\text{Gauss}}$ \citep{mighell_1999}. In the case of the well-known radio source located at the edge of the $145\,\upmu$m  map of HH\,1, this method could not be applied since parts of the source lie outside the map, and therefore the aperture covers a region of  unspecified   continuum pixels. Instead, we  analytically integrated the fitted 2D Gaussian within an aperture of radius $1.5\,\sigma_{\text{Gauss}}$ to estimate the continuum flux. All continuum fluxes are reported in Table \ref{table:continuum_fluxes}. The listed values are in line with expected values from the SIMBAD database.  
 
\noindent  
The observed FWHM  of the continuum sources are $10-13\arcsec$ in the blue channel and $ 20-22\arcsec$ in the red channel. We therefore conclude that all observed sources are extended.

\begin{figure*}   
\centering
\subfloat{\includegraphics[width=0.470\textwidth]{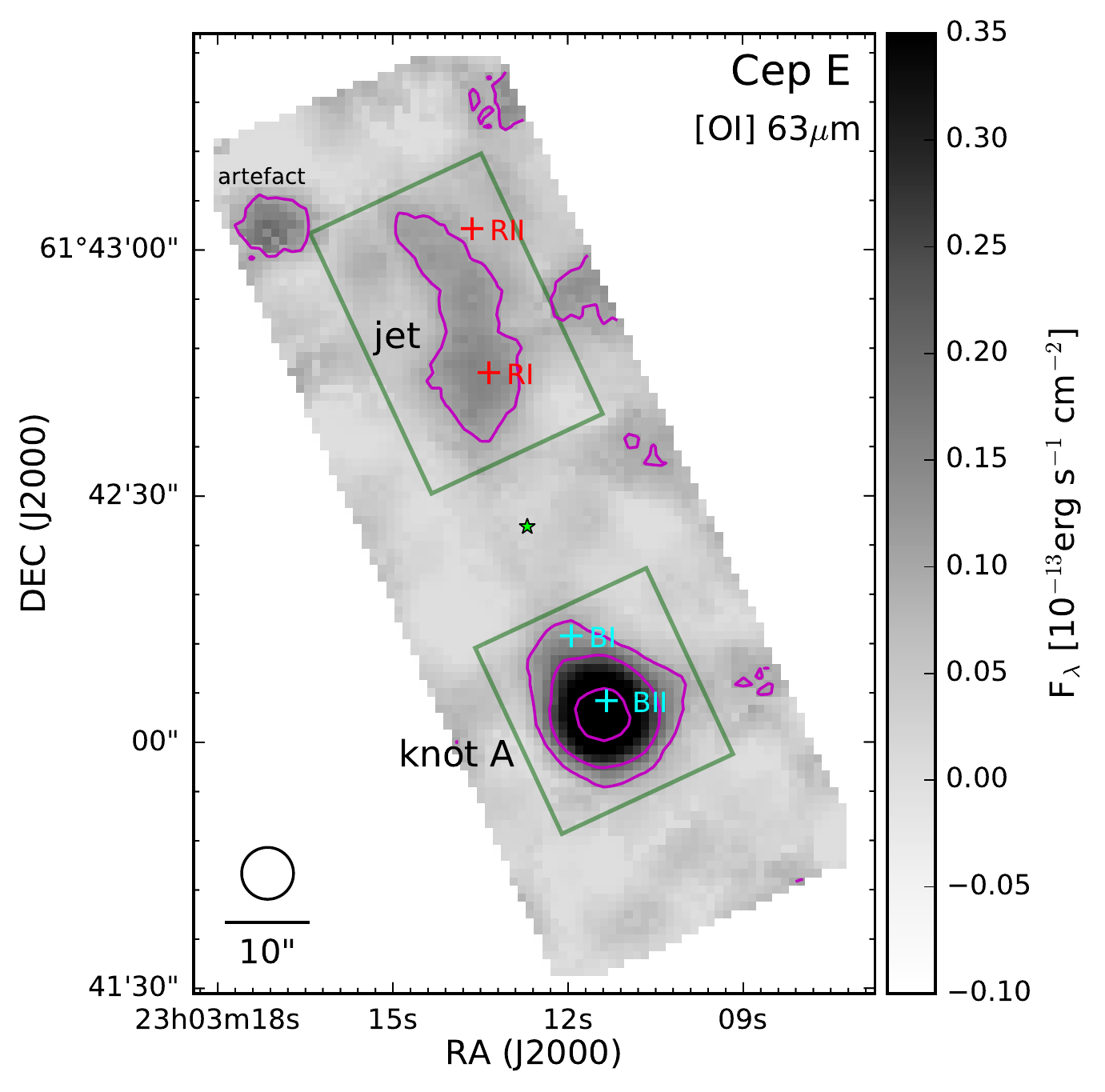}} 
\hfill
\subfloat{\includegraphics[width=0.470\textwidth]{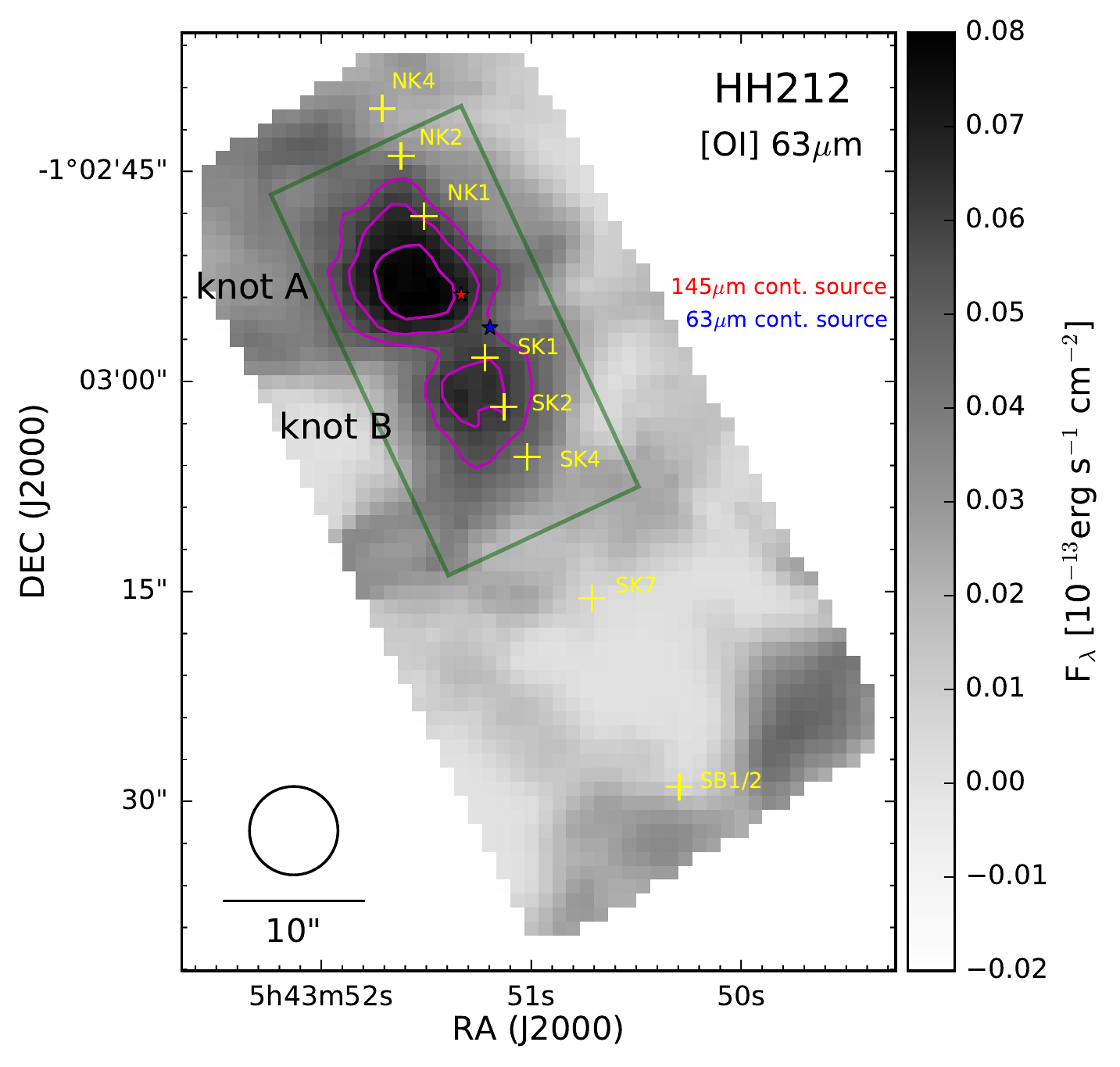}} 
\hfill
\subfloat{\includegraphics[width=0.47\textwidth]{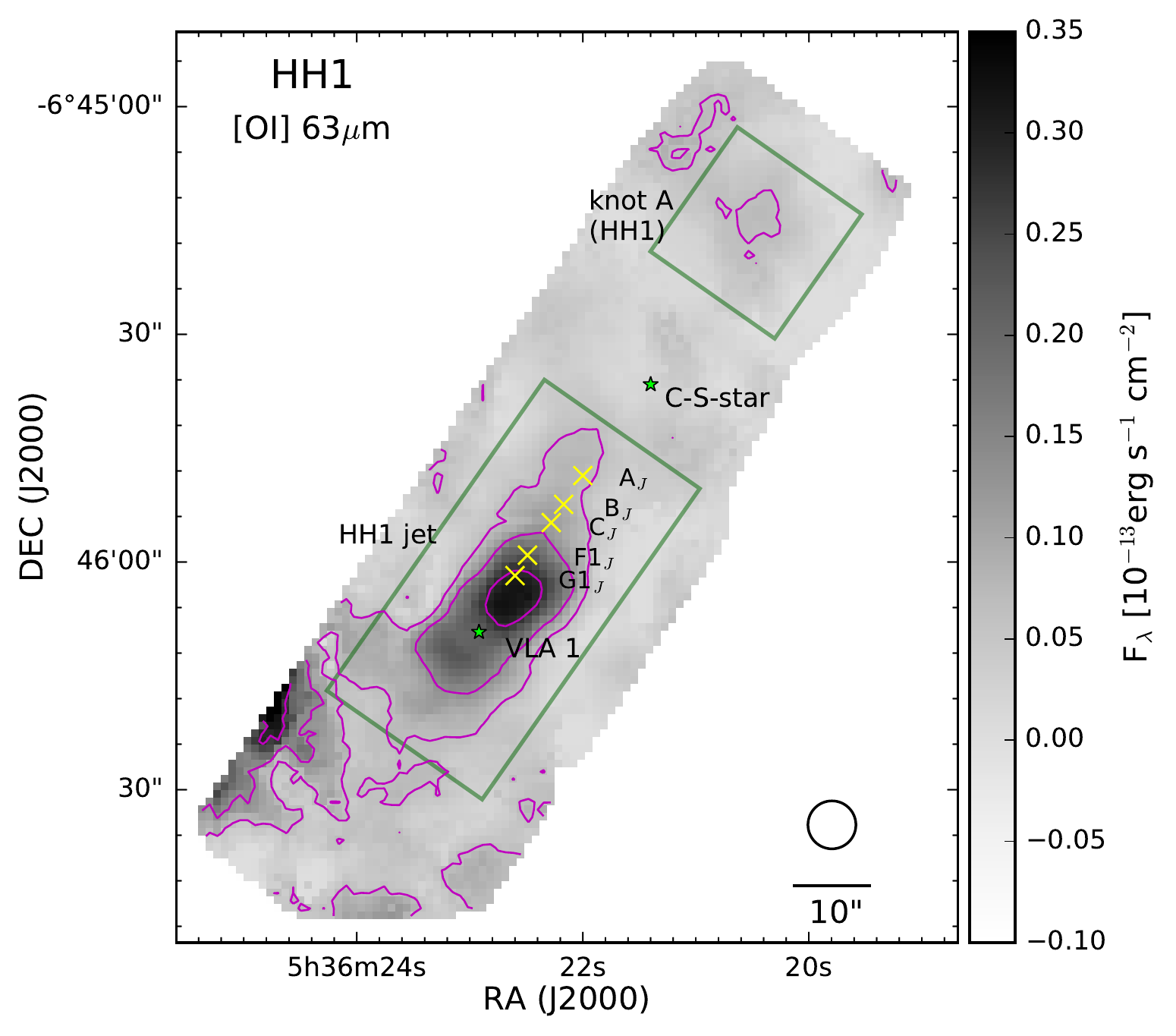}} 
\hfill
\subfloat{\includegraphics[width=0.47\textwidth]{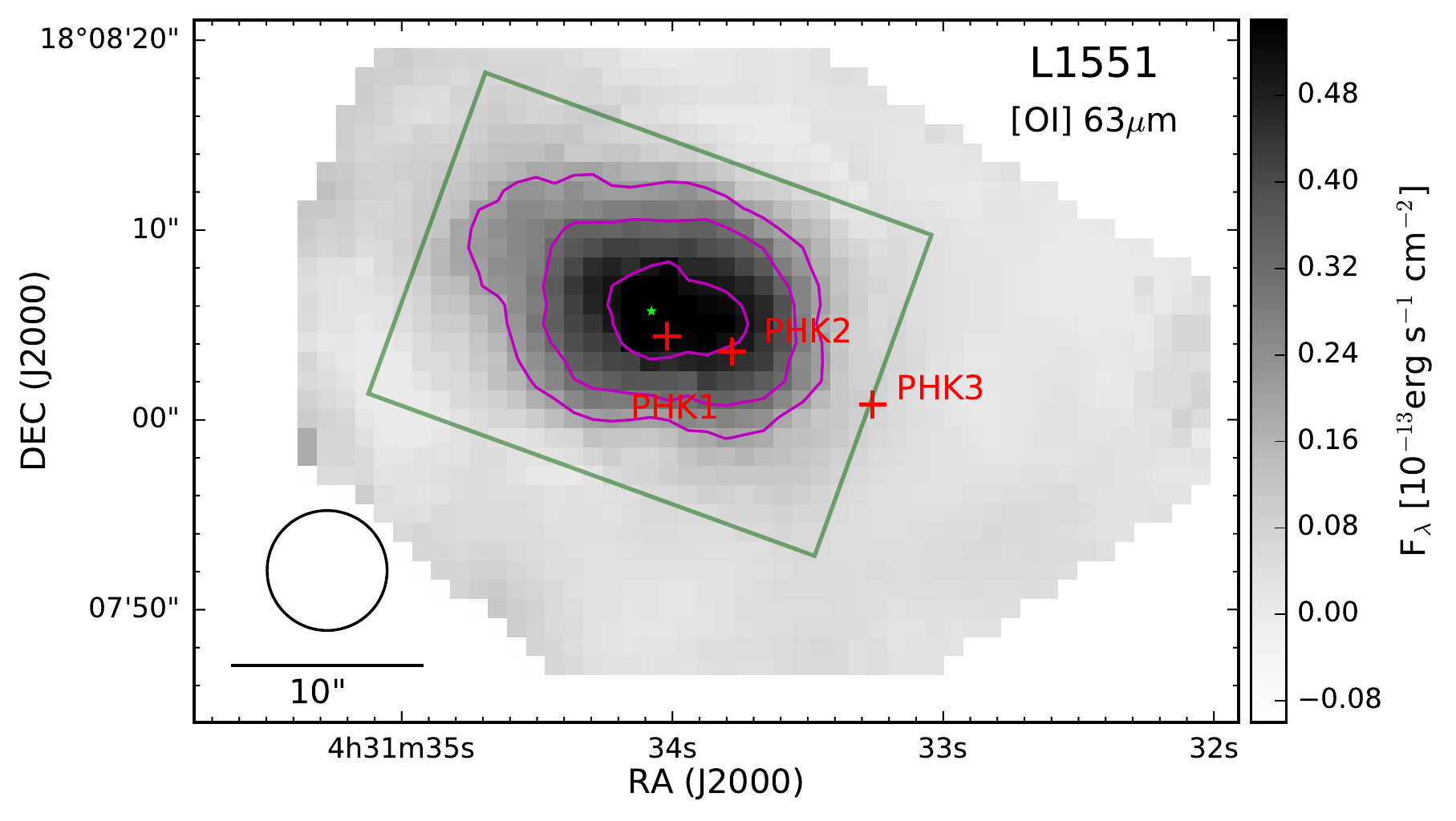}}
\caption{\small{Continuum-subtracted [O\,I]$_{63}$ maps of Cep\,E, HH\,1, HH\,212, and L1551\,IRS5. The black circle shows the FWHM spatial beam size in the blue channel of the FIFI-LS instrument. The light green stars indicate the position of the respective  jet-driving  source.  Schematic views are presented in Fig.\,\ref{fig:schematicss} of the Appendix. Contour lines are drawn in magenta in logarithmic scale at three intensity levels (IL). \textbf{Cep\,E:} RI, RII, BI, and BII are the infrared knots seen in the Spitzer/IRAC band-two (4.5\,$\upmu$m) image of Cep\,E (see \citet{gusdorf_2017}). IL: $(0.10, 0.21, 0.45)\times 10^{-13}\,\text{erg}\,\text{s}^{-1}\,\text{cm}^{-2}$.  \textbf{HH\,1:} Yellow crosses indicate the positions of a few selected optical knots \citep{bally_2002_hh1_2}. IL:  $(0.06, 0.13, 0.28)\times 10^{-13}\,\text{erg}\,\text{s}^{-1}\,\text{cm}^{-2}$. \textbf{HH\,212:} Yellow crosses indicate the positions of the prominent H$_2$ knots \citep{zinnecker_1998}. IL: $(0.053, 0.062, 0.074)\times 10^{-13}\,\text{erg}\,\text{s}^{-1}\,\text{cm}^{-2}$. \textbf{L1551:} The brightest near-infrared [Fe\,II] knots (PHK1, PHK2, PHK3) are marked as red crosses \citep{pyo_2002}. IL: $(0.20, 0.32, 0.50)\times 10^{-13}\,\text{erg}\,\text{s}^{-1}\,\text{cm}^{-2}$. }}\label{fig:emission_cepE_hh1}
\end{figure*} 

\subsection{[O\,I] Morphology and line fluxes}
  
In this section, we briefly describe the morphology of the obtained [O\,I]$_{63}$ maps of the observed targets (Fig.\,\ref{fig:emission_cepE_hh1}). The [O\,I]$_{145}$ is not presented here since the [O\,I]$_{145}$  line was detected only in a few regions at low signal-to-noise. A detailed analysis of the detected [O\,I]$_{63}$ emission and  schematic views are presented in the Appendix.
  
\noindent  
The continuum-subtracted [O\,I]$_{63}$ map of Cep\,E reveals a bright and extended [O\,I]$_{63}$ emission knot A located about 20$\arcsec$ south of the continuum source IRAS 23011+6126. 
Adopting the knot denotation from \citet{gomez_ruiz_2012}, for example, we find that this emission region coincides with the blueshifted lobe, BII, and partly with BI. On the opposite side to knot A in a north-easterly direction, a prominent 35$\arcsec$ long jet-like structure is detected. This emission region matches with the location of the redshifted lobes RII and RI  fairly well \citep{gusdorf_2017}. We note that the emission region at RI and RII is slightly curved towards the eastern direction, meaning in the opposite direction to that one would expect to see from H$_2$ observations presented in \citet{eisloeffel_1996}.  At the location of  IRAS 23011+6126 no significant [O\,I]$_{63}$ emission is detected. The [O\,I]$_{145}$ line is detected at the Cep\,E jet and knot A in only a few spaxels.

\noindent
The HH\,1 jet \citep[e.g.][]{eisloeffel_1994_hh1, hester_1998} is prominently seen in [O\,I]$_{63}$ and  features one bright emission knot about 5$\arcsec$ north-west of VLA 1. Yellow crosses mark the positions of the prominent optical knots \citep{bally_2002_hh1_2}. We also detect strong [O\,I]$_{63}$ emission at the driving source VLA 1 and faint [O\,I]$_{63}$ emission    about 65$\arcsec$ away from VLA\,1 at HH\,1. No [O\,I]$_{63}$ is detected at the location of the C-S-star. The [O\,I]$_{145}$ is detected in only  a few spaxels at VLA\,1 and the HH\,1 jet. 
 
\noindent 
Serving as orientation, we mark the positions of the knots seen in H$_2$ as yellow crosses \citep{zinnecker_1998}. Two  bright [O\,I]$_{63}$ emission regions (knot A and B) are located  opposite each other with IRAS 05413--0104 roughly in the middle, and together they portray a symmetric constellation.  The emission knot A is brighter and more extended than knot B. Both knots and the continuum source lie on a projected line at P.A. $\sim$ 25$^\text{o}$.  Since the astrometry is only accurate to a few arcseconds and the HH\,212 jet reflects a remarkably high symmetry, we conclude that the [O\,I]$_{63}$ emission of the northern  knot A  coincides with NK1 and the southern knot B with SK1.   No significant [O\,I]$_{63}$ emission is detected along the jet axis towards SK7 and SB1/2. \\
In the [O\,I]$_{145}$  map,  a very bright emission knot C is present a few arcseconds downstream of SB1/2 (see Fig.\,\ref{fig:hh212_red_emission_figure} and  Fig.\,\ref{fig:hh212_red_emission_figure_B} in the Appendix). Unfortunately, this region is at the edge of the [O\,I]$_{63}$ map and thus not seen there.  

\noindent  
We record bright [O\,I]$_{63}$ emission at the detected continuum source. The peak of this extended, blob-like emission is located a few arcseconds from the source position. The [O\,I] emission region appears to be stretched out alongside an axis at P.A. $\sim 250^\text{o}$. This position angle is consistent with the jet directions of the northern and southern jets seen in [Fe\,II]  \citep{pyo_2009} or H$_2$ \citep{davis_2002}. Serving as orientation, we marked the position of the distinctive [Fe\,II] emission knots PHK2 and PHK3, as well as the brightest parts of the northern and southern jets \citep{pyo_2002}. The innermost parts of the two marked jets emerging from IRS5 roughly match with the brightest [O\,I] emission region. At PHK3, almost no [O\,I] emission is detected. North-east of the source, the observed [O\,I]$_{63}$ line is redshifted, and towards south-west, the line is blueshifted (see Fig.\,\ref{fig:all_minispectra_D}). Thus, both lobes of the outflow are prominently detected in [O\,I].\[\]
We determined [O\,I] line fluxes for all relevant outflow regions (green boxes in Fig.\,\ref{fig:emission_cepE_hh1})  using the same method presented in \citet{sperling_2020}. Measured [O\,I] line fluxes and luminosities are listed in Table\,\ref{table:main_results}. We consider the impact of foreground absorption by cold gas in the line of sight to be negligible \citep[see discussion in][]{nisini_2015}. 


\subsection{Mass-loss rates}

 We wish to estimate mass-loss rates from the obtained [O\,I]$_{63}$ maps of the observed outflows. Basically, two distinct physical approaches are worth considering here: a) $\dot{M}_\text{out}$ from the jet luminosity; b) $\dot{M}_\text{out}$ from a shock model. As described in detail in the corresponding Sections, \,\ref{sec:mdot_from_collisional_diagram} and \ref{sec:mdot_from_shock_model}, both methods are based on different model assumptions. Common to both is the underlying premise that any PDR \citep{goldsmith_2019} or disc \citep{gorti_2008} contribution to the observed [O\,I]$_{63}$ line luminosity is negligible. If in turn  a substantial amount of emission originates from PDRs or disc surfaces,  mass-loss rates derived from these methods only represent upper limits.  
 In this regard,  complementary observations at other wavelengths can be insightful. One such extensive discussion can be found in the Appendix. In summary (see Table\,\ref{table:main_results_II}), the observed [O\,I]$_{63}$ line luminosity for Cep\,E and HH\,1 is very likely contaminated by the presence of a PDR (strong contribution in Cep\,E and less strong in HH\,1). Thus, their mass-loss rates represent upper limits at best. \\
 Conceptually, the two methods put forward to estimate $\dot{M}_\text{out}$ from the [O\,I]$_{63}$ line luminosity stand for two distinct perspectives of potentially the same physical situation. 
 From the first perspective (Eq.\,\ref{equ:sperling_2020_formula}), the oxygen atoms along the ejected outflow material are assumed to be collisionally excited in an almost neutral, dense, and warm enviroment. Thus, all oxygen atoms are assumed to be in neutral from. We note that shocks as a potential physical origin of the [O\,I]$_{63}$ line emission are not explicitly mentioned, and other physical processes such as chemical networks or cooling by other emission lines are deliberately ignored. Only the emitting oxygen atoms in the flow are put into focus, since only they contribute to the detected [O\,I]$_{63}$ emission. Ultimately, the outflow geometry, which is the extent of the flow and the velocity of the flowing material, determines the amount of ejected material per unit time.\\    
The second perspective (Eq.\,\ref{equ:HM89_formula}) is to explicitly attribute the [O\,I]$_{63}$ line emission to a single wind shock. In the HM89 model, the [O\,I]$_{63}$ line is also collisionally excited, but the physical setting causing the particle collisions is determined via shock physics (e.g. Rankine-Hugoniot relations). Assumptions on the shock type, shock velocities, gas densities, shock chemistry etc. are inferred, putting the shock as a whole in focus, whereby the emitting oxygen atoms are only one constituent among many.  Compared to the first method,  the shock model includes higher energy levels (higher than the lowest $^3$P$_{0,1,2}$ levels) in the atomic oxygen system and allows the presence of ionised oxygen.   
Line intensities from various species (molecular and atomic) are calculated in this framework and as a numerical result: a) the [O\,I]$_{63}$ line dominates the cooling in these shocks, and b) Eq.\,\ref{equ:HM89_formula} holds approximately true.

\subsubsection{Mass-loss rates from the [O\,I]$_{63}$ jet luminosity}\label{sec:mdot_from_collisional_diagram} 

We estimate mass-loss rates from jet luminosity in the [O\,I]$_{63}$ line, that is $L(\text{[O\,I]}_{63})$, and specific jet parameters such as the tangential jet velocity $v_\text{t}$ and the projected jet length $\theta$ \citep[e.g.][]{dionatos_2018}.  \citet{sperling_2020} derive the following useful relation:
\begin{equation}\label{equ:sperling_2020_formula}
\left( \frac{\dot{M}_\text{out}^\text{lum}}{M_\odot\,\text{yr}^{-1} } \right) =  \left(3.3-6.7\right)  \times 10^{-3}    \cdot   \left( \frac{v_\text{t} }{\text{km}\,\text{s}^{-1}} \right) \left( \frac{ '' }{ \theta  }\right)   \left( \frac{\text{pc}}{ D} \right)\left( \frac{L(\text{[O\,I]}_{63})}{L_\odot}\right),
\end{equation}
whereby $v_\text{t}$ is the component of the jet velocity on the plane of the sky, and $\theta$ is the angular size of the jet. Equation\,\ref{equ:sperling_2020_formula} was derived assuming a range of temperatures ($T\sim 300$--$8000$\,K) and a density equal to the critical density at these temperatures. Fundamentally, this assumption  constrains  the (unknown) level population in the atomic oxygen system and ultimately results in an uncertainty of a factor of $\sim 10$  in the derived mass-loss rates  \citep{dionatos_2017, sperling_2020}.   Mass-loss rates estimated via Eq.\,\ref{equ:sperling_2020_formula} are listed in Table\,\ref{table:main_results_II}.

\subsubsection{Mass-loss rates from shock model}\label{sec:mdot_from_shock_model} 
 
Alternatively, instantaneous mass-loss rates might be  obtained from the [O\,I]$_{63}$ line luminosity  applying the \citet{hollenbach_1989} shock model (hereafter HM89). As one of the main results, this model predicts that the [O\,I]$_{63}$ line luminosity   is proportional to the mass-loss rate in a single wind shock; that is,
\begin{equation}\label{equ:HM89_formula}
\dot{M}_\text{out}^\text{shock}  = 10^{-4}\,L(\text{[O\,I]}_{63})/L_\odot\, M_\odot\,\text{yr}^{-1}.
\end{equation}
In the underlying wind shock scenario, the ejected material from the driving source of the outflow is fast enough to produce a dissociative J-shock, and the [O\,I]$_{63}$ line is the dominant coolant in the post-shock gas over a wide range of shock parameters ($n_0\times v_\text{shock} \leq 10^{12}\,\text{cm}^{-2}\,\text{s}^{-1}$, $n_0$ pre-shock density, $v_\text{shock}$ shock velocity). Other emission lines such as [Si\,II]$_{35}$ and  [Fe\,II]$_{26}$ might even be used as proxies for the [O\,I]$_{63}$ line measuring the mass-outflow rate \citep{watson_2016}.\\
However, as discussed in detail by \citet{sperling_2020}, it can be quite difficult to rigorously prove the applicability of Eq.\,\ref{equ:HM89_formula} to specific emission regions where [O\,I]$_{63}$ is prominently detected; hence, the HM89 method is affected by large uncertainties. In addition, if the detected [O\,I]$_{63}$ emission is caused by multiple, potentially unresolved shocks, mass-loss rates derived by the HM89 method will be overestimated.

 \subsection{Accretion rates}\label{sec:accretion_rates}

Accretion rates of protostellar sources may be determined via different methods \citep[e.g.][and references therein]{beltran_2016}. In the widely accepted scenario of magnetospheric accretion \citep{camenzind_1990, bouvier_2007, hartmann_2016} material flows along stellar magnetic field lines from the accretion disc onto the forming star, causing potentially detectable emission lines  to appear in the spectrum \citep{muzerolle_1998, rigliaco_2012, antoniucci_2014_macc} or an excess of emitted UV flux \citep{gullbring_1998, herczeg_2008}. \\
In this framework, the instantaneous mass accretion rate $\dot{M}_\text{acc}$ is calculated from the accretion luminosity  $L_\text{acc}$ and additional stellar parameters such as the stellar mass $M_\star$ and the stellar radius $R_\star$.  
Typically, in the case of Class 0/I sources, fiducial values for their stellar parameters are inferred, since observationally based estimates using evolutionary tracks are considered unreliable \citep[see discussion in][]{contreras_2017}.  The most direct measurements of $L_\text{acc}$ may be obtained from  UV spectra, which would probably fail for any of the 14 considered outflow sources, since they are highly embedded.\\
In order to have a comparable set of accretion rates, we followed \citet{mottram_2017} and calculated them from 
\begin{equation}\label{equ:accretion_rates}
\dot{M}_\text{acc} = \frac{L_\text{acc}R_\star}{GM_\star}.
\end{equation}  
Depending on the evolutionary state of the outflow source, we adopted the  following values  \citep[see discussion in][]{mottram_2017}:
\begin{itemize}
\item[] Class 0: \quad  $M_\star = 0.2\,M_\odot$, $R_\star = 4\,R_\odot$, $L_\text{acc} =L_\text{bol}$
\item[] Class I: \quad $M_\star = 0.5\,M_\odot$, $R_\star = 4\,R_\odot$, $L_\text{acc} =0.5 L_\text{bol}$.
\end{itemize}
Bolometric luminosities are taken primarily from \citet{karska_2018}, and if not listed therein from \citet{dishoeck_2011}.  As a caveat, we note that bolometric luminosities for the considered targets often vary by a factor of few in the literature since obtaining exact measurements is  challenging \citep[e.g.][]{dunham_2013}.  As a result, mass-accretion rates estimated via Eq.\,\ref{equ:accretion_rates} may feature uncertainties up to one order of magnitude. The calculated values for the accretion rates are listed in  Tables\,\ref{table:main_results_II} and \ref{table:my_main_results}. \\   
 In the case of Cep\,E, we decided to estimate its accretion rate using the same approach but with different stellar parameters. The reason for this is that  Cep\,E-mm is an intermediate-mass protostar \citep[$M_\star \sim 2-5\,M_\odot$, e.g.][]{velusamy_2011}, thus appearing as scaled-up version of the remaining low mass protostars of our sample \citep[see references in][]{frank_2014, hartmann_2016}. Assuming the stellar parameters for  Cep\,E-mm to be $M_\star \sim 3\,M_\odot$ \citep{velusamy_2011}, $L_\text{bol} \sim 80\,L_\odot$ \citep{froebrich_2003}, and $R_\star \sim 20\,R_\odot$ \citep{velusamy_2011}, and further assuming that $L_\text{acc}\approx L_\text{bol}$, we estimated the accretion rate via \citep{mckee_2007}: 
\begin{equation}
\left(\frac{\dot{M}_\text{acc}}{10^{-6}\,M_\odot\,\text{yr}^{-1}}\right) \approx \frac{1}{3.1\,f_\text{acc}}\left(\frac{0.25\,M_\odot}{M_\star}\right)\left(\frac{R_\star}{2.5\,R_\odot}\right)\left(\frac{L_\text{acc}}{L_\odot}\right).
\end{equation}
With $f_\text{acc}\approx 1$ (fraction of the gravitational potential energy released by accretion), we obtain $\dot{M}_\text{acc}\sim 1.7\times 10^{-5}\,M_\odot\,\text{yr}^{-1}$ for Cep\,E-mm. 

\section{Discussion}

\subsection{Other outflow components}\label{sec:other_components}

 In order to evaluate the importance of the atomic outflow component in relation to the total outflow, we compiled mass-loss rates that were estimated via other species (Table\,\ref{table:main_results_II}). Tendentiously, mass-loss rates in our
sample determined via the [O\,I]$_{63}$ emission line are higher than values obtained from near-infrared emission lines (H$_2$ or [Fe\,II]) and lower or comparable to values obtained from sub-millimetre molecular species such as CO. \\ 
The mass-loss rate for Cep\,E derived from the jet geometry is about one order of magnitude lower than the measurements from low-J CO observations undertaken by \citet{moro_martin_2001} and \citet{lefloch_2015}. This points to the conclusion that the collimated CO jet analysed in both mentioned studies represents the main outflow component (not entrained material); that is, the jet is mainly molecular.  \\   
  \citet{nisini_2005} estimated mass-flux rates along the HH\,1 jet using near-infrared and optical emission lines. They found that amongst the considered jet components,  the bulk mass loss resides in the partially ionised [Fe\,II] outflow and quantifies about $6 \times 10^{-7}\,M_\odot\,\text{yr}^{-1}$ (summing over all knots). This is almost one order of magnitude lower than the mass loss determined via [O\,I], indicating that the bulk contribution resides in  a low-excited atomic jet, as suspected by \citet{nisini_2005}. Recent observations of the molecular component traced by low-J CO support this conclusion \citep{tanabe_2019}.\\
 The mass loss residing in the molecular (in particular CO, SO, SiO) high-velocity component of the innermost part of the HH\,212 jet was estimated as $\sim 10^{-6}\,M_\odot\,\text{yr}^{-1}$ \citep{lee_2007, podio_2015, lee_2015_hh212jet}. Surprisingly, an almost identical mass-loss rate is found in the atomic component seen in [O\,I]$_{63}$. Both components therefore contribute almost equally to the total mass-loss rate from the driving source. \\
In H$_2$, \citet{davis_2000} provided mass-loss rates of the order of $\sim 10^{-7}\,M_\odot\,\text{yr}^{-1}$ for the inner two knots (NK1 and SK1). Therefore, even though the HH\,212 jet is most prominently seen in H$_2$, the bulk mass of the outflow does not reside in that component of the molecular gas mixture.  \\ 
With regard to L1551, the mass-loss connected to the neutral wind component is seen in the 21\,cm HI line emission and was estimated as $8.6\times 10^{-7}\,M_\odot\,\text{yr}^{-1}$ \citep{giovanardi_2000}; that is, about the same value derived from the [O\,I]$_{63}$ emission. This points to the conclusion that at least parts of the neutral gas seen in the 21\,cm line is cooled, entrained material due to the partially ionised wind \citep{pyo_2002}, and an unknown amount might be directly connected to the driving atomic jet seen in [O\,I]$_{63}$ \citep{giovanardi_2000}. From near-infrared observations, \citet{davis_2003} estimated $\dot{M}_\text{[Fe\,II]}\sim 2\times 10^{-7}\,M_\odot\,\text{yr}^{-1}$ and $\dot{M}_{\text{H}_2}\sim 4\times 10^{-8}\,M_\odot\,\text{yr}^{-1}$ , indicating that the atomic component is comparably more important in transporting  outflow material. 
  \\ 
Surprisingly high outflow rates have been measured in molecular transitions of, for example, CO and are of the order of $10^{-5}\,M_\odot\,\text{yr}^{-1}$ \citep[e.g.][]{hogerheijde_1998, fridlund_2002, yildiz_2015}. That is to say they are at least a factor of 10 higher compared to the outflow rate determined here. However, this material is rather connected to entrained material, since it is has been traced by low-velocity, low-J CO emission. Conclusively, the L1551 outflow is predominantly atomic.

\begin{sidewaystable*}
\caption{\small{Mass-loss rates and accretion rates for outflow sources that have been mapped extensively in [O\,I]$_{63}$.}}\label{table:my_main_results}
\centering
\begin{tabular}{c c c c c c |c c c |c|c  }
\hline\hline
Source & Class &    $D$ & $\theta$ & $v_\text{t}$ &  $L(\text{[O\,I])}$ & $\dot{M}_\text{out}^\text{lum}(\text{[O\,I]})$  &$\dot{M}_\text{out}^\text{shock}(\text{[O\,I]})$ & $\dot{M}_\text{other}$  $\&$ component &   $\dot{M}_\text{acc}$\tablefootmark{a} &  Main ref.    \\[1.5pt]
 & &   (pc) & ('') & (km\,$\text{s}^{-1}$) &  $L_\odot$  &  & \small{$(10^{-7}\,M_\odot\,\text{yr}^{-1})$} & &  \small{$(10^{-7}\,M_\odot\,\text{yr}^{-1})$} &   \\
\hline \hline   
\tiny{HH\,111 IRS} & \small{I} &  \small{420}  & \small{20} & \small{270} & \small{$2.5\times 10^{-2}$} &\small{$26-53$} & \small{$22.9-26.4$} &  \small{$4$ $\&$ CO}\tablefootmark{b} & \small{$27$}\tablefootmark{k}  &  \small{\citet{sperling_2020}}\\
\tiny{HH\,111 jet} &   &  \small{420}  & \small{45} & \small{260} & \small{$1.3 \times 10^{-2}$} &\small{$6-12$} & \small{$\lesssim 10-16$} &  \small{$2-6$ $\&$ [O\,I]$\lambda$6300}\tablefootmark{r} &    &  \\ [1.5pt]
\hline  
\tiny{SVS13A} & \small{I}   & \small{235} & \small{22} &  \small{270} & \small{$1.5\times 10^{-2}$}  &\small{$25-51$} & \small{$13.1-16.0$} &  \small{$8.9$ $\&$ [Fe\,II]}\tablefootmark{c}  &\small{$140-170$}\tablefootmark{l} & \small{\citet{sperling_2020}}\\ 
  &    &   &   &   &   &  &  &   \small{$7.0$ $\&$ H$_2$}\tablefootmark{c}  &  &   \\
    &    &   &   &   &   &  &  &   \small{$30$ $\&$ HI}\tablefootmark{d}  &  &   \\ 
        &    &   &   &   &   &  &  &   \small{$90$ $\&$ low-J CO}\tablefootmark{e}, swept-up gas  &  &   \\  [1.5pt]
\hline   
\tiny{HH\,34 IRS} & \small{I}   & \small{430} & \small{26} & \small{160} & \small{$2.4\times 10^{-2}$ } &\small{$11-23$} & \small{$20.7-27.5$} & \small{$0.7$ $\&$ [Fe\,II]}\tablefootmark{c} &\small{$35-115$}\tablefootmark{m} &  \small{\citet{sperling_2020}}\\
  &   &   &   &   &  & &   & \small{$0.03$ $\&$ H$_2$}\tablefootmark{c} &  &   \\ 
    &   &   &   &   &  & &   & \small{$\sim 1.5$ $\&$ [O\,I]$\lambda$6300}\tablefootmark{r} &  &   \\ [1.5pt]
 \hline  
 \hline  
\tiny{L1448-C} & \small{0} & \small{232} & \small{45} &  \small{170} &  \small{$1.8\times 10^{-3}$}&\small{$1-2$} & \small{$2-4$} &  \small{$\sim 24 $ $\&$ SiO, SO, CO}\tablefootmark{f} &\small{35}  &  \small{\citet{nisini_2015}} \\ [1.5pt]
\hline  
\tiny{IRAS4A} &\small{0} & \small{235} & \small{38}&  \small{$100-140$} & \small{$9.1\times 10^{-4}$} &\small{$0.3-1.0$}   & \small{$1-2$}   &  \small{$\gtrsim  18 $ $\&$ SiO, SO, CO}\tablefootmark{g} & \small{58}  & \small{\citet{nisini_2015}}\\ [1.5pt]
\hline  
\tiny{HH\,46} & \small{I} &  \small{450} & \small{59} & \small{300} & \small{$2.0\times 10^{-2}$} &\small{$7-15$}  & \small{$20-40$} & \small{$15-28$ $\&$ CO}\tablefootmark{v} &\small{34}\tablefootmark{n}  &\small{\citet{nisini_2015}}   \\ [1.5pt]
 \hline  
\tiny{BHR\,71} & \small{0} & \small{200} & \small{33} & \small{$50-100$} & \small{$3.2\times 10^{-3}$} &\small{$1-3$}  & \small{$3-6$} & \small{$21$ $\&$ CO}\tablefootmark{t} &\small{73}\tablefootmark{o}  &  \small{\citet{nisini_2015}}\\ [1.5pt]
 \hline   
\tiny{VLA 1623} & \small{0} & \small{120} & \small{78} & \small{60} & \small{$2.1\times 10^{-3}$} &\small{$0.5-1$}  & \small{$2-4$}   & \small{$16-160$ $\&$ CO} & \small{21}   &   \small{\citet{nisini_2015}}\\ [1.5pt]
 \hline  \hline
\tiny{HH\,211 SE lobe} & \small{0} & \small{250} & \small{51} & \small{115}  & \small{$3.92\times10^{-3}$} &\small{$1.2-2.4$} &  \small{$3.9$}  &  &  &     \\
\tiny{HH\,211 NW lobe} & \small{0} & \small{250} & \small{45} &\small{115} & \small{$3.57\times10^{-3}$}  &\small{$1.2-2.4$} &  \small{$3.6$}  &  &  &    \\
\tiny{HH\,211 both lobes} &   &   & & & &\small{$2.4-4.8$} &  \small{$7.5$}  & \small{$7-28$ $\&$ SiO, CO, SO}\tablefootmark{h}   &\small{$14$}\tablefootmark{p}   & \small{\citet{dionatos_2018}}    \\
  &   &   & & & &  &     & \small{$\sim 20-28$ $\&$ H$_2$}\tablefootmark{i}   &  &     \\
 \hline\hline
\tiny{IRAS 2A, blue lobe SN} & \small{0}  & \small{235} & \small{188} & \small{50} & \small{$3.5\times 10^{-3}$} &\small{$0.1-0.3$} & \small{$3.5$} & \small{$200$ $\&$ CO}\tablefootmark{u}  & &     \\
\tiny{IRAS 2A, red lobe SN} & \small{0}  & \small{235} & \small{133} & \small{50} &  \small{$4.1\times 10^{-3}$} &\small{$0.2-0.4$} & \small{$4.1$} &  \small{$400$ $\&$ CO}\tablefootmark{u}  & &      \\
\tiny{IRAS 2A, both lobes SN} &  &  & & & &\small{$0.3-0.7$} & \small{$7.6$} & \small{$6$ $\&$ H$_2$}\tablefootmark{j} &\small{$234$}\tablefootmark{q}  &   \small{\citet{dionatos_2017}}   \\
  &  &  & & & &  &   &  \small{$> 6.7$ $\&$ SiO, SO, CO}\tablefootmark{s} &  &   \\
 \hline\hline
\end{tabular}
\tablefoot{  
\tablefoottext{a}{\small{Calculated as described in Section\,\ref{sec:accretion_rates}.}} 
\tablefoottext{b}{\small{\citet{lefloch_2007}, the high-velocity outflowing gas is detected therein in the CO  J=7--6  transition. We therefore think that the stated mass-loss rate is connected to the jet itself and not to swept-up gas.}}
\tablefoottext{c}{\small{\citet{davis_2003}.}}
\tablefoottext{d}{\small{\citet{lizano_1988}.}}
\tablefoottext{e}{\small{Calculated from \citet{knee_2000}, from Table 1 therein we adopt the relevant values for the blue lobe of HH\,7-11 ($\dot{P} = \dot{M}_\text{out}v =2.8\times 10^{-4}\,M_\odot\,\text{km}\,\text{s}^{-1}\,\text{yr}^{-1}$, $v=31\,\text{km}\,\text{s}^{-1}$).}}
\tablefoottext{f}{\small{\citet{podio_2020, lee_2020_molecular_review}. The stated value is in good agreement with measurements of \citet{yoshida_2021}.}}
\tablefoottext{g}{\citet{podio_2020}; \small{\citet{yildiz_2015} measure $\dot{M}_\text{out}>160\times 10^{-7}\,M_\odot\,\text{yr}^{-1}$ based on CO J=6--5 observations (sum of both lobes).}}
\tablefoottext{h}{\small{\citet{lee2007_hh211, lee_2010_hh211}; \citet{lee_2020_molecular_review}.}}
\tablefoottext{i}{\small{\citet{dionatos_2010}.}}
\tablefoottext{j}{\small{\citet{maret_2009}.}}
\tablefoottext{k}{\small{The stated accretion rate is in good agreement with \citet{lee_2010_hh111}. \citet{yang_1997} estimate a higher value of $6.9\times 10^{-6}\,M_\odot\,\text{yr}^{-1}$.}} 
\tablefoottext{l}{\small{We estimate this accretion rate from the Br$\gamma$ line (see Appendix\,\ref{appendix:svs13_accretion_rate}).}} 
\tablefoottext{m}{\small{Based on measurements of accretion-induced emission lines \citet{antoniucci_2008} estimate an accretion rate of $\dot{M}_{\text{acc}} \sim 41.1 \times 10^{-7}\,M_\odot\,\text{yr}^{-1}$, whereas \citet{nisini_2016} $\dot{M}_{\text{acc}} \sim 75\pm 40 \times 10^{-7}\,M_\odot\,\text{yr}^{-1}$. \citet{hartigan_1994} state  $\dot{M}_{\text{acc}} \sim 110\times 10^{-7}\,M_\odot\,\text{yr}^{-1}$. We combine the mentioned values to a range of $(35-115) \times 10^{-7}\,M_\odot\,\text{yr}^{-1}$.}} 
\tablefoottext{n}{\small{\citet{antoniucci_2008} estimate $2.2\times 10^{-7}\,M_\odot\,\text{yr}^{-1}$ based on a substantially lower bolometric luminosity.}}  
\tablefoottext{o}{\small{\citet{yang_2017} estimate $1.2\times 10^{-5}\,M_\odot\,\text{yr}^{-1}$.}}
\tablefoottext{p}{\small{The stated accretion rate is about a factor of six lower than the value estimated by \citet{lee_2007_HH211_A}, who however assume a substantially lower mass for the HH\,211 protostar  $M_\star = 0.06\,M_\odot$.}}
\tablefoottext{q}{\small{The stated value lies inbetween the values estimated by \citet{brinch_2009} and \citet{hsieh_2019}, that are $94\times 10^{-7}\,M_\odot\,\text{yr}^{-1}$ and $500\times 10^{-7}\,M_\odot\,\text{yr}^{-1}$ respectively.}}
\tablefoottext{r}{\small{\citet{hartigan_1994}.}}
\tablefoottext{s}{\small{\citet{podio_2020}.}}
\tablefoottext{t}{\small{\citet{yang_2017}; \citet{yildiz_2015} measure  $76\times 10^{-7}\,M_\odot\,\text{yr}^{-1}$ based on CO J=6--5 observations.}}
\tablefoottext{u}{\small{Based on CO J=6--5 observations \citep{yildiz_2015}.}}
\tablefoottext{v}{\small{The stated value is taken from \citet{nisini_2015}, \citet{yildiz_2015} measure $\dot{M}_\text{out} \sim 200  \times 10^{-7}\,M_\odot\,\text{yr}^{-1}$ from CO J=6--5 observations.}}
}
\end{sidewaystable*}

\subsection{Efficiencies of sources with spatially resolved [O\,I]$_{63}$ outflows} 
 
Until now, only a handful of protostellar outflows had been mapped to a great extent in the far-infrared [O\,I]$_{63,145}$ emission lines. However, spatially resolving the [O\,I] emission from young outflow sources and analysing the spectral properties and overall morphology are crucial for drawing further conclusions on the energy budget of the outflow. High resolution maps provide valuable insights into the physical origin of the seen emission, and together with observations at other wavelengths the notion of the shock origin of the [O\,I] line can be supported or refuted (see e.g. discussion in \citet{sperling_2020}.\\   
The two main instruments that have been used for mapping protostellar outflows in far-infrared emission lines are SOFIA/FIFI-LS and Herschel/PACS. We compiled a set of 14 young sources (nine Class 0 and five Class I) that have been mapped along their outflow in [O\,I]$_{63}$. Sources observed with only one footprint of the $5\times 5$ spaxel  array of Herschel/PACS are excluded from this sample but are considered in Section\,\ref{sec:other_sources}.

\noindent
Three additional sources from SOFIA/FIFI-LS: \citet{sperling_2020} probed five Class I/II outflows (HH\,111, HH\,34, SVS13\,A, HH\,26, and HH\,30) with FIFI-LS aboard SOFIA. A detailed analysis showed that only for three of the sources (HH\,111 IRS, SVS13\,A, HH34\,IRS) could outflow rates be determined.  
Seven additional sources from Herschel/PACS: \citet{nisini_2015} observed five Class 0/I sources and their outflows, namely L1448-C (Class 0), NGC 1333-IRS4 (Class 0), HH\,46 (Class I), BHR\,71 (Class 0), and  VLA\,1623 (Class 0).  
 \citet{dionatos_2018} mapped the  HH\,211 protostellar system driven by HH\,211-MM (Class 0)  in [O\,I]$_{63}$ and various molecular transitions (CO, H$_2$O, and OH).
 \citet{dionatos_2017} mapped the NGC 1333 star forming region in [O\,I]$_{63}$ and [C\,II]$_{157}$ covering several known outflow sources such as SVS13\,A, IRAS\,4A, IRAS\,2A S-N, SK1, and SK 14. The atomic jets connected to IRAS\,4A and SVS13\,A were already mapped by   \citet{nisini_2015} and \citet{sperling_2020}, respectively. Since the IRAS\,2A S-N outflow in the \citet{dionatos_2017} map is clearly  distinguishable from other outflows in the crowded field, we only include this source in our sample.

\begin{figure} 
\resizebox{\hsize}{!}{\includegraphics[trim=0 0 0 0, clip, width=0.9\textwidth]{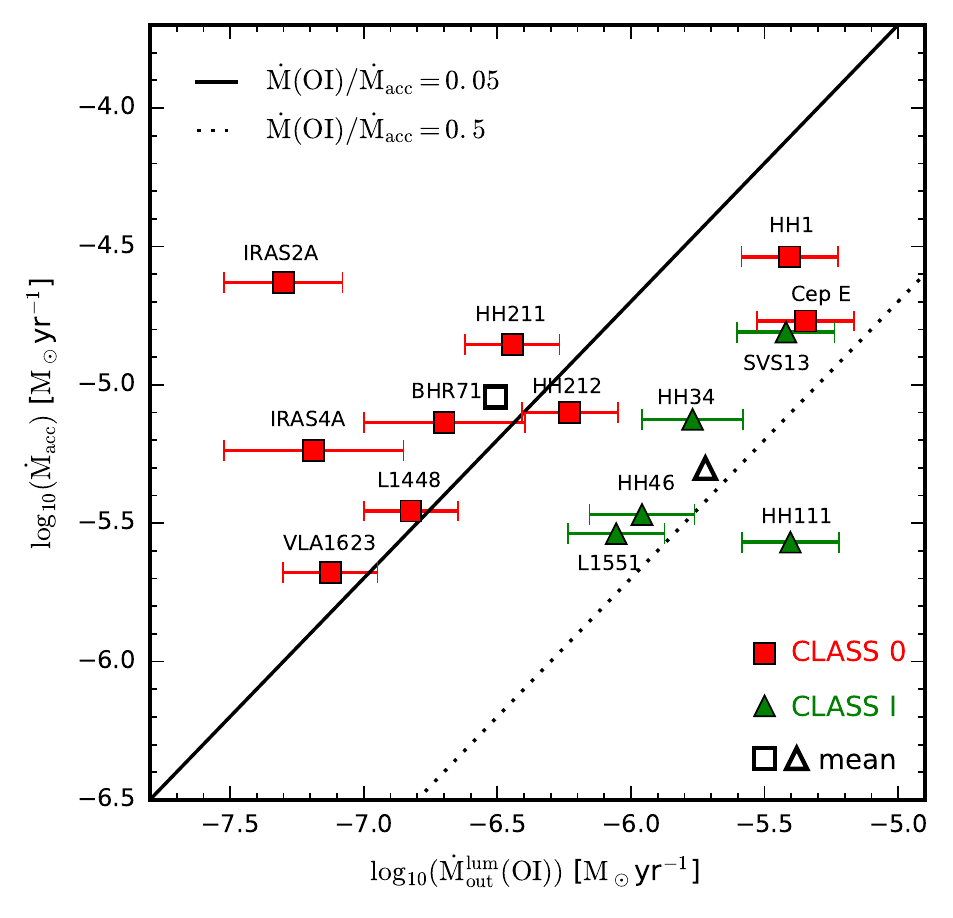}}
\caption{\small{ $\dot{M}_\text{out}^\text{lum}(\text{[O\,I]})$ (from Eq.\,\ref{equ:sperling_2020_formula}) versus $\dot{M}_\text{acc}$ diagram for the extensively mapped outflow sources. The adopted mass-loss rate for Cep\,E is the sum of both contributions from knot A and the jet; that is, $(29.6-60.2)\times 10^{-7}\,M_\odot\,\text{yr}^{-1}$. In the case of HH\,1, we considered the measurement at the HH\,1 jet to be the best representation of the instantaneous mass-loss rate. Geometric means for Class 0/I outflows are plotted as non-filled black markers. Error bars in accretion rates are not depicted, but can be of the order of one magnitude.}}\label{fig:accretion_rates_II} 
\end{figure}

\noindent
Given the  uncertainty of the applicability of the HM89 shock model (Eq.\,\ref{equ:HM89_formula}), we present newly calculated mass-loss rates (with the exception of HH\,34, HH\,111, and SVS13\,A)  based on the [O\,I]$_{63}$ line luminosity (Eq.\,\ref{equ:sperling_2020_formula}) in Table\,\ref{table:my_main_results}. With only one exception (SVS13\,A), all mass-loss rates calculated via the HM89 formula are higher compared to the values from Eq.\,\ref{equ:sperling_2020_formula}. This points to the conclusion that the HM89 formula indeed overestimates mass-loss rates because parts of the [O\,I]$_{63}$ emission are not connected to the impact of the original protostellar wind with the ambient medium. Several other internal shocks in the flow (spatially unresolved) and the presence of a PDR are the main interfering influences on the detected [O\,I]$_{63}$ emission. We therefore based our further analysis on mass-loss rates derived from Eq.\,\ref{equ:sperling_2020_formula}.\\
Accretion rates of the additional outflow sources (Table\,\ref{table:my_main_results}) were calculated as described in Section\,\ref{sec:accretion_rates}. Only in the cases of HH\,34 and SVS13\,A are accretion-induced emission lines used as a proxy to determine their accretion luminosity and mass-accretion rates (both sources were not part of the WISH, WILL, DIGIT, or GASPS surveys). Thus, for all other sources mass-accretion rates were estimated rather indirectly and might feature high uncertainties. 
However, as long as further direct measurements are not yet available, the cited accretion rates in Tables\,\ref{table:main_results_II} and \ref{table:my_main_results} may be considered adequate estimates.\\
In this context, there is growing evidence that episodic accretion  and ejection may be an important factor in protostellar evolution  \citep{audard_2014}. In the considered outflow sample, IRAS2A, for instance, shows a high accretion rate.  IRAS2A  is likely undergoing  an  accretion  burst \citep{hsieh_2019} potentially explaining the high accretion rate as compared to the mass-ejection rate.\\
Figure\,\ref{fig:accretion_rates_II} shows the mass-loss rate $\dot{M}_\text{out}^\text{lum}(\text{[O\,I]})$ derived from the  [O\,I]$_{63}$ line luminosity as a function of the accretion rate $\dot{M}_\text{acc}$ for the fully mapped outflows. Half of the  sources feature efficiency ratios of $\dot{M}_\text{out}^\text{lum}(\text{[O\,I]})/\dot{M}_\text{acc} \sim 0.05-0.5$. This  range for the outflow efficiency ratio  is consistent with a) proposed jet launching models; that is, the X-wind scenario \citep{shu_1988, shu_1994},  or magnetohydrodynamical disc wind models \citep{ferreira_1997, casse_2000}; and b) other observational studies such as \citet{ellerbroek_2013}, \citet{mottram_2017}, \citet{podio_2020}, \citet{yoshida_2021}, and \citet{lee_2020_molecular_review}. Similar efficiency ratios are also found in massive protostellar outflows \citep[e.g.][]{beuther_2002}.  Since $\dot{M}_\text{out}^\text{lum}(\text{[O\,I]})/\dot{M}_\text{acc} \lesssim 1.0 $ for all our observed targets except for HH\,111, we conclude that they are accretion-dominated. Geometric means of the mass-accretion rates for Class 0 and Class I outflow sources are almost the same; that is, $\sim 10^{-5}\,M_\odot\,\text{yr}^{-1}$.  The geometric mean of the mass-ejection rates for the fully mapped Class 0 and Class I outflows are $3 \times 10^{-7}\,M_\odot\,\text{yr}^{-1}$ and $2\times 10^{-6}\,M_\odot\,\text{yr}^{-1}$, respectively.  
In Fig.\,\ref{fig:accretion_rates_II}, Class I outflow sources are clustered rather close together, whereas the Class 0 sources are scattered over a huge region in the diagram. Class 0 outflow sources are predominantly scattered in regions of the diagram where $\dot{M}_\text{out}^\text{lum}(\text{[O\,I]})/\dot{M}_\text{acc} \lesssim 0.05$. In contrast, Class I outflow sources tend to cluster towards higher efficiency rates of the order of 0.5. \\
However, the   separation seen in Class 0 and Class I outflow sources in Fig.\,\ref{fig:accretion_rates_II} might be an artefact from   calculating the individual accretion rates. In fact, accretion rates attributed to Class 0 sources in Eq.\,\ref{equ:accretion_rates} are per se five times higher than accretion rates from Class I outflow sources with the same  bolometric luminosity. A physical explanation for the trend seen in Fig.\,\ref{fig:accretion_rates_II} could be that the mass-ejection rates traced by [O\,I]$_{63}$ in Class 0 sources are underestimated. This conclusion is supported by an analysis of the other outflow components and their contributions to the total mass-loss rate (Section\,\ref{sec:dominant_component}).     \\
 Given the low number of analysed outflow targets, a deep statistical cluster analysis of Fig.\,\ref{fig:accretion_rates_II} would not be very meaningful, and in addition to that our SOFIA sample may be  biased since we selected the brightest and actively accreting known Class I outflow sources. However, it would be interesting to have   similar observations of Class II outflows. So far, only very few Class II outflows have been observed spatially resolving the [O\,I]$_{63}$ emission along the outflow \citep{podio_2012}. Fortunately, jets from Class II sources are usually very compact ($\sim 10-20\arcsec$) potentially rendering extensive mapping redundant. 
Additionally, for these more evolved sources a substantial amount of [O\,I]$_{63}$ emission at the driving source is expected to be connected to the formed disc and not a wind shock. Properly disentangling both contributions is challenging but necessary to determine mass-loss rates \citep{aresu_2014}.

\subsection{Dominant outflow components}\label{sec:dominant_component} 

In Tables\,\ref{table:main_results_II} and \ref{table:my_main_results}, we compare mass-loss rates derived from the [O\,I]$_{63}$ line with mass-loss rates estimated via other tracers for the fully mapped outflow sources. In seven out of nine Class 0 outflows, the bulk mass-loss resides in the molecular component. Only in two cases of Class 0 outflows is the atomic component traced by [O\,I]$_{63}$  either dominant (HH\,1) or comparable (HH\,212) to the molecular component.   \\
Among the five Class I outflows, four are predominantly atomic (HH\,111, HH\,34, SVS13\,A, L1551),  and in the case of HH\,46 both components contribute a comparable amount to the mass loss. In this analysis, we ignored mass losses that are likely associated with entrained material.\\
In fact, mass-loss rates derived from CO emission may trace entrained material. In this regard, we point out that Tables \ref{table:main_results_II} and \ref{table:my_main_results} contain recent mass-loss rate measurements for the molecular component at high angular resolution \citep{podio_2020, yang_2017, yoshida_2021, lee_2020_molecular_review}. In detail, the stated mass-loss rates for the CO, SiO, and SO outflow components of L1448-C, IRAS\,4A, BHR\,71, IRAS\,2A, HH\,211, HH\,212, and Cep\,E can be considered as reliable estimates.  
They give a more reliable determination of the molecular mass-loss rate because they can better disentangle the compact high-velocity gas in the  jet and the extended entrained emission. \\
In comparison,  \citet{yildiz_2015} utilised CO observations  (CO J=3--2 and CO J=6--5) at lower angular resolution to estimate mass-loss rates. Of both CO transitions, the CO J=3--2 emission line traces entrained material, and mass-loss rates derived from it are usually a factor of $\sim$ 100 higher than the values stated in \citet{podio_2020}, \citet{yang_2017}, \citet{yoshida_2021}, and \citet{lee_2020_molecular_review}. \citet{yildiz_2015} speculated that the CO J=6--5 transition potentially traces non-entrained material and may allow more reliable estimates of mass-loss rates. However, mass-loss rates derived from the CO J=6--5 line are almost always of the same order of magnitude as mass-loss rate estimates based on the CO J=3--2 transition. We therefore think that mass-loss rates derived from CO J=6--5 observations are still heavily affected by entrained gas. \\
In conclusion, there is evidence that the mass-loss rates of the considered Class 0 outflow sources in Fig.\,\ref{fig:accretion_rates_II} are significantly underestimated since the molecular component contributes predominantly to the mass loss.

\begin{figure} 
\resizebox{\hsize}{!}{\includegraphics[trim=0 0 0 0, clip, width=0.9\textwidth]{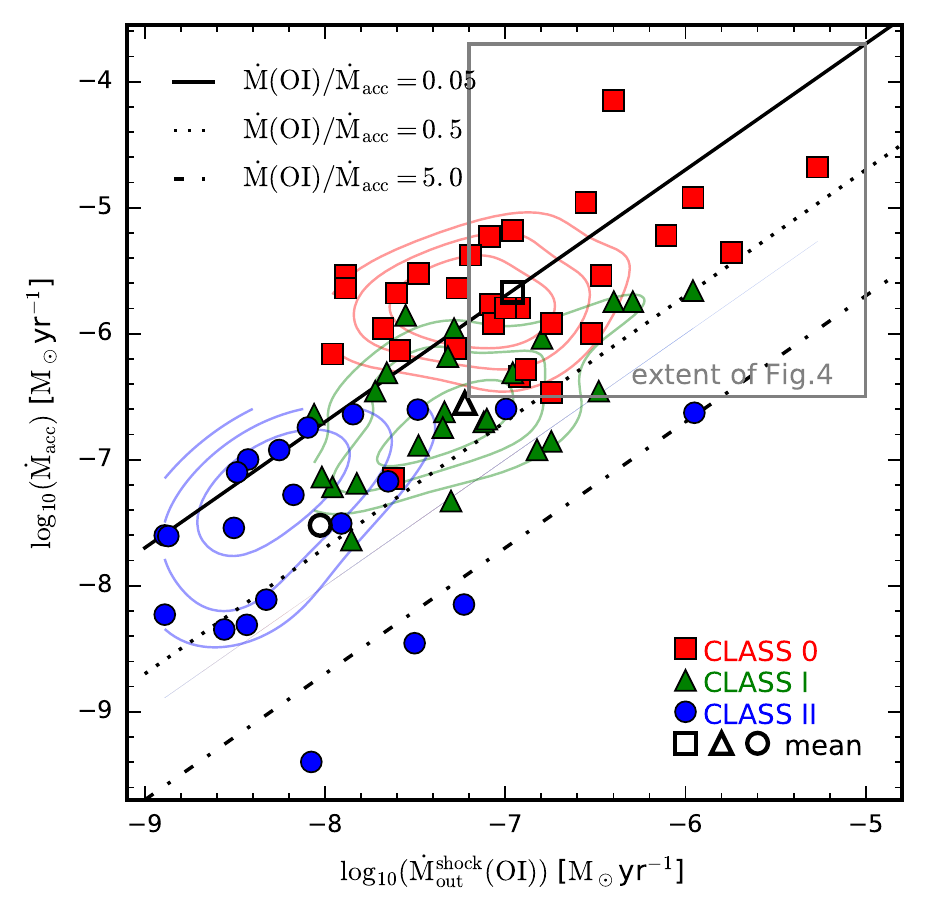}}
\caption{\small{Mass-loss rates $\dot{M}_\text{out}^\text{shock}(\text{[O\,I]})$ (from Eq.\,\ref{equ:HM89_formula}) versus accretion rates $\dot{M}_\text{acc}$ of protostellar outflow sources at different evolutionary stages observed with a single Herschel/PACS footprint. Mass-outflow rates were determined assuming that the HM89 shock model conditions prevail. Geometric means for Class 0/I/II outflows are plotted as non-filled black markers. Thin coloured lines are  contour lines based on a kernel-density estimate using Gaussian kernels.}}\label{fig:accretion_rates_III} 
\end{figure}

\subsection{A comparison to sources with spatially unresolved [O\,I]$_{63}$ outflows}\label{sec:other_sources}

 In order to compare the results from the fully mapped outflows with other studies, we compiled a list of Class 0/I/II outflow sources that have been observed with Herschel/PACS as part of the WISH+DIGIT+WILL+GASPS surveys, thus providing a single footprint with a $47\arcsec\times 47\arcsec$ field of view in $5\times 5$ spatial pixels of $9.4\arcsec\times 9.4\arcsec$ each.  For these outflow sources, the [O\,I]$_{63}$ line is either spatially unresolved or not resolved enough \citep[e.g. the sample of][]{podio_2012} to infer relevant properties of the jet geometry. Under the given circumstances, mass-loss rates utilising the [O\,I]$_{63}$ line are derived from the HM89 shock model and not from the [O\,I]$_{63}$ jet luminosity.  \\
 \citet{watson_2016} did a similar analysis in their survey of 84 YSOs using the Spitzer Infrared Spectrograph. Their mass-loss rate mesurements are, however, not directly based on the HM89 shock model, since the [Si\,II]$_{35}$ and [Fe\,II]$_{26}$ lines are used as a proxy for the [O\,I]$_{63}$ line. This introduces new uncertainties as discussed in their paper. Therefore, we do not include their sources here.\\
 From the Herschel/PACS  surveys, we selected an additional 72 outflow sources (28 Class 0, 23 Class I, and 21 Class II; see Tables\,\ref{literature_I}, \ref{literature_II}, and \ref{literature_III}) for which the [O\,I]$_{63}$ line is prominently detected but mostly spatially unresolved. In detail, our selection comprises the following.  
 \citet{mottram_2017} compiled a list of 91 protostellar sources that were observed as part of the WILL, WISH, and DIGIT surveys. From the 49 WILL sources, we only selected the 26 (15 Class 0, 11 Class I) outflow sources for which $\dot{M}_\text{out}$ and $\dot{M}_\text{acc}$ are specified in Table A.7 of that study. From the remaining 42 DIGIT+WISH sources, we excluded the sources a) for which  $\dot{M}_\text{out}$ is not given in Table A.7; b) that were analysed in much more detail in the spatially resolved sample; c) that could not be classified fairly unambiguously as Class 0 or Class I sources. This results in  25 (13 Class 0, 12 Class I) supplementary sources. 
To complete our sample with even more evolved outflow sources, we included 21 Class II targets from the  GASPS survey \citep{alonso_martinez_2017}. In this study, 26 outflow sources are listed. We excluded four sources for which no accretion rates were specified or the classification was unsure. Among the selected 21 targets, the \citet{podio_2012} sample is included, with the exception of DG Tau B \citep[TAU04 in][]{mottram_2017}. However, we took the newly determined [O\,I]$_{63}$ line fluxes specified in \citet{alonso_martinez_2017} to recalculate the mass-loss rates via Eq.\,\ref{equ:HM89_formula}. The [O\,I]$_{63}$ line fluxes have changed since the Herschel/PACS data were reduced using HIPEv10 in \citet{alonso_martinez_2017}, whereas \citet{podio_2012} used HIPE 4.0.1467. For a discussion of this data reduction issue, we invite the reader to consult, for example, \citet{howard_2013}. The stated accretion rates for these more evolved sources are determined from the U-band excess. \\
Figure\,\ref{fig:accretion_rates_III}  shows $\dot{M}_\text{out}^\text{shock}(\text{[O\,I]})$ versus  $\dot{M}_\text{acc}$ for the selected 72 sources. For a better comparison,  we mark the coverage of the same plot for  the fully mapped outflow sources (Fig.\,\ref{fig:accretion_rates_II}) as grey rectangle. Mass-loss rates $\dot{M}_\text{out}^\text{shock}(\text{[O\,I]})$ in Figure\,\ref{fig:accretion_rates_III} were determined via the HM89 shock model, meaning they potentially suffer from large uncertainties.
From Fig.\,\ref{fig:accretion_rates_III}, four principal tendencies can be recognised:  
(1) Most sources are roughly located within the efficiency stripe of $f\sim 0.05-0.5$, which is in line with our previous findings. However, four Class I sources feature $f\gtrapprox 1.0,$ and four Class II sources show even higher efficiencies $f\gtrapprox 5.0,$ indicating that they are outflow-dominated. In these sources, a contribution from PDRs or discs could be more relevant;
(2) Many Class 0 outflows feature conspicuously low efficiency ratios $f<0.05$. The same trend is seen in the outflow sample of Fig.\,\ref{fig:accretion_rates_II}. For these targets, the total mass-loss rate from the [O\,I]$_{63}$ emission line could have been underestimated due to a significant molecular contribution. In this context, \citet{yildiz_2015} and \citet{mottram_2017} utilised CO observations to measured mass-loss rates residing in the molecular outflow component (values are listed in Tables\,\ref{literature_I} and \ref{literature_II}). In particular, the mass-loss rates traced by CO J=6--5 in \citet{yildiz_2015} are insightful, since  this gas may not be entrained material but part of the actual outflow. However, the analysis in Sect.\,\ref{sec:dominant_component} indicates that mass-loss rates based on CO J=6--5 observations are still substantially affected by entrained material. In this regard, observations at higher angular resolution as undertaken by \citet{podio_2020}, \citet{yang_2017}, \citet{yoshida_2021}, and \citet{lee_2020_molecular_review} may provide more robust mass-loss rate estimates. A comparison of both outflow components (atomic and molecular) is presented in Table\,\ref{literature_I}. The notion that in Class 0 outflow sources the molecular component is the dominant compared to the atomic is well supported by these data. However, the comparison also shows that limitedly mapped Class I outflow sources do not necessarily feature a dominant atomic outflow component. Extensive mapping in [O\,I]$_{63}$ might be necessary to draw meaningful conclusions; 
(3) Sources from the same Class are grouping in overlapping clusters as indicated by coloured contour lines in Fig.\,\ref{fig:accretion_rates_III}, which  represent a kernel-density estimate. In agreement with the findings of \citet{ellerbroek_2013} and \citet{watson_2016}, there is a trend for which both mass-accretion and mass-loss rates evolve from higher to lower values passing from Class 0 to Class II. A similar conclusion cannot be drawn from the fully mapped outflows alone, since the sample is too small and biased towards brightest outflow sources;
(4) There is a broad scatter in the $\dot{M}_\text{out}$ versus $\dot{M}_\text{acc}$ diagram. Various factors play a role here: a) accretion rates of the younger sources might have been underestimated due to higher obscuration \citep{bacciotti_2011}; b) time variability \citep{watson_2016}; c) the [O\,I] emission line is exclusively used as a  proxy to estimate the total instantaneous mass-loss rate. However, several other outflow components might contribute an unknown amount to the total mass-loss rate, or the underlying shock conditions might not entirely prevail \citep{sperling_2020}. \\
 We note that Fig.\,\ref{fig:accretion_rates_III} has some obvious shortcomings compared to Fig.\,\ref{fig:accretion_rates_II}.  In Fig.\,\ref{fig:accretion_rates_III}, the mass-loss rates are purely based on the HM89 shock model. However, the applicability of the HM89 shock model has not been evaluated for any of the included sources. It is possible that for many sources in Fig.\,\ref{fig:accretion_rates_III} the derived mass-loss rates are only of limited informative value, if the HM89 shock conditions do not prevail. 
In this regard, Fig.\,\ref{fig:accretion_rates_II} is more meaningful since the [O\,I]$_{63}$ jet luminosity and the individual geometric properties of the outflows are taken into account. Such a detailed analysis became possible with our new SOFIA observations, which spatially resolve the [O\,I]$_{63}$ line in the extended jets. However, the comparision with other outflow contributions clearly demonstrated that [O\,I]$_{63}$ maps alone are not enough to evaluate the evolution of protostellar outflows. 

\subsection{The role of the atomic outflow component in the evolutionary picture}\label{sec:why_discrepancy} 

There is a growing number of observational studies showing that the low-excitation, atomic outflow component traced by [O\,I]$_{63}$ plays a crucial role in the overall outflow evolution \citep[e.g.][]{nisini_2015, watson_2016, alonso_martinez_2017}. In fact, as protostellar outflow sources evolve from Class 0 to Class II, their accretion rates and mass-loss rates tend to decrease \citep{watson_2016}, while the decisive ratio $\dot{M}_\text{out}/\dot{M}_\text{acc}$ may even remain constant \citep{podio_2012}. In this context, it is sometimes speculated that the mass-loss rate determined via the [O\,I]$_{63}$ emission line may  truly  represent the bulk ejected material reflecting the anticipated outflow evolution \citep{dionatos_2017}. However, our observations and the comparison with other surveys suggest, as already pointed out by \citet{nisini_2015}, that this hypothesis is barely true in outflows from Class 0 sources, where the bulk of ejected material resides mainly in the molecular component. Thus, for these outflows mass-loss rates determined via [O\,I]$_{63}$ can be significantly underestimated. Qualitatively, these jets are largely not fast enough and too dense to be dissociative, and therefore they show very low excitation. As the source evolves towards the Class II stage, temperatures increase, densities decrease, winds become faster, and dissociative shocks connected to internal or wind shocks cause prominent atomic and ionic emission features depending on the specific excitation conditions in the flow material. These more evolved outflows become prominently detectable in,  for example, [Fe\,II], [S\,II], [Si\,II], and [O\,I], effectively tracing the same gas component at low excitation. As a result, the low-excitation outflow component associated with Class I/II sources and traced by [O\,I]$_{63}$  potentially (but not necessarily) represents the dominant contribution to the total mass loss.   
It would be interesting to see a $\dot{M}_\text{out}$ versus $\dot{M}_\text{acc}$ diagram that considers all relevant outflow components in order to obtain a clear picture on the importance of the mass loss traced by [O\,I]$_{63}$.\\
It becomes clear that the importance of [O\,I] emission as a tracer of mass loss changes over time during the formation of the protostar, and that the far-infrared [O\,I]$_{63}$ emission alone cannot be used as a sole tracer to study the evolution of  $\dot{M}_\text{out}/\dot{M}_\text{acc}$ during the whole star formation process.

\section{Summary}
 
This paper presents SOFIA/FIFI-LS observations, that is, spectroscopic maps in the [O\,I]$_{63,145}$ transitions, of four protostellar outflows associated with Cep\,E, HH\,1, HH\,212, and L1551\,IRS5. For each associated continuum source, we determined flux values at $63\,\upmu\text{m}$ and $145\,\upmu\text{m}$. Notable [O\,I]$_{63}$ emission was detected in all outflow regions, however with different morphologies. \\
For L1551\,IRS5, HH\,1, and HH\,212, most [O\,I]$_{63}$ is detected close to their respective driving sources, whereas at Cep\,E-mm  no [O\,I]$_{63}$ is seen. A detailed analysis indicates that in the cases of L1551\,IRS5 and HH\,212, the \citet{hollenbach_1989} shock conditions most likely prevail, meaning that the detected [O\,I]$_{63}$ emission is connected to dissociative J-shocks. In the other cases, an unknown amount of [O\,I]$_{63}$ emission can be associated to PDR regions or non-dissociative shocks.\\
Utilising the [O\,I]$_{63}$ line luminosity (Eq.\,\ref{equ:sperling_2020_formula}), we calculated mass-loss rates for Cep\,E, HH\,1, HH\,212, and L1551 that are in the range of $(5-50)\times 10^{-7}\,M_\odot\,\text{yr}^{-1}$. A comparison with mass-loss rates connected to other species showed that a) for HH\,1, the bulk mass ejected from the driving source resides in the atomic [O\,I] component; b) the outflows associated with Cep\,E and L1551 are predominantly molecular; and c) the ejected mass in the HH\,212 outflow is almost equally partitioned in an atomic and molecular component.\\
Taking accretion rates from former studies, we estimate efficiency ratios ($f=\dot{M}_\text{out}/\dot{M}_\text{acc}$) that are largely in the range of $ f\sim 0.05-0.5$. This finding is consistent with theoretical predictions (e.g. \citet{shu_1988, shu_1994, ferreira_1997, casse_2000}) and other observational studies \citep{ellerbroek_2013, mottram_2017}.\\
We compared our findings with survey data of 72 Class 0/I/II outflow sources that have been observed in [O\,I]$_{63}$ using a single  Herschel footprint ($5\times 5$ spatial pixels covering a $47\arcsec \times 47\arcsec$ FOV) as part of the DIGIT, WILL, WISH, and GASPS surveys. \\
Most of these outflow sources show efficiency ratios that are consistent with our observations ($f\sim 0.05-0.5$). Surprisingly, many outflows from Class 0 sources feature low efficiency ratios: $f < 0.05$. We conclude that for Class 0 outflow sources the bulk ejected material resides in the molecular component, and mass-loss rates exclusively estimated from the [O\,I]$_{63}$ emission line  underestimate the total mass-loss substantially. In the case of more evolved outflows from Class I and Class II sources, the low-excitation, atomic gas component can represent the main outflow component. There is a trend of gradually decreasing accretion and mass-loss rates as sources evolve from the Class 0 to the Class II stage. However, this depicted trend needs further confirmation considering all main outflow components.

\begin{acknowledgements}  
This research is based on observations made with the NASA/DLR Stratospheric Observatory for Infrared Astronomy (SOFIA). SOFIA is jointly operated by the Universities Space Research Association, Inc. (USRA), under NASA contract NNA17BF53C, and the Deutsches SOFIA Institut (DSI) under DLR contract 50 OK 0901 to the University of Stuttgart.
This work has been supported by the German Verbundforschung grant 50OR1717 to JE.

Teresa Giannini and Brunella Nisini acknowledge the support of the project PRIN-INAF-MAIN-STREAM 2018 \textit{Protoplanetary disks seen through the eyes of new-generation instruments}.

We thank Bill Vacca for providing us with the ATRAN-models.
\end{acknowledgements}

\bibliographystyle{aa} 
\bibliography{papers}  

\newpage

\section*{Appendix A: Continuum maps}\label{appendix:continuum_maps}

 The obtained continuum maps of Cep\,E, HH\,1, HH\,212, and L1551 at the two wavelengths $63.18\,\upmu\text{m}$ ($\Delta \lambda = 0.3\,\upmu\text{m}$) and $145.53\,\upmu\text{m}$ ($\Delta \lambda = 1.0\,\upmu\text{m}$) are shown in Fig.\,\ref{fig:continuum_maps} (bandwidths are in parentheses). Continuum flux measurements are presented in Table\,\ref{table:continuum_fluxes}. 
\begin{figure*} 
  \centering
  \subfloat{\includegraphics[trim=0 0 0 0, clip, width=0.5\textwidth]{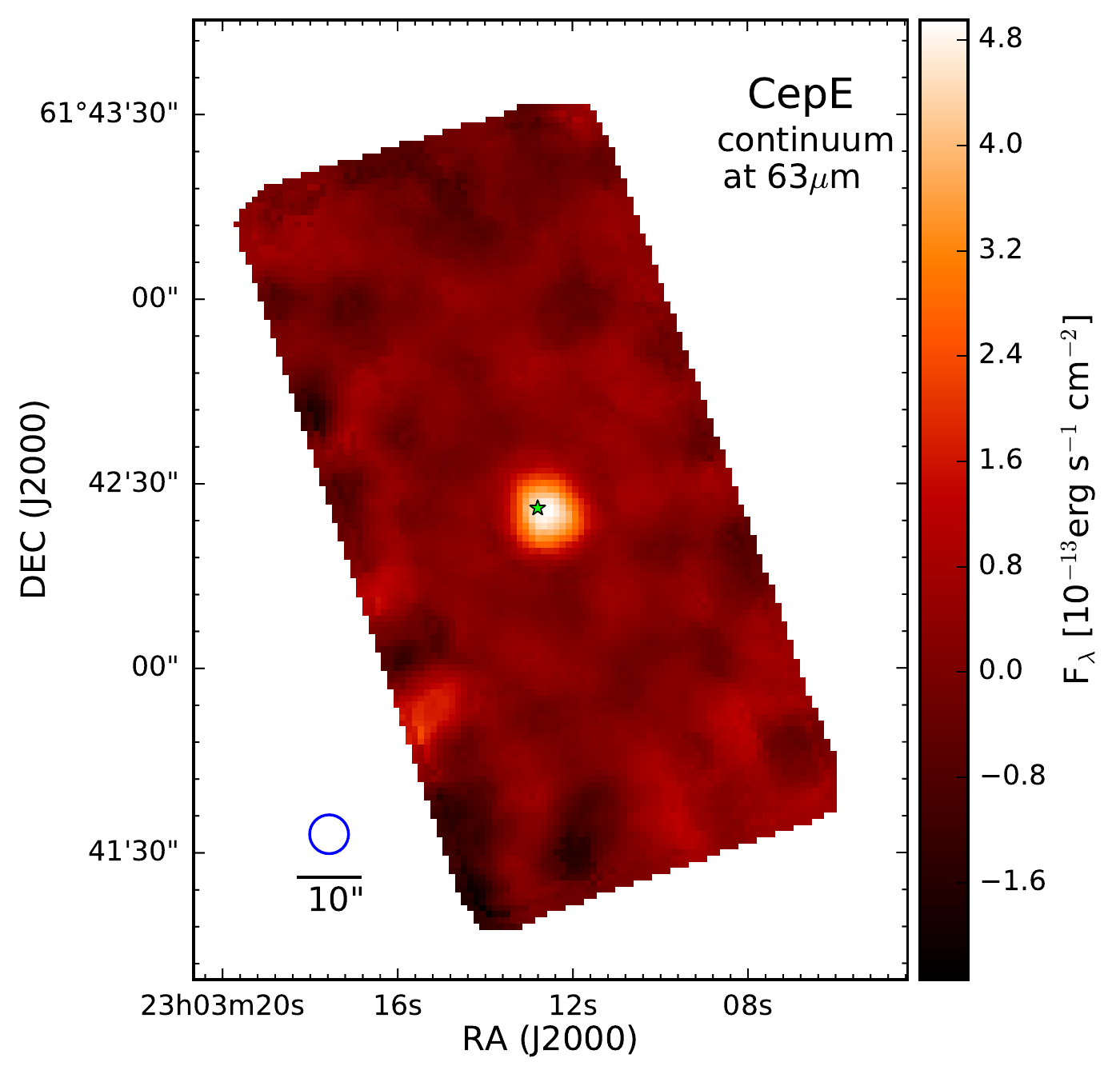}}
  \hfill
  \subfloat{\includegraphics[trim=0 0 0 0, clip, width=0.5\textwidth]{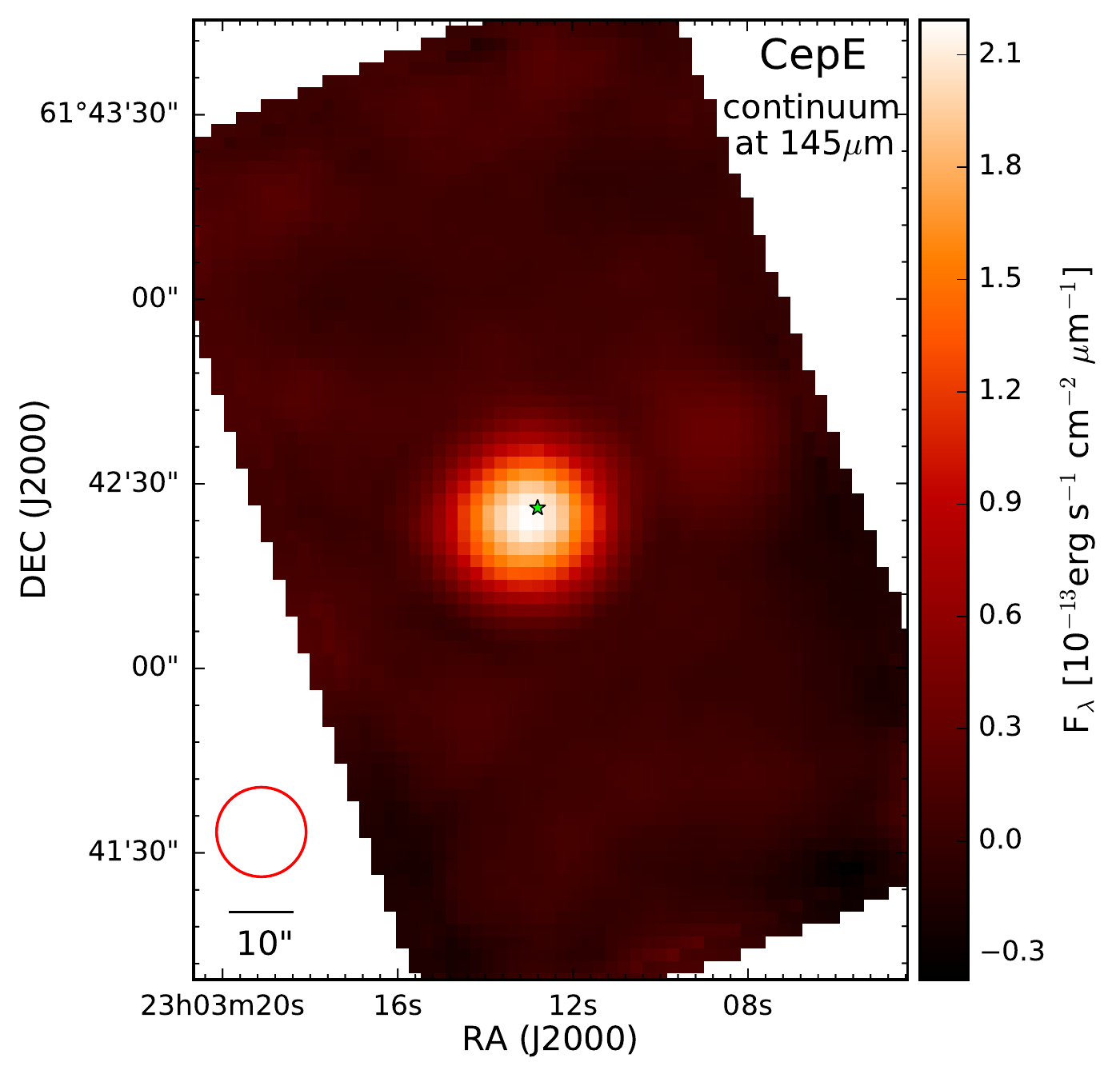}}
  \hfill
  \subfloat{\includegraphics[trim=0 0 0 0, clip, width=0.5\textwidth]{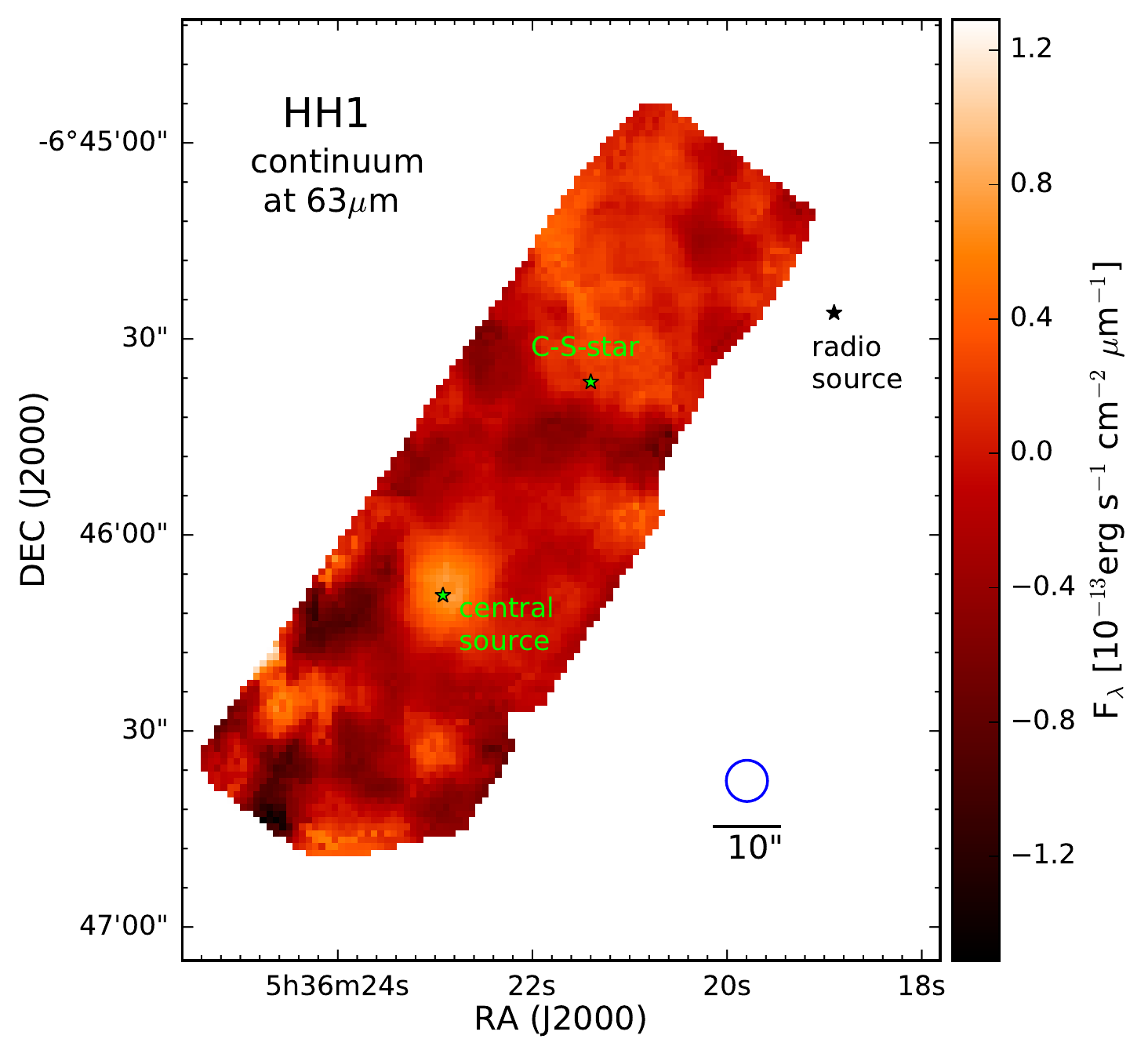}}
  \hfill
  \subfloat{\includegraphics[trim=0 0 0 0, clip, width=0.5\textwidth]{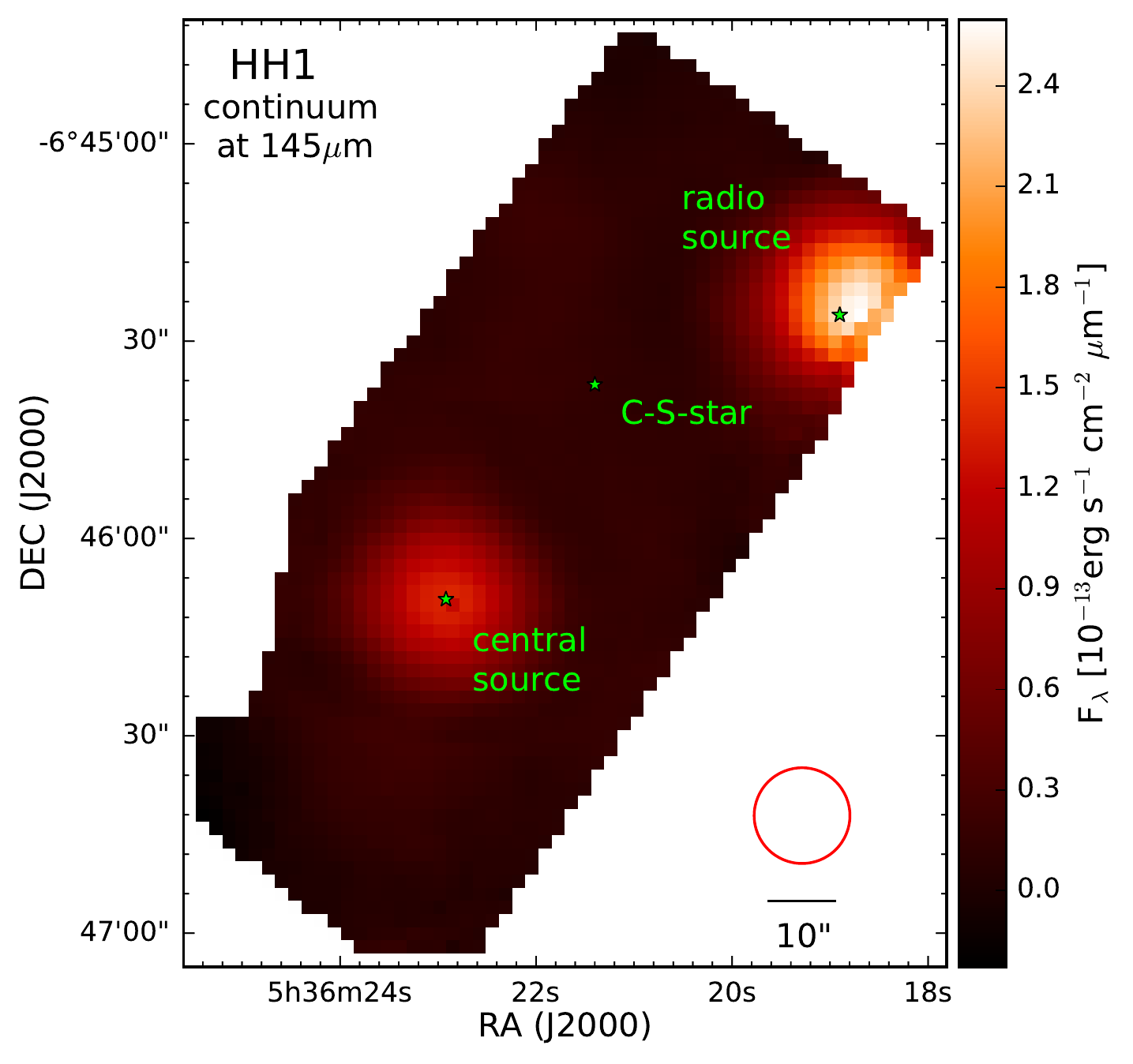}}
  \caption{Continuum maps of Cep\,E, HH\,1, HH\,212, and L1551 obtained from SOFIA/FIFI-LS observations in the two FIFI-LS channels at 63.18\,$\upmu\text{m}$ and 145.53\,$\upmu\text{m}$. Green stars mark the positions   of the continuum sources taken from the 2MASS catalogue (Table\,\ref{table:objects}), with the exeption of HH\,212. In the case of HH\,212, the green stars indicate the position of the centroid of the fitted 2D Gaussian (see discussion in Section\,\ref{sec:continuum_sources}). }\label{fig:continuum_maps}
   \end{figure*}
   
\begin{figure*}
 \ContinuedFloat 
  \centering   
  \subfloat{\includegraphics[trim=0 0 0 0, clip, width=0.5\textwidth]{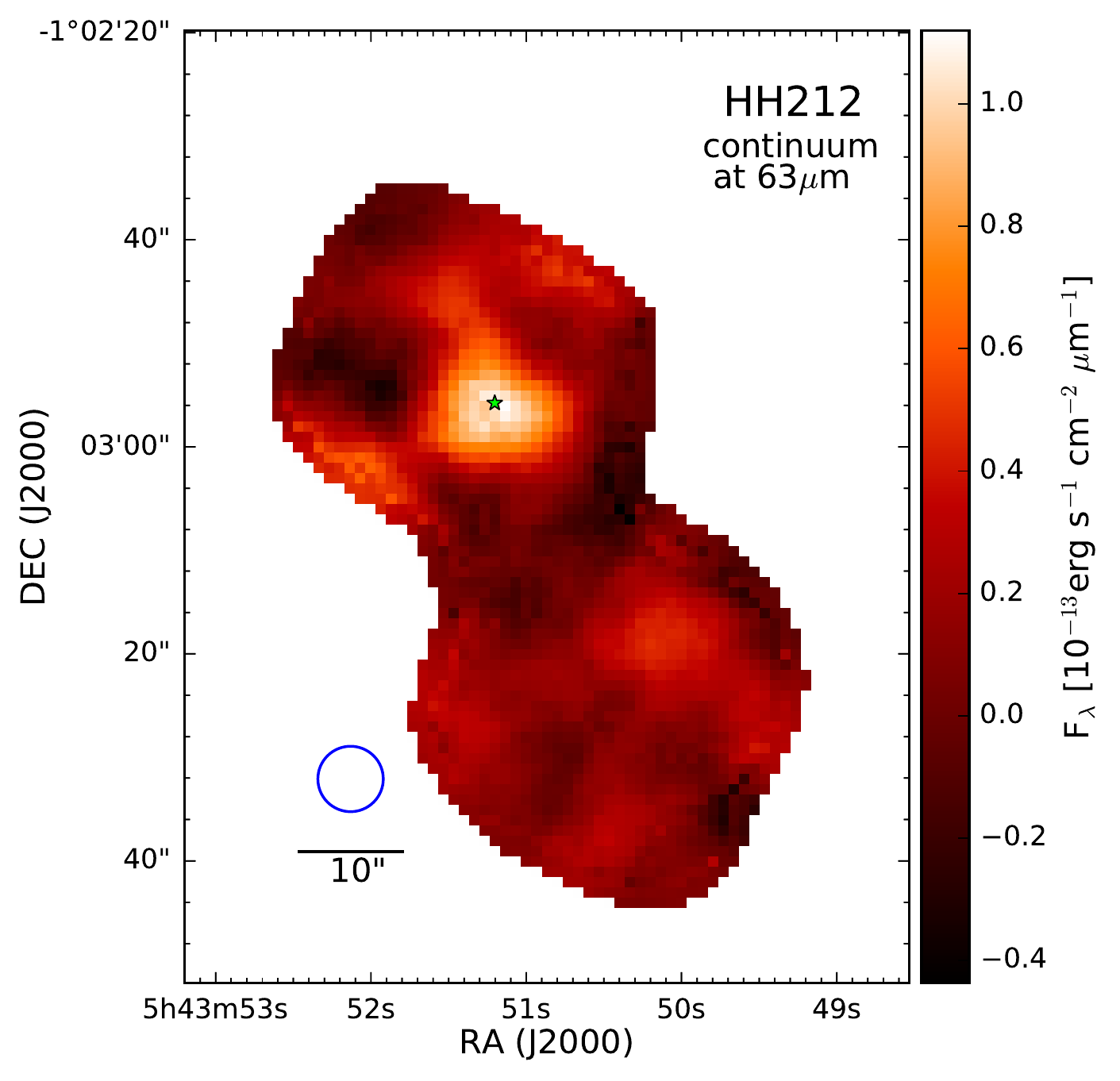}}
  \hfill
  \subfloat{\includegraphics[trim=0 0 0 0, clip, width=0.5\textwidth]{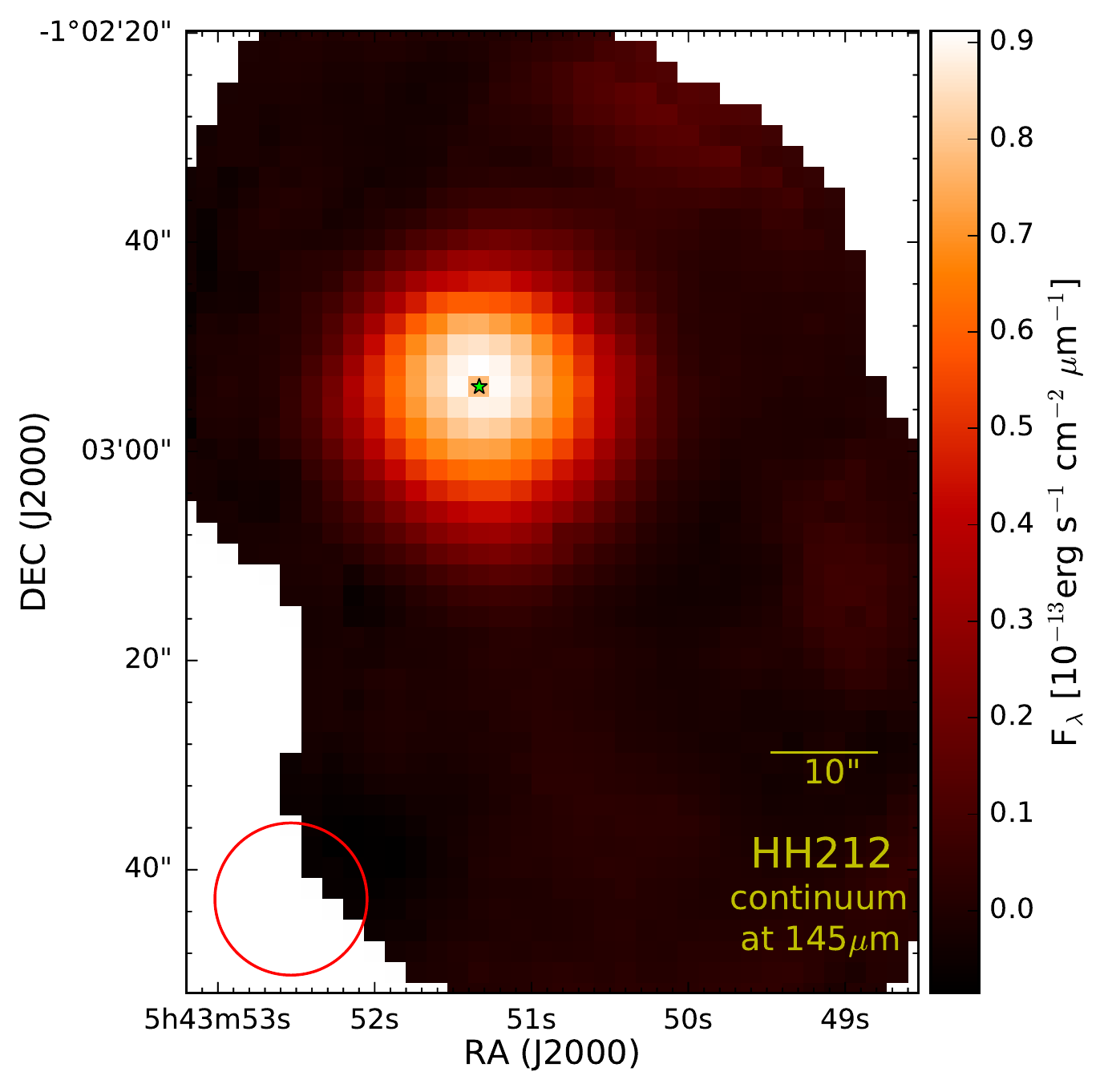}}
   \hfill
   \subfloat{\includegraphics[trim=0 0 0 0, clip, width=0.5\textwidth]{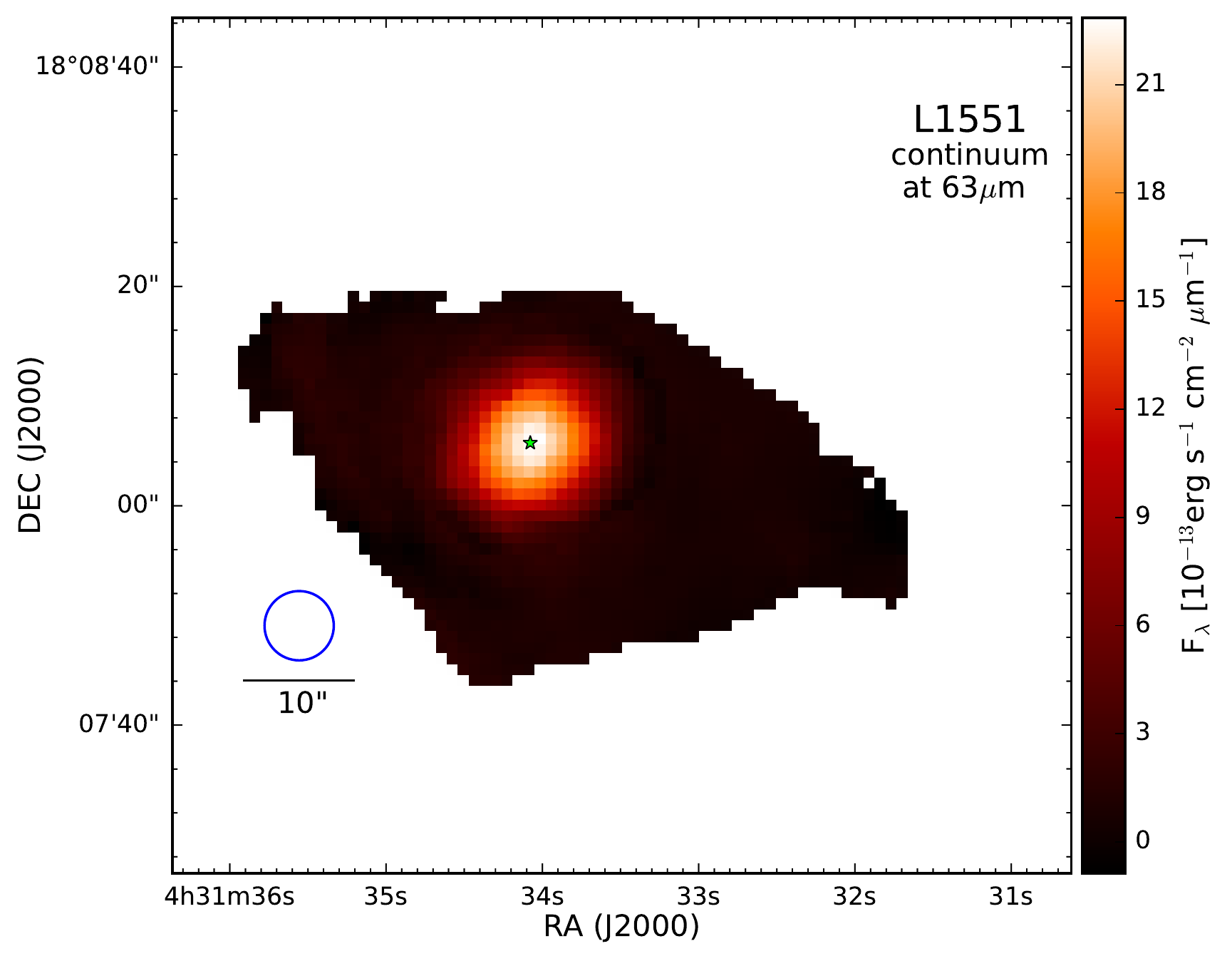}}
  \hfill
  \subfloat{\includegraphics[trim=0 0 0 0, clip, width=0.5\textwidth]{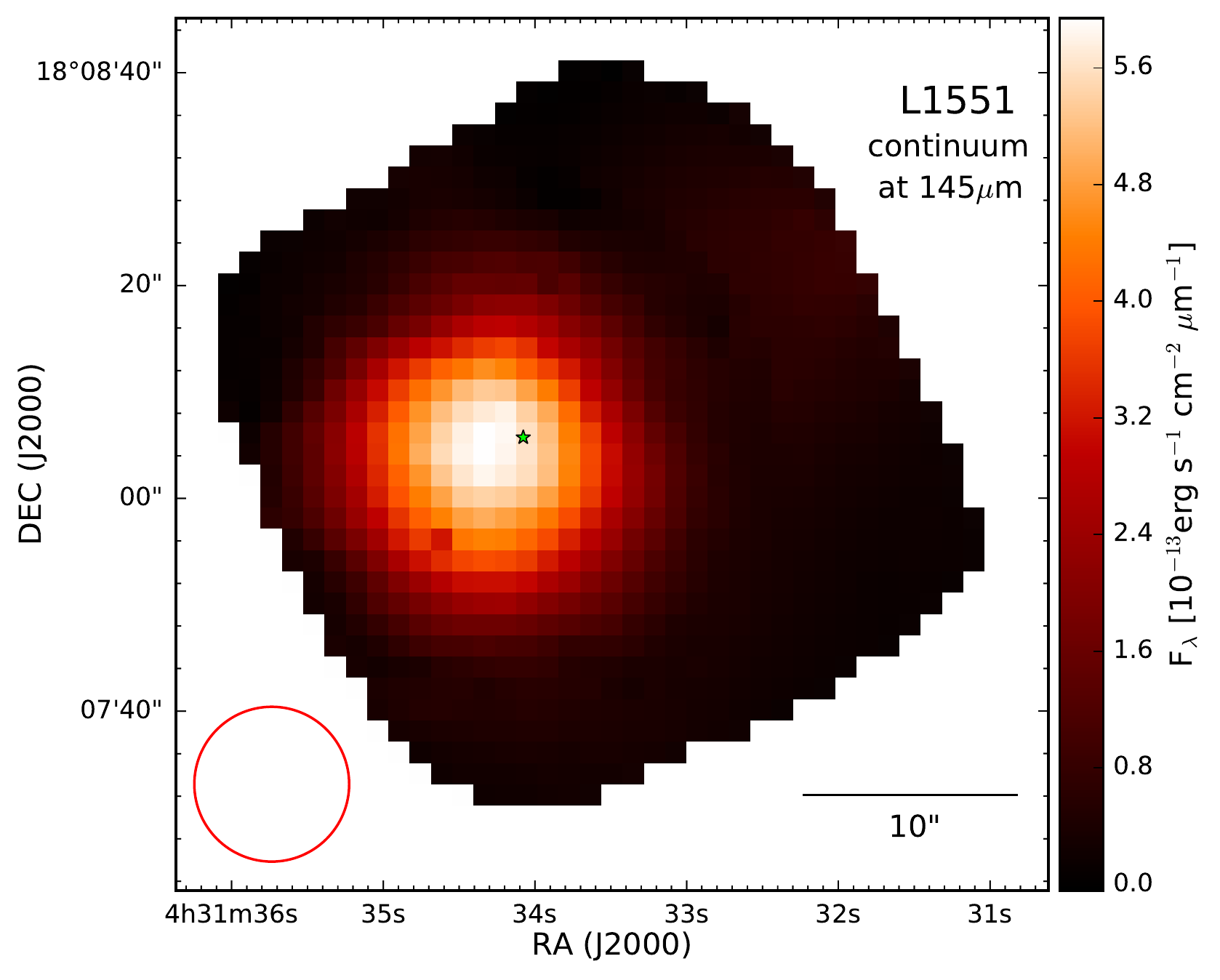}}
  \caption{Continued.}\label{fig:continuum_maps}
\end{figure*}


\section*{Appendix B: [O\,I]$_{145}$ map of HH\,212 }\label{appendix:hh212_OI_145mum_map}
 
 We present the continuum-subtracted [O\,I]$_{145}$ map of HH\,212 in Fig.\,\ref{fig:hh212_red_emission_figure}. A bright emission feature (knot C), very likely shocked material, is detected in a region not covered by the corresponding [O\,I]$_{63}$ map. A sample spaxel from that specific region at the [O\,I]$_{145}$  transition   is shown in Fig.\,\ref{fig:hh212_red_emission_figure_B}.  

\begin{figure}[htb!] 
\resizebox{\hsize}{!}{\includegraphics[trim=0 0 0 0, clip, width=1.0\textwidth]{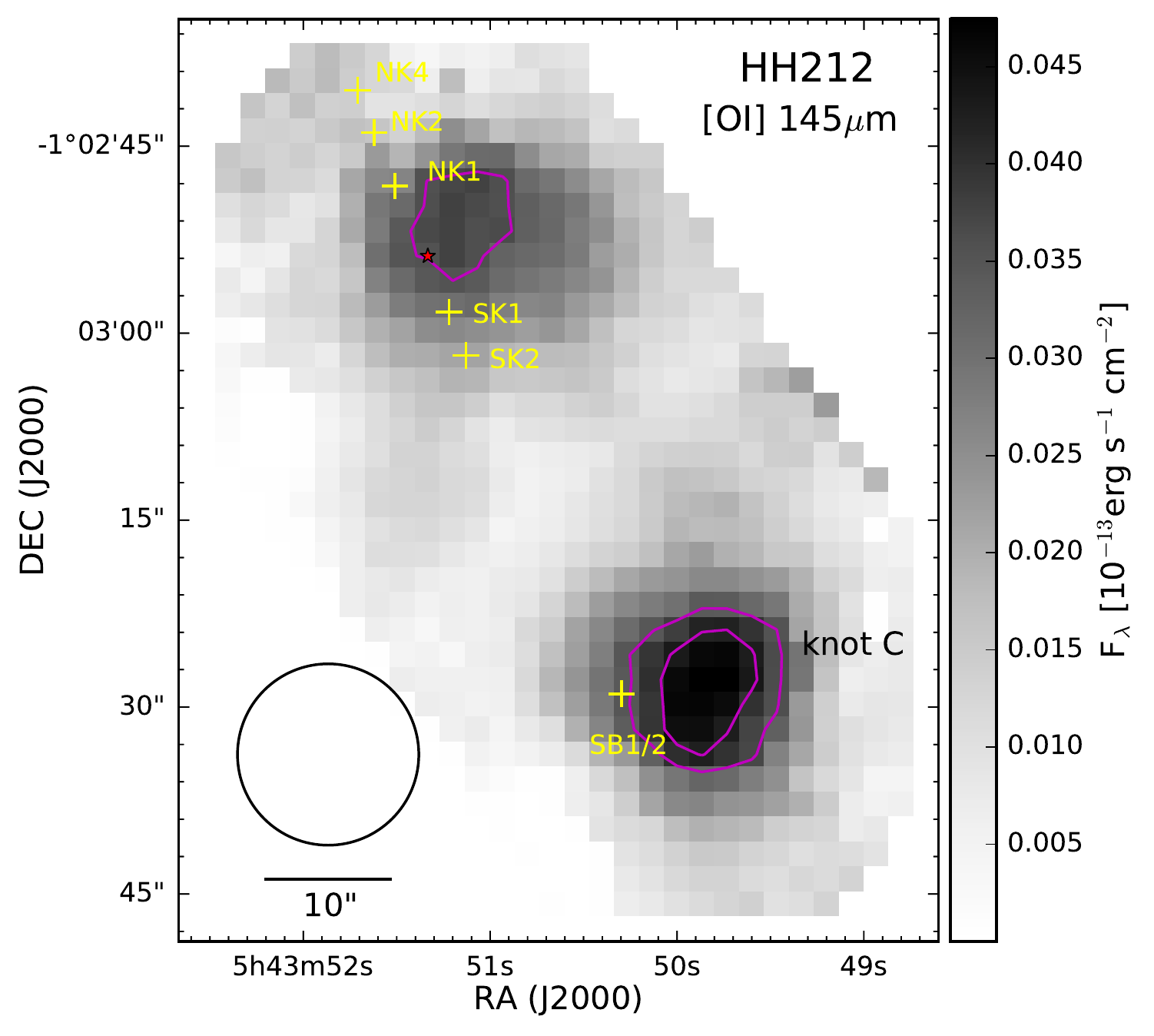}}
\caption{\small{The continuum-subtracted [O\,I]$_{145}$ emission map of HH\,212. The bright emission knot C is not seen in the [O\,I]$_{63}$ map, since it lies outside the mapped region (see Fig.\ref{fig:hh212_red_emission_figure_B} for the [O\,I]$_{145}$ line detection).}}\label{fig:hh212_red_emission_figure} 
\end{figure}

\begin{figure} 
\centering
\subfloat{\includegraphics[trim=0 0 0 0, clip, width=0.49\textwidth]{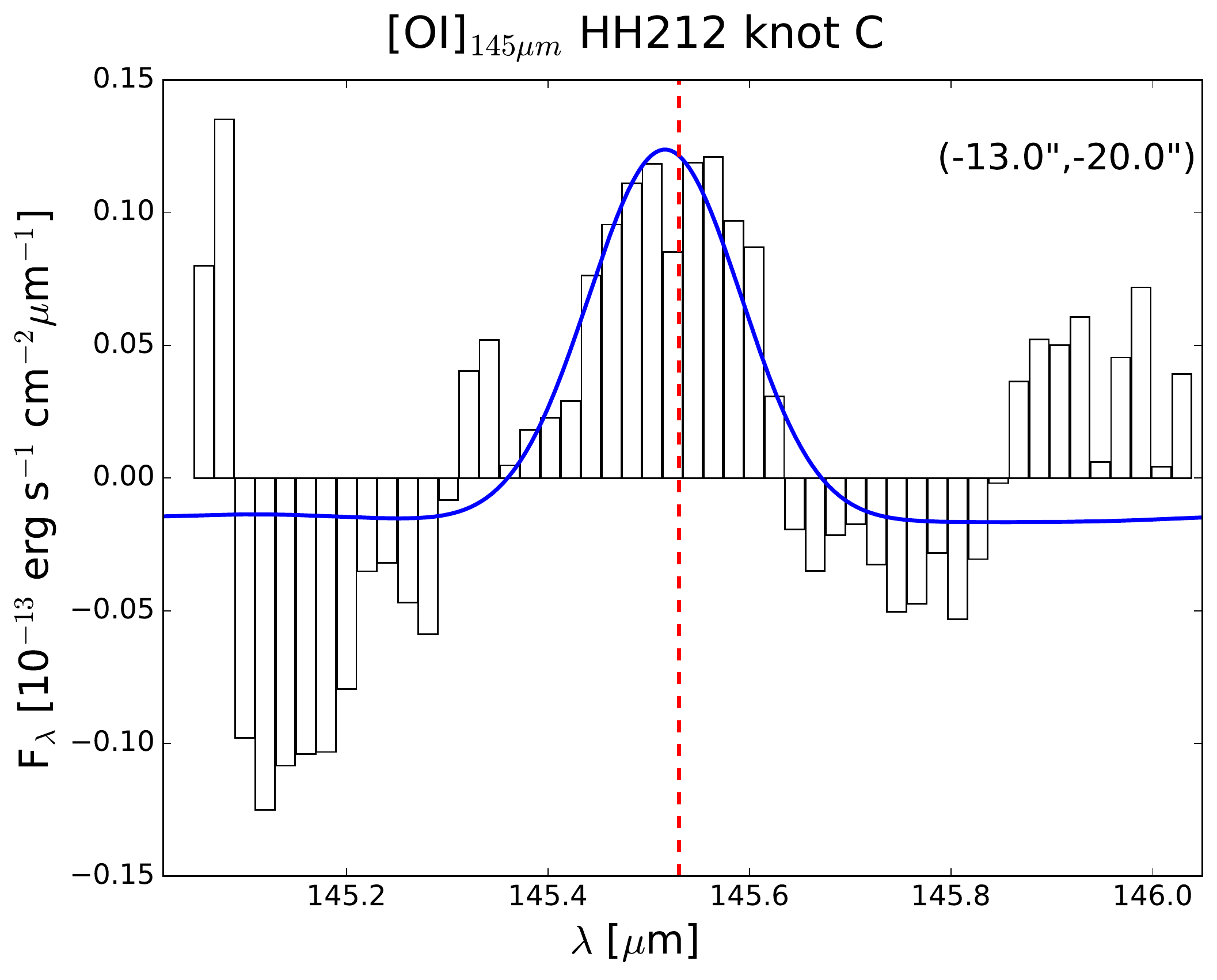}}  
\caption{\small{Detected [O\,I]$_{145}$ line at knot C towards HH\,212.}}\label{fig:hh212_red_emission_figure_B}
\end{figure} 


\section*{Appendix C: Estimate of the SVS13A accretion rate}\label{appendix:svs13_accretion_rate}

The accretion rate of SVS13A can be estimated via the emission in the Br$\gamma$ line at $\lambda = 2.166\,\upmu$m.
\citet{davis_2011} measured a line flux of $F_{\text{Br}\gamma}^\text{obs} = (249\pm 5) \times 10^{-17}\,\text{W}\,\text{m}^{-2}$. We corrected this line flux assuming the extinction law \citep{rieke_1985}:  
\begin{equation}
A_\lambda/A_\text{v} = \left( \lambda_\text{v}/\lambda\right)^{1.6},
\end{equation}
and 
\begin{equation}
F_{\text{Br}\gamma}^0 = F_{\text{Br}\gamma}^\text{obs} \times 10^{A_\lambda/2.5},
\end{equation}
whereby $A_\text{v}$ represents the visual extinction at $\lambda_\text{v} = 0.55\,\upmu$m and $F_{\text{Br}\gamma}^0$ is the extinction-corrected line flux. We assume $A_\text{v} = 10$ \citep{davis_2011_svs13}. Thus,  at $D = 235\,\text{pc,}$ we obtain a line luminosity of $L(\text{Br}\gamma) \approx 1.2\times 10^{-2}\,L_\odot $. 
 Therefore, the accretion luminosity estimated via  
\[\text{log}(L_\text{acc}/L_\odot) = a   \times\text{log}(L(\text{Br}\gamma)/L_\odot) + b , \]
with $a = 1.19\pm 0.10,$ and $b =4.02\pm 0.51$ \citep{alcala_2017} is $L_\text{acc} \approx  54\,L_{\odot} $. This value is consistent with the bolometric luminosity stated in \citet{davis_2011_svs13}; that is, $L_\text{bol} = 50-80\,L_\odot$. Inferring the fiducial values of the protostellar mass and radius of $M_\star = 0.5\,M_\odot$ and
 $R_\star = 4\,R_\odot$ respectively, we calculated a mass accretion rate of \citep{gullbring_1998}
\[\dot{M}_\text{acc} = \frac{L_\text{acc} R_\star}{GM_\star} \approx 1.4 \times 10^{-5}\,M_\odot\,\text{yr}^{-1}. \]
The accretion rate might be even higher  by a factor of $\left(1- R_\star/R_i \right)^{-1}$. In consideration of $R_i=5\,R_\star,$   the corresponding mass accretion rate is  $\dot{M}_\text{acc} = 1.7\times 10^{-5}\,M_\odot\,\text{yr}^{-1} $.  

 
\newpage
    
\section*{Appendix D: Inspecting the [O\,I] origin -- a schematic view}\label{sec:OI_inspection} 
    
In this section, we discuss the possible physical origin of the detected [O\,I]$_{63}$ emission towards the observed protostellar outflows. Associated mass-loss rates might be determined from the [O\,I]$_{63}$ line luminosity if the bulk of the detected [O\,I]$_{63}$ line emission is collisionally excited in shocks. Previous  observations at other transitions may prove useful to test this crucial assumption. All schematics are shown in Fig.\,\ref{fig:schematicss}.

\noindent  
\textbf{Cep\,E:}  The brightest [O\,I]$_{63}$ emission region knot A coincides with the Herbig-Haro object HH\,377 located in the southern lobe of the Cep\,E outflow \citep{devine_1997}. HH\,377 is a compact knot, detected prominently in [S\,II] and less bright in H$\alpha$ indicating the presence of a weak bow-shock region. No optical [O\,III] or [N\,II] line was detected at HH\,377, and therefore it must be a low-excitation object \citep{ayala_2000}.  \\
The presence of a J-shock at the apex of the HH\,377 bow shock can be justified,  since it is also very bright in near-infrared H$_2$ \citep[e.g.][]{eisloeffel_1996, noriega_crespo_1998}. In this scenario, the luminous H$_2$ emission at HH\,377 may be attributed to C-shocks in bow-shock wings \citep{smith_2003}, and the optical emission arises from J-shocks at a potentially present Mach disc \citep{ayala_2000, moro_martin_2001}. Therefore, the observed [O\,I]$_{63}$ line emission is most likely excited via collisions in a J-shock \citep{noriega_crespo_2002}. Since almost no near-infrared [Fe\,II] was detected at HH\,377 \citep{eisloeffel_1996}, we conclude that the shocks are predominantly non-dissociative \citep[see discussion in e.g.][]{bally_2007}. The very faint near-infrared [Fe\,II] emission at HH\,377 might hint at a partly dissociative reverse shock region, that is, the Mach disc. On the other hand, the [O\,I]$_{63}$ line is the dominant far-infrared coolant in both lobes supporting its connection to dissociative J-shocks \citep{giannini_2001}. \\  
However, the [C\,II]$_{158}$/[O\,I]$_{63}$ line ratios at the northern and  southern lobes of Cep E are of the order of $\sim 1$ \citep{moro_martin_2001, giannini_2001, smith_2003}, indicating that some fraction of the observed [O\,I]$_{63}$ emission is not connected to shocks. \citet{moro_martin_2001} estimated that about 20 \%\ of the observed [O\,I]$_{63}$ emission (in both lobes) comes from the presence of
 a PDR region, and $90-95$ \% of the residual [O\,I]$_{63}$  shock component is due to a J-shock. UV illumination by the driving protostar was discussed in \citet{gusdorf_2017}.\\
 Fortunately, the Cep\,E outflow could be fully mapped at the 1670 GHz (179.5\,$\mu$m) H$_2$O emission in the Herschel-WISH survey \citep{tafalla_2013}.  From our maps, we see that far from the source [O\,I] peaks where H$_2$O and H$_2$ peak. Additionally, a few OH lines in the far-infrared spectral range are detected towards the southern and the northern lobes \citep{moro_martin_2001}. Therefore, we conclude that water is copiously formed at the shocked regions (knot A, jet region). Hence, the outflow shock reaches temperatures higher than about 400\,K (low-velocity shock) leading to a chemical conversion of oxygen atoms into H$_2$O by their reaction with H$_2$ \citep[e.g.][]{kaufman_1996, bergin_1998, giannini_2001, flower_2010, tafalla_2013_H2O}. \\
 At the driving source, Cep E mm, we detect no [O\,I]$_{63,145}$. In fact, the lack of emission at other emission lines such as H$_2$, [Fe\,II], and H$\alpha$ towards Cep E mm points to the conclusion that it must be an extremely deeply embedded source \citep{ladd_1997} or system of two sources \citep{moro_martin_2001, ospina_zamudio_2018}. Close to Cep E mm, \citet{lefloch_2011} and \citet{ospina_zamudio_2018} detected a bipolar SiO emission along the north-south flow. This emission roughly matches with the H$_2$O emission \citep[see also][]{tafalla_2013_H2O}, indicating the presence of low-velocity C-shocks potentially interacting with dust grains \citep[e.g.][]{flower_1996, nisini_2007}. \\
  An outflow cavity is seen in CO \citep{lefloch_2015} which is in line with the schematic view of the southern outflow lobe of Cep E presented in \citet{moro_martin_2001}. \citet{ospina_zamudio_2019} detected hot, dense knots along the almost symmetric  CO-jet that can be attributed to internal jet shocks. However, an atomic jet component towards knot A is not seen in our maps, indicating that the jet is mainly molecular \citep{ospina_zamudio_2019}. The perpendicular, small-scale outflow discovered by \citet{eisloeffel_1996} in H$_2$ is not detected in our maps, but it is potentially connected to the second SiO flow detected by \citet{ospina_zamudio_2018}. We suspect that it  is either too faint to be detected in our study or unfavourable shock conditions prevail (e.g. small shock regions, a mainly molecular jet, high extinction). In CO, \citet{ladd_1997} report another, but large-scale (fossil), outflow that is not seen in our [O\,I] maps either.    \\
 At the spatial resolution of our maps, the northern redshifted lobe of the Cep\,E outflow is detected in [O\,I] as one jet-like structure. The same region reveals a highly complex and knotty internal structure seen prominently in H$_2$ \citep{ladd_1997}. No optical H$\alpha$, [S\,II], or near-infrared [Fe\,II] is detected there. Therefore, we conclude that this region is associated with multiple, primarily non-dissociative bow shocks. \citet{velusamy_2011} identified three consecutive bow shocks along the flow at the northern emission region. In [O\,I], only one of them is present at the outermost tip of the emission region.  \\
The [O\,I] emission is slightly curved towards the east and matches the H$_2$ emission in the lower part of the region. Further upstream curvatures on [O\,I] and H$_2$  differ slightly, indicating that both emissions arise from different environments. \citet{eisloeffel_1996} proposed that a precession of the outflow is responsible for the wiggling structure seen in H$_2$. In this context, it seems interesting that in [O\,I] a different wiggle curvature is seen. Considering that the wiggling of the jet was observed 23 years ago, the different wiggle morphology can be interpreted as a temporal evolution of the precessing jet. Alternatively, deflection regions along the jet axis - in particular at the tip of the northern bow shock - mimic precession of the outflow source.   \\
Fortunately, the [O\,I]$_{145}$ line is detected in the Cep\,E jet region (signal-to-noise ratio  $\sim 3$), yielding a line ratio of [O\,I]$_{63}$/[O\,I]$_{145}$ $\sim 23$ in perfect agreement with \citet{giannini_2001}. A similar ratio was found in the Cep\,E blue lobe \citep{giannini_2001}. Depending on pre-shock densities and shock velocities, the \citet{hollenbach_1989} shock model predicts [O\,I] line ratios to be between 10 and 35, which is consistent with our measurement. From the collision diagram presented in \citet{nisini_2015}, we measure collider densities of $n\sim 10^{4.3} - 10^{5.4}\,\text{cm}^{-3}$.\\ 
\textbf{HH\,1:} Most detected [O\,I]$_{63}$ coincides with the location of the well-studied optical HH\,1 jet and its associated chain of shock-excited emission knots \citep[e.g.][]{hester_1998, bally_2002_hh1_2, hartigan_2011}. As in the optical, the [O\,I]$_{63}$ emission fades out about 30$\arcsec$ downstream of the driving source VLA1. Naturally, the HH\,1 jet is interpreted as collimated outflow  featuring a series of internal shocks with bow-shock morphology \citep[e.g.][]{ray_1996}. We therefore conclude that the bulk of the observed [O\,I]$_{63}$ emission is connected to the HH\,1 jet, its shocks, and its associated  internal working surfaces. The detected [O\,I]$_{63}$ emission south-east of VLA1 can be attributed to the counter jet \citep{noriega_crespo_2012}.
Interestingly, in [O\,I]$_{63}$ only one very bright clumpy knot about 5$\arcsec$ north-west of VLA1 is seen roughly coinciding with the innermost optical knot G1$_J$ and the two bright knots of HH\,501 \citep{hester_1998}. So, it may be the case that only one shock region, such as a strong disc wind or a few shock-excited knots are responsible for the main [O\,I]$_{63}$ emission at the HH\,1 jet base. In this context, it is worth mentioning that VLA\,1 is a highly obscured source,  so the lack of [O\,I]$_{63}$ emission - and near-infrared [Fe\,II], H$_2$ emission - at the jet base might be attributed to the dense molecular envelope \citep{davis_2000, reipurth_2000}.  \\
Furthermore, towards VLA\,1 and HH\,1 line ratios of [O\,I]$_{63}$/[C\,II]$_{157}$ $, \sim$ $1$-$2$  were measured, indicating  that a substantial amount of the observed [O\,I]$_{63}$ emission (about 50 \%) may not originate from shocks but from the impact of PDR regions \citep{giannini_2001, molinari_2002}.  
However, at VLA\,1 and HH\,1, far-infrared line ratios of [O\,I]$_{63}$/[O\,I]$_{145}$  $\sim  30$ and [O\,I]$_{63}$/[O\,I]$_{145}$  $\sim  25$ \citep{giannini_2001, molinari_2002} are consistent with predictions from the \citet{hollenbach_1989} J-shock model. Nevertheless, the presence of C-shocks cannot be ruled out from that finding alone \citep{giannini_2001}. More importantly, the [O\,I]$_{63}$ dominates the far-infrared line cooling in both regions, which can  indirectly be interpreted as a confirmation of the assumption that the bulk [O\,I]$_{63}$ emission originates from a dissociative J-shock \citep{giannini_2001}. Indeed, dissociative shock signatures are certified, since near-infrared [Fe\,II] emission about 3$\arcsec$  north-west of VLA1 is prominently detected \citep{davis_2000}.  \\
We detect some clumpy [O\,I]$_{63}$ emission about 65$\arcsec$ north-west of VLA1, that is, at the location of the feature-rich HH\,1 complex. In the optical, HH\,1 is prominently seen in H$\alpha$, [S\,II], and [O\,III], indicating the presence of multiple bow shocks, internal emission knots, filametary structures, and an extended cooling zone \citep[e.g.][]{hartigan_2011}. The [O\,I]$_{63}$ emission is mostly seen behind the leading bow shocks, and therefore this emission is likely connected to the cooling zone behind dissociative shocks.\\ 
\textbf{HH\,212:} The highly symmetric HH\,212 outflow with a projected size of about 240$\arcsec$ ($\sim 0.5\,\text{pc}$) is most prominently seen in molecular transitions; that is, in  H$_2$ \citep[e.g.][]{zinnecker_1998, mcCaughrean_2002, smith_2007}, SiO \citep[e.g.][]{codella_2007, lee_2007, cabrit_2007}, or CO \citep[e.g.][]{gueth_1999, lee_2000, lee_2006}. Surprisingly, no optical counterpart of the innermost part of the jet has been detected in  H$\alpha$ and [S\,II] \citep{reipurth_2019}, indicating that the jet's nature might be mainly molecular \citep{zinnecker_1996} or that high extinction prevents such a detection in the optical. Nevertheless, the atomic component of the HH\,212 jet seen in our [O\,I]$_{63}$ maps can be traced back to the driving source, IRAS 05413--0104,  and consolidates the presence of strong shocks in a dense environment  \citep{gibb_2004, takami_2006}. The detected [O\,I]$_{63}$ emission matches the mid-infrared [Fe\,II]$_{26}$ and [S\,I]$_{25}$ emission very well \citep{anderson_2013}, supporting the notion of dissociative J-shocks and slow C-shocks being present at NK1 and SK1 \citep{hollenbach_1989, kaufman_1996}. Furthermore, \citet{caratti_2006} and \citet{smith_2007} detected strong near-infrared [Fe\,II] emission at both of these knots, supporting the dissociative J-shock assumption. A potential bow-shock morphology featuring a Mach disc at NK1 and SK1 is unresolved in our maps. Strong disc wind signatures have been detected at the driving source \citep{Lee_2018_disk_wind}. Thus, some of the detected [O\,I]$_{63}$ emission might be connected to this disc wind.    \\
Further downstream in both lobes, that is, towards knots SK2 and NK2 and beyond, the [O\,I] emission rapidly faints out, indicating that  only at NK1 and SK1 are the shocks strong enough to excite [O\,I]$_{63}$ \citep{lee_2007}. At the other knots not seen in [O\,I], non-dissociative shocks are present. Since we detect strong [O\,I]$_{145}$ emission in front of SB1/2 - a region not fully mapped here in [O\,I]$_{63}$ - we suspect strong [O\,I]$_{63}$ emission to be detected there as well, indicating another dissociative J-shock region.\\ 
\textbf{L1551\,IRS5:} In the case of L1551\,IRS5, only the innermost arcmin was observed in [O\,I]$_{63 }$ , although this source is connected to multiple spectacular Herbig-Haro objects \citep[e.g.][]{devine_1999, hayashi_2009}.  \citet{white_2000} obtained an infrared spectrum towards IRS5 (ISO LWS, beamsize $\sim 80\arcsec$), indicating that the observed [O\,I]$_{63}$ line originates mainly from shocks and not a PDR region ([O\,I]$_{63}$/[C\,II]$_{157}$ $\sim 7.3$). \\
The two continuum sources associated with L1551\,IRS5 are embedded in a large-scale dusty envelope \citep{looney_1997, rodriguez_1998, fridlund_2002, takakuwa_2020}, and each source features a circumstellar disc, from which [O\,I]$_{63}$ emission can potentially originate.\\
From L1551\,IRS5 a prominent optical jet \citep[HH\,154,][]{fridlund_1998} emerges featuring a knotty internal structure, underpining the assumption of their shock origin \citep[e.g.][]{fridlund_1994, hartigan_2000, fridlund_2005}. Deep optical observations proved that there are actually two westwards-directed jets at the location of HH\,154, that is, the fast northern and slow southern jets \citep{hartigan_2000}. \\ Indeed, most of the (blueshifted) [O\,I]$_{63}$ emission arises between IRS 5 VLA and PHK2, meaning it is connected to the 5$\arcsec$ outset of the northern jet. This part of the jet is prominently seen in the near-infrared [Fe\,II] lines \citep{itho_2000, pyo_2002}, H$_{\alpha}$ and [S\,II] \citep{fridlund_2005}, indicating that the detected [O\,I] emission is most likely connected to dissociative shocks in the outflow.
This conclusion is supported by the noted alignment of the [O\,I]$_{63}$ emission at P.A.$\sim 250^\text{o}$ , which is consistent with the stated P.A. $\sim 260^\text{o}$ for the northern jet \citep{pyo_2009}.  \citet{pyo_2002} recognised two velocity components in the northern [Fe\,II] jet of which the high-velocity ($v_\text{LSR} \sim -300\,\text{km}\,\text{s}^{-1}$) component potentially originates from a collimated stellar jet, whereas the low-velocity ($v_\text{LSR} \sim -100\,\text{km}\,\text{s}^{-1}$) component is connected to a disc wind. The observed spectral [O\,I]$_{63}$ line shifts in both lobes correspond to radial velocities as high as $v_\text{rad}\sim 100 - 150\,\text{km}\,\text{s}^{-1}$, which is consistent with measurements in \citet{lee_2014}, and therefore it is more likely to  originate from disc winds.  \\
The knot PHK2 is peculiar since the northern [Fe\,II] outflow  changes direction at this location \citep{pyo_2009}. Thus, at PHK2 a deflection shock might be present causing parts of the detected [O\,I]$_{63}$ emission.\\ 
The brightest optical emission knot, i.e. knot D in \citet{neckel_1987} or knot 3 in \citet{stocke_1988}, is located about $10\arcsec$ southwest of the source, and it is interpreted as terminal bow shock of the L1551 IRS5 jet with a Mach disc \citep{hartigan_2000, fridlund_2005}. This knot is prominently seen in H$\alpha$, [S\,II] (e.g. Fridlund et al 1994, Fridlund et al 2005) and coincides with the near-infrared [Fe\,II] knot PHK3 \citep{pyo_2002}. Surprisingly, at the position of this prominent emission knot we detect almost no [O\,I]$_{63}$. However, the detection of the optical [O\,III] line only at knot 3 \citep{cohen_1985} hints at the presence of a strong shock \citep[e.g.][]{hartigan_2000, fridlund_2005} ionising almost the whole medium behind a high-velocity shock. \citet{fridlund_1998} estimated an extremely high ionisation fraction close to 100 \%\ at this knot. We therefore conclude that at PHK3 a strong bow shock from HH\,154 ionises almost all oxygen atoms, leaving no neutral oxygen atoms to take part in the cooling in the cooling zone. \\ 
A very faint counter jet northeast of IRS 5 was first detected in [Fe\,II] by \citet{hayashi_2009}. Since there are some spaxels at this location showing a redshifted [O\,I]$_{63}$ line, we conclude that this small-scale counter jet is tentatively detected in our maps as well. The observed [O\,I]$_{63}$ emission seems to follow the [Fe\,II] emission, indicating that both lines are connected to the outflow and their shocks. \\
At L1551 IRS 5, very weak H$_2$O and OH emission are detected \citep[e.g.][]{tafalla_2013_H2O}, indicating a complete photodissociation of these molecules \citep{lee_2014}. Herschel/PACS observations indicate that dissociative J-shocks connected to disc winds and/or both jets are responsible for the observed low water emission \citep{lee_2014}. However, C-shocks  attributed to cavity shocks might also be present \citep{lee_2014}. At  L1551 IRS 5, line ratios of [O\,I]$_{63}$/[O\,I]$_{145}$ $\sim 10-20$ have been determined \citep{lee_2014} and are in line with predictions from the \citep{hollenbach_1989} shock model.\\
In conclusion, the HM89 shock conditions very probably prevail at L1551 IRS5 and near to the HH\,212 driving source (Table\,\ref{table:main_results}). For Cep\,E and HH\,1, a significant (and still unknown) amount of [O\,I]$_{63}$ is connected to a  PDR region, as well as to non-dissociative J-shocks.

\begin{figure*}   
\centering
\subfloat{\includegraphics[trim=55 30 160 23, clip, width=0.4\textwidth]{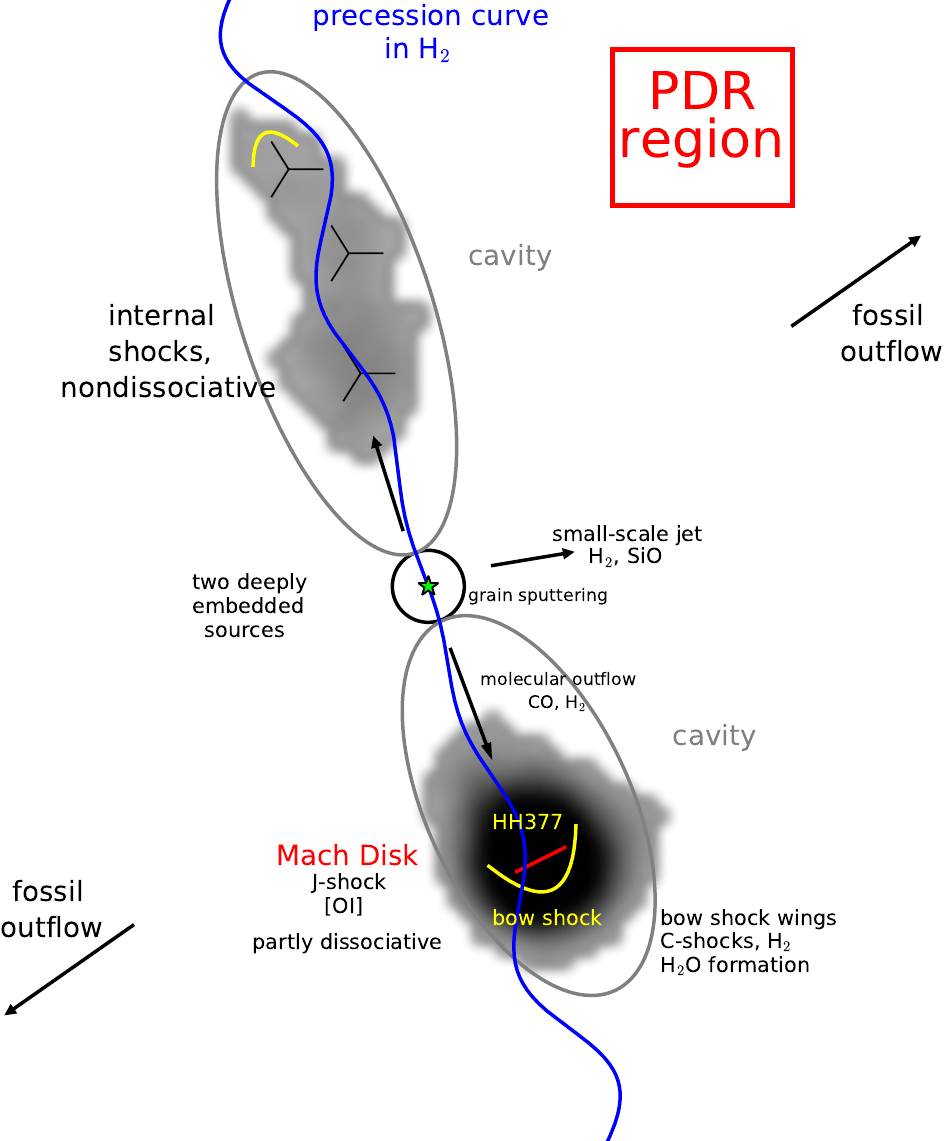}} 
\hfill
\subfloat{\includegraphics[trim=68 62 4 3, clip, width=0.4\textwidth]{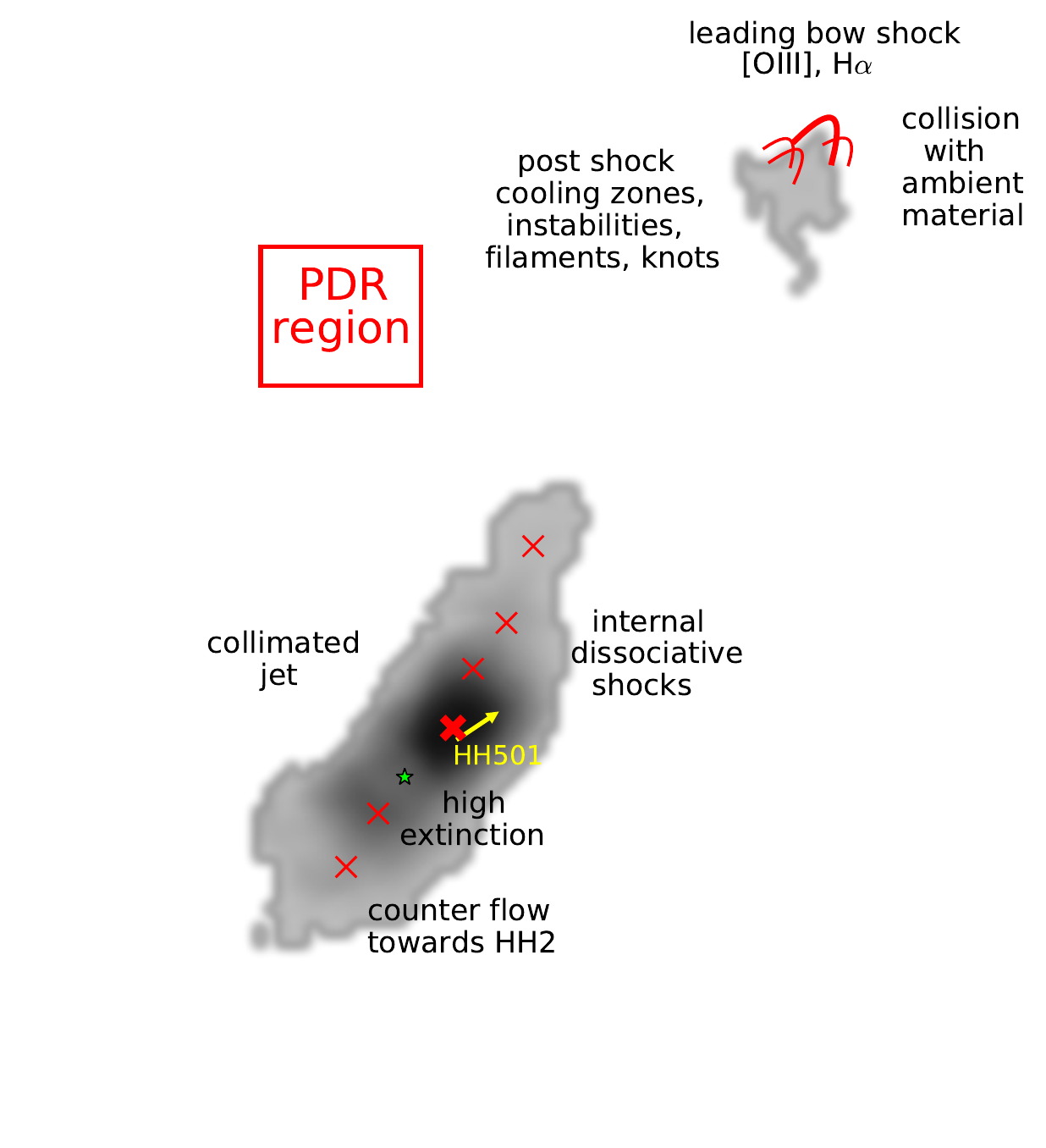}}
\end{figure*}
   
\begin{figure*}
\centering
\subfloat{\includegraphics[trim=50 170 45 25, clip, width=0.49\textwidth]{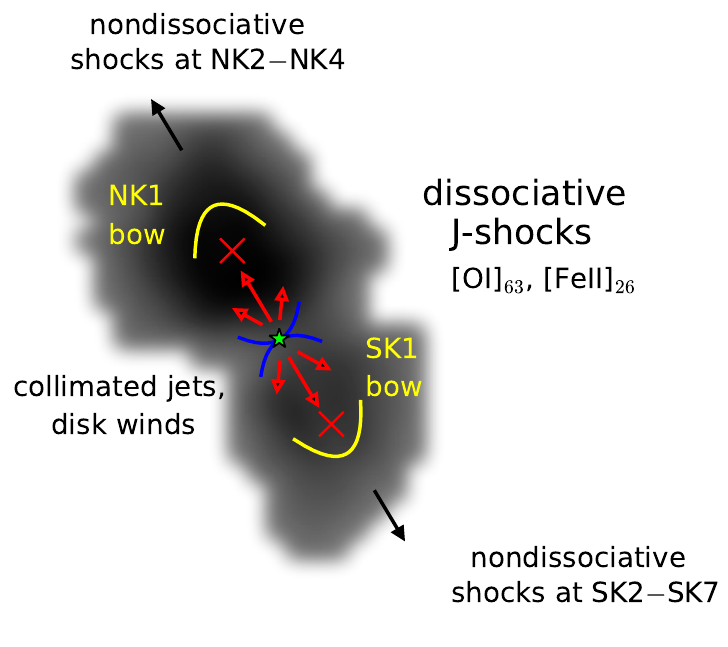}} 
\hfill
\subfloat{\includegraphics[trim=33 85 30 75, clip, width=0.4 \textwidth]{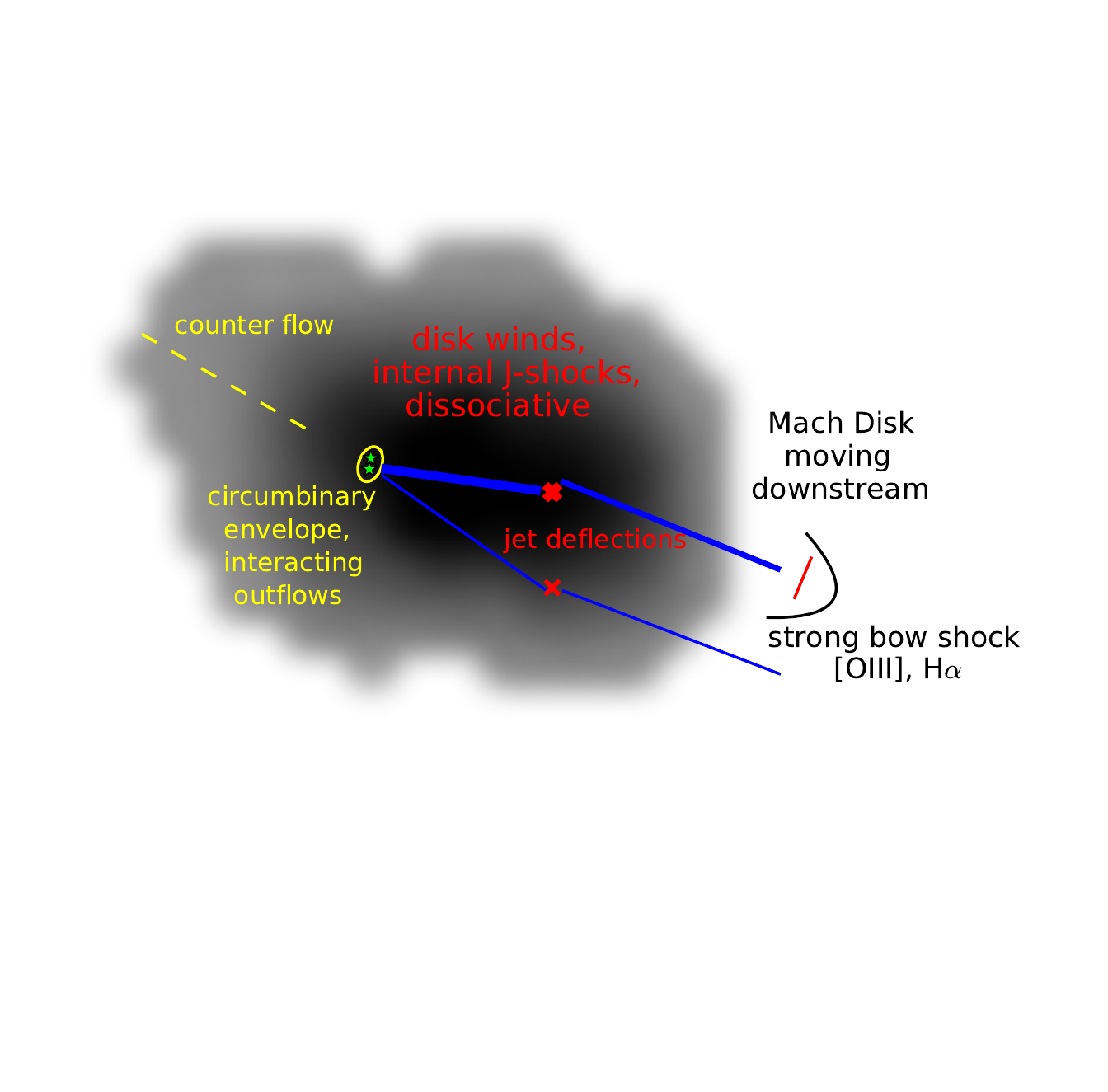}} 
\caption{\small{Schematics of the Cep\,E (\textit{upper left}), HH\,1 (\textit{upper right}), HH\,212 (\textit{lower left}), and L1551\,IRS5 (\textit{lower right}) outflow. The continuum-subtracted [O\,I]$_{63}$ emission maps from the SOFIA/FIFI-LS observations are plotted in a grey colour scale. Source positions are marked as green stars. Our interpretations of the observed  [O\,I]$_{63}$ emission are based on previous studies briefly discussed in Appendix\,D.}}\label{fig:schematicss}
\end{figure*}

\newpage

\section*{Appendix E: Table of all objects}\label{appendix:all_objects}
  
Tables\,\ref{literature_I}--\ref{literature_III} list 72 protostellar outflow  sources that have been observed   in the [O\,I]$_{63}$ transition with a single footprint with Herschel/PACS; that is, a field of view of $5\times 5$ spatial pixels, extending $47\arcsec\times 47\arcsec$. Thus, these observations feature limited mapping compared to the presented SOFIA data.  We compared their jet efficiencies $\dot{M}_\text{out}/\dot{M}_\text{acc}$ with the ones determined in this study.

{\renewcommand{\arraystretch}{1.2}
\begin{table*}[!htb]
\caption{\small{Source sample from the DIGIT and WISH surveys. We took 25 sources (13 Class 0 and 12 Class I) from  Table A.7 in  \citet{mottram_2017}, hereafter abbreviated as Mo17,  with the following selection criteria: a) $\dot{M}_\text{acc}$ and $\dot{M}_\text{out}$  are given in Table A.7; b) sources are not part of the spatially resolved \citet{nisini_2015} sample, which were analysed in much more detail; c) flat sources are excluded (IRAS03301+3111, IRAS12496, RNO91); d) we  excluded RCRA IRS5A, RCrA-I7A, and RCrA-I7B  since they are ambiguously classified.}}\label{literature_I} 
\centering\tiny
\begin{tabular}{ c c c c c c c c c c c}  \\
\hline\hline
Source & Cloud & Class\tablefootmark{a} & RA(J2000)\tablefootmark{b} &  DEC(J2000)\tablefootmark{b}  & D & $\dot{M}_\text{out}(\text{[O\,I]})$  & $\dot{M}_\text{out}(\text{mol})$ & $\frac{\dot{M}_\text{out}(\text{[O\,I]})}{\dot{M}_\text{out}(\text{mol})}$\tablefootmark{i} & $\dot{M}_\text{acc}$\tablefootmark{g}  & Main Ref.  \\[1.5pt]
 & &  & (h m s)   &  ($^\text{o}$\,'\,'')  &  (pc) &     $(M_\odot\,\text{yr}^{-1})$ & $(M_\odot\,\text{yr}^{-1})$&  &  $(M_\odot\,\text{yr}^{-1})$   &     \\   
\hline  
 IRAS03245+3002  & Per & 0 & 03 27 39.09 & $+$30 13 03.0 & 235 & $6.4\times 10^{-8}$ & &   & $4.2\times 10^{-6}$ & Mo17  \\ 
 L1455-ISR3      & Per & I & 03 28 00.40 &  $+$30 08 01.3 & 235     & $5.0\times 10^{-8}$ &   & & $4.7\times 10^{-8}$ & Mo17    \\
 NGC 1333-IRAS 4B & Per & 0 & 03 29 12.04 & $+$31 13 01.5 & 235     & $1.3\times 10^{-8}$ & $\sim  3.5 \times 10^{-7}$\tablefootmark{k}  & $3.7\times 10^{-2}$ & $2.9\times 10^{-6}$ & Mo17   \\
 B1a             & Per & I & 03 33 16.66 & 31 07 55.2 &    235 & $1.9\times 10^{-8}$ &  & &  $3.5\times 10^{-7}$ & Mo17  \\
 L1489           & Tau & I &  04 04 42.9 &  $+$26 18 56.3 &  140  & $2.2\times 10^{-8}$ & $\sim 6.1\times 10^{-7}$\tablefootmark{e}  & $3.6\times 10^{-2}$ & $4.9\times 10^{-7}$ & Mo17   \\
 TMR1            & Tau & I & 04 39 13.9 &  $+$25 53 20.6 &  140  & $1.1\times 10^{-7}$ & $\sim 4.1\times 10^{-7}$\tablefootmark{e} & $2.6\times 10^{-1}$ & $4.9\times 10^{-7}$ & Mo17  \\
 TMC1A           & Tau & I & 04 39 35.0 &  $+$25 41 45.5 &  140  & $3.3\times 10^{-7}$ & $\sim 7.3\times 10^{-7}$\tablefootmark{e}  &  $4.5\times 10^{-1}$ & $3.5\times 10^{-7}$ & Mo17  \\
 L1527           & Tau & 0 & 04 39 53.9 & $+$26 03 09.8 &  140  & $1.8\times 10^{-7}$ & $\sim 4.5\times 10^{-6}$\tablefootmark{e} & $4.0\times 10^{-2}$ & $1.2\times 10^{-6}$ & Mo17  \\
 TMC1            & Tau & I & 04 41 12.7  &  $+$25 46 35.9 &  140   & $1.5\times 10^{-7}$ & $\sim 1.6\times 10^{-6}$\tablefootmark{e}  & $9.4\times 10^{-2}$& $1.2\times 10^{-7}$ & Mo17\\
 IRAM04191+1522  & Tau & 0 &  04 21 56.9 &  $+$15 29 45.9 &  140     & $2.4\times 10^{-8}$ &  &  & $7.1\times 10^{-8}$ & Mo17  \\
 Ced110-IRS4     & ChaI & 0 & 11 06 47.0 &  $-$77 22 32.4 &  150   & $1.3\times 10^{-7}$ & $\sim 3.3\times 10^{-7}$\tablefootmark{e} &  $3.4\times 10^{-1}$ & $5.2\times 10^{-7}$ & Mo17  \\
 IRAS15398-3359  & Lup & 0 &  15 43 01.29 &  $-$34 09 15.4 &  130\tablefootmark{c} & $3.0\times 10^{-7}$ & $\sim 3.2\times 10^{-6}$\tablefootmark{e} & $9.4\times 10^{-2}$ & $1.0\times 10^{-6}$ & Mo17 \\
 GSS30 IRS1      & Oph & I &  16 26 21.48 &  $-$24 23 04.2 &   125   & $5.1\times 10^{-7}$ & $\sim 1.3\times 10^{-5}$\tablefootmark{e}  & $3.9\times 10^{-3}$ & $1.8\times 10^{-6}$ & Mo17  \\
 WL12            & Oph & I & 16 26 44.2 &  $-$24 34 48.4 &  125  & $7.9\times 10^{-8}$ &  &   & $2.1\times 10^{-7}$ & Mo17\\
 Elias29         & Oph & I &  16 27 09.36 & $-$24 37 18.4 &   125   & $4.0\times 10^{-7}$ & $\sim 1.6\times 10^{-6}$\tablefootmark{e} & $2.5\times 10^{-1}$ & $1.8\times 10^{-6}$ & Mo17   \\
 IRS44           & Oph & I &   16 27 28.1 & $-$24 39 33.4 &  125     & $4.8\times 10^{-8}$ &  &  & $6.6\times 10^{-7}$ & Mo17 \\
 IRS46           & Oph & I &   16 27 29.4 &  $-$24 39 16.1 &  125    & $1.5\times 10^{-8}$ &   & & $6.5\times 10^{-8}$ & Mo17\\
 IRS63           & Oph & I &   16 31 35.76 &  $-$24 01 29.2 &  125  & $3.3\times 10^{-8}$ & $\sim 1.3\times 10^{-5}$\tablefootmark{f} &   & $1.3\times 10^{-7}$ & Mo17 \\
 L483 MM         & Aqu & 0 & 18 17 29.9 &  $-$04 39 39.5 &   200\tablefootmark{c} & $1.1\times 10^{-7}$ &  $\sim 1.1\times 10^{-5}$\tablefootmark{e} & $1.0\times 10^{-2}$ & $6.6\times 10^{-6}$ & Mo17\\
Ser-SMM1        & Ser & 0 &   18 29 49.56 &  $+$01 15 21.9 &  436 & $4.0\times 10^{-7}$ & $\sim 2.2\times 10^{-5}$\tablefootmark{e} & $ 1.8\times 10^{-2}$ & $7.1\times 10^{-5}$ & Mo17    \\
Ser-SMM4        & Ser & 0 &   18 29 56.7 &  $+$01 13 17.2 &  436 & $1.8\times 10^{-6}$ & $\sim 2.6\times 10^{-5}$\tablefootmark{e} &  $6.9\times 10^{-2}$& $4.4\times 10^{-6}$ & Mo17\\
Ser-SMM3        & Ser & 0 &   18 29 59.3 &  $+$01 14 01.7 &  436 & $1.1\times 10^{-6}$ & $\sim 3.9\times 10^{-5}$\tablefootmark{e} &  $2.8\times 10^{-2}$ & $1.2\times 10^{-5}$ & Mo17 \\
L723 MM         & Core & 0 & 19 17 53.7 &  $+$19 12 20.0 &  300\tablefootmark{c}  & $1.3\times 10^{-8}$ & $\sim 1.5\times 10^{-5}$\tablefootmark{e}  &  $8.7\times 10^{-4}$ & $2.3\times 10^{-6}$ & Mo17   \\
B335            & Core & 0 & 19 37 00.9 &  $+$07 34 09.6 & 106\tablefootmark{d}   & $2.5\times 10^{-8}$ & $\sim  1.0 \times 10^{-7}$\tablefootmark{j}  & $ 2.5\times 10^{-1}$ & $2.1\times 10^{-6}$ & Mo17\\
L1157           & Core & 0 & 20 39 06.3 & $+$68 02 16.0 &  325\tablefootmark{d} & $3.3\times 10^{-8}$  &  $\sim  4.5 \times 10^{-7}$\tablefootmark{k} & $ 7.3\times 10^{-2}$ & $3.0\times 10^{-6}$ & Mo17   \\
 \hline
\end{tabular}
\tablefoot{  
\tablefoottext{a}{\small{Classification from \citet{nisini_2002}, \citet{green_2013}, \citet{karska_2013}, and \citet{yang_2018}.}}
\tablefoottext{b}{\small{Coordinates from \citet{karska_2013}.}} 
\tablefoottext{c}{\small{\citet{karska_2013}.}} 
\tablefoottext{d}{\small{\citet{green_2013}.}}
\tablefoottext{e}{\small{\citet{yildiz_2015} from CO J=6--5 observations; the listed values are the sum of the mass-loss rates of both lobes.}}
\tablefoottext{f}{\small{\citet{tanabe_2019} from CO J=1--0 observations, i.e. rather entrained material.}}
\tablefoottext{g}{\small{From \citet{mottram_2017} we adopt their method calculating the mass-accretion rates listed in Tables\,\ref{table:main_results_II} and \ref{table:my_main_results}.}}
\tablefoottext{h}{\small{\citet{yildiz_2015} from CO J=3--2 observations; the listed values are the sum of the mass-loss rates of both lobes. As stated therein,  CO J=3--2 traces entrained material.}}
\tablefoottext{i}{\small{Ratio of column 7 and 8, not calculated in cases where the swept-up material is traced by CO.}} 
\tablefoottext{j}{\small{\citet{lee_2020_molecular_review}.}}
\tablefoottext{k}{\small{\citet{podio_2020}.}}
}
\end{table*}
 

{\renewcommand{\arraystretch}{1.2}
\begin{table*}[!htb]
\caption{\small{Source sample from the WILL survey. We only took the 26 (15 Class 0 and 11 Class I) outflow sources, for which $\dot{M}_\text{acc}$ is presented in Table A.7 in Mo17.}}\label{literature_II} 
\centering\tiny
\begin{tabular}{ c c c c c c c c c c}  \\
\hline\hline
Source\tablefootmark{a} & Cloud & Class & RA(J2000) &  DEC(J2000)  & D & $\dot{M}_\text{out}(\text{[O\,I]}_{63})$ & $\dot{M}_\text{out}(\text{mol} )$\tablefootmark{b} & $\dot{M}_\text{acc}$\tablefootmark{c} & Main Ref.  \\[1.5pt]
 & &  & (h m s)   &  ($^\text{o}$\,'\,'')  &  (pc) &     $(M_\odot\,\text{yr}^{-1})$ &       $(M_\odot\,\text{yr}^{-1})$ &$(M_\odot\,\text{yr}^{-1})$   &     \\   
\hline  
AQU 01  &  Aqu  & 0 &  18 29 03.82 & $-$01 39 01.5 & 436 & $8.3\times 10^{-8}$ & $8.5\times 10^{-5}$ & $1.7\times 10^{-6}$ & Mo17  \\
AQU 02  &  Aqu  & 0 &  18 29 08.60 & $-$01 30 42.8 & 436 & $8.2\times 10^{-8}$ &  $1.3\times 10^{-5}$  & $5.9\times 10^{-6}$ & Mo17 \\
CRA 01  &  CrA  & 0 &  19 02 58.67 & $-$37 07 35.9 & 130 & $1.2\times 10^{-7}$ & $ 2.3\times 10^{-6}$   & $1.6\times 10^{-6}$ & Mo17\\
OPH 02  &  Oph  & I &  16 32 00.99 & $-$24 56 42.6 & 125 & $5.2\times 10^{-8}$ & $ 2.5\times 10^{-8}$  & $1.1\times 10^{-6}$ & Mo17 \\
PER 01  &  Per  & 0 &  03 25 22.32 & $+$30 45 13.9 & 235 & $3.4\times 10^{-7}$ & $ 8.1\times 10^{-5}$ & $2.9\times 10^{-6}$ & Mo17 \\
PER 02  &  Per  & 0 &  03 25 36.49 & $+$30 45 22.2 & 235 & $7.8\times 10^{-7}$ &  -- & $6.0\times 10^{-6}$ & Mo17  \\
PER 04  &  Per  & 0 &  03 26 37.47 & $+$30 15 28.1 & 235 & $5.3\times 10^{-8}$ & $ 7.1\times 10^{-7}$ & $7.6\times 10^{-7}$ & Mo17\\
PER 05  &  Per  & I &  03 28 37.09 & $+$31 13 30.8 & 235 & $2.8\times 10^{-8}$ &$1.2\times 10^{-7}$  & $1.4\times 10^{-6}$ & Mo17   \\
PER 06  &  Per  & I &  03 28 57.36 & $+$31 14 15.9 & 235 & $1.6\times 10^{-7}$ & -- & $9.2\times 10^{-7}$ & Mo17  \\
PER 07  &  Per  & 0 &  03 29 00.55 & $+$31 12 00.8 & 235 & $1.2\times 10^{-7}$ & $ 3.8\times 10^{-6}$ & $4.6\times 10^{-7}$ & Mo17  \\
PER 08  &  Per  & I &  03 29 01.56 & $+$31 20 20.6 & 235 & $1.1\times 10^{-6}$ & $ 1.4\times 10^{-5}$ & $2.2\times 10^{-6}$ & Mo17   \\
PER 14  &  Per  & I &  03 30 15.14 & $+$30 23 49.4 & 235 & $4.6\times 10^{-8}$ & $3.7\times 10^{-6}$ & $2.4\times 10^{-7}$ & Mo17  \\
PER 15  &  Per  & 0 &  03 31 20.98 & $+$30 45 30.1 & 235 & $2.1\times 10^{-8}$ & $2.4\times 10^{-6}$ & $1.1\times 10^{-6}$ & Mo17   \\
PER 16  &  Per  & 0 &  03 32 17.96 & $+$30 49 47.5 & 235 & $2.6\times 10^{-8}$ & $7.6\times 10^{-6}$ & $7.4\times 10^{-7}$ & Mo17   \\
PER 17  &  Per  & I &  03 33 14.38 & $+$31 07 10.9 & 235 & $1.4\times 10^{-8}$ & $1.5 \times 10^{-5}$ & $2.3\times 10^{-8}$ & Mo17  \\
PER 18  &  Per  & 0 &  03 33 16.44 & $+$31 06 52.5 & 235 & $1.8\times 10^{-7}$ & $ 3.4\times 10^{-6}$ & $3.4\times 10^{-7}$ & Mo17  \\
PER 19  &  Per  & I &  03 33 27.29 & $+$31 07 10.2 & 235 & $1.8\times 10^{-7}$ & $ 1.7\times 10^{-6}$ & $1.4\times 10^{-7}$ & Mo17  \\
PER 21  &  Per  & 0 &  03 43 56.84 & $+$32 03 04.7 & 235 & $8.6\times 10^{-8}$ & $ 7.8\times 10^{-6}$ & $1.2\times 10^{-6}$ & Mo17  \\
PER 22  &  Per  & 0 &  03 44 43.96 & $+$32 01 36.2 & 235 & $1.0\times 10^{-7}$ & $ 1.5\times 10^{-6}$ & $1.6\times 10^{-6}$ & Mo17  \\
SERS 01 &  Serp & 0 &  18 29 37.70 & $-$01 50 57.8 & 436 & $2.8\times 10^{-7}$ & $ 1.8\times 10^{-5}$ & $1.1\times 10^{-5}$ & Mo17  \\
TAU 01  &  Tau  & I &  04 19 58.40 & $+$27 09 57.0 & 140 & $7.6\times 10^{-8}$ & $ 1.7\times 10^{-6}$ & $2.0\times 10^{-7}$ & Mo17  \\
TAU 02  &  Tau  & I &  04 21 11.40 & $+$27 01 09.0 & 140 & $1.1\times 10^{-8}$ & $ 2.1\times 10^{-7}$ & $6.1\times 10^{-8}$ & Mo17  \\
TAU 04  &  Tau  & I &  04 27 02.60 & $+$26 05 30.0 & 140 & $4.5\times 10^{-8}$ & $7.1 \times 10^{-7}$ & $1.8\times 10^{-7}$ & Mo17   \\
TAU 06  &  Tau  & I &  04 27 57.30 & $+$26 19 18.0 & 140 & $9.6\times 10^{-9}$ & $1.8\times 10^{-7}$ & $7.3\times 10^{-8}$ & Mo17  \\
W40 02  &  W40  & 0 &  18 31 10.36 & $-$02 03 50.4 & 436 & $5.4\times 10^{-6}$ & $1.0\times 10^{-5}$ & $2.1\times 10^{-5}$ & Mo17   \\
W40 07  &  W40  & 0 &  18 32 13.36 & $-$01 57 29.6 & 436 & $5.4\times 10^{-8}$ & $ 6.5\times 10^{-6}$ & $2.3\times 10^{-6}$ & Mo17   \\
 \hline
\end{tabular}
\tablefoot{  
\tablefoottext{a}{\small{Source names adopted from \citet{mottram_2017}.}} 
\tablefoottext{b}{\small{Calculated via the same method presented in \citet{yildiz_2015} but based on low-J CO observations; that is, CO J=3--2 measuring the entrained material.}}
\tablefoottext{c}{\small{From \citet{mottram_2017}, we adopted their method calculating the mass-accretion rates listed in Tables\,\ref{table:main_results_II} and \ref{table:my_main_results}.}} 
}
\end{table*}

{\renewcommand{\arraystretch}{1.2}
\begin{table*}[!htb]
\caption{\small{Source sample from \citet{alonso_martinez_2017}, hereafter abbreviated as AM17. Sources were selected with the following selection criteria: a) Class II outflow sources \citep[see Table 2 in][]{howard_2013}; b)  $\dot{M}_\text{acc}$ is given in Table A.1  in AM17; c) integrated [O\,I]$_{63}$  line fluxes over $3\times 3$ spaxels are stated with error margins in Table C.3.}}\label{literature_III} 
\centering\tiny
\begin{tabular}{ c c c c c c c c c  }  \\
\hline\hline
Source & Cloud & Class & RA(J2000)\tablefootmark{a} &  DEC(J2000)\tablefootmark{a}  & D & $\dot{M}_\text{out}(\text{[O\,I]}_{63})$  & $\dot{M}_\text{acc}$\tablefootmark{g} & Main Ref.  \\[1.5pt]
 & &  & (h m s)   &  ($^\text{o}$\,'\,'')  &  (pc) &     $(M_\odot\,\text{yr}^{-1})$ &  $(M_\odot\,\text{yr}^{-1})$   &     \\   
\hline  
AA Tau   & Tau & II & 04 34 55.42  & $+$24 28 53.16 & 140 &    $1.29\pm 0.37\times 10^{-9}$\tablefootmark{b}     & $2.51\times 10^{-8}$  & AM17 \\ 
CW Tau   & Tau & II & 04 14 17.00  & $+$28 10 57.80 & 140 &    $6.68\pm  1.72\times 10^{-9}$\tablefootmark{b}     & $5.27\times 10^{-8}$  & AM17\\ 
DF Tau & Tau & II & 04 27 03.08 & $+$25 42 23.30 & 140 & $3.74\pm 0.37 \times 10^{-9}$\tablefootmark{d} &   $10.05\times 10^{-8}$  & AM17 \\  
DG Tau   & Tau & II & 04 27 04.70  & $+$26 06 16.30 & 140 & $1.01\pm 0.04 \times 10^{-7}$\tablefootmark{b}  & $2.53\times 10^{-7}$ & AM17 \\ 
DL Tau      & Tau & II & 04 33 39.06 & $+$25 20 38.23 & 140 & $1.35\pm 0.12\times 10^{-9}$\tablefootmark{d}  &  $2.48\times 10^{-8}$  & AM17 \\ 
DO Tau   & Tau & II & 04 38 28.58  & $+$26 10 49.44 & 140 &     $1.23\pm 0.29 \times 10^{-8}$\tablefootmark{b}     & $3.12\times 10^{-8}$  & AM17\\
DP Tau   & Tau & II & 04 42 37.70  & $+$25 15 37.50 & 140  &  $8.39\pm 1.23 \times 10^{-9}$\tablefootmark{b}     & $0.04\times 10^{-8}$  & AM17\\
DQ Tau & Tau & II & 04 46 53.05 & $+$17 00 00.20 & 140 &  $1.29\pm 0.25  \times 10^{-9}$\tablefootmark{d} &    $0.59\times 10^{-8}$ & AM17 \\ 
FS Tau   & Tau & II & 04 22 02.18  & $+$26 57 30.50 & 140  & $3.27\pm 0.12 \times 10^{-8}$\tablefootmark{b}  &  $(2-3)\times 10^{-7}$\tablefootmark{c} & AM17 \\
GG Tau   & Tau & II & 04 32 30.35  & $+$17 31 40.60  & 140  & $ 3.25\pm 0.74 \times 10^{-9}$\tablefootmark{b}  &  $7.95 \times 10^{-8}$ & AM17 \\
Haro 6-13   & Tau & II & 04 32 15.41 & $+$24 28 59.70 & 140 & $3.12\pm 0.86 \times 10^{-9}$\tablefootmark{b}  & $2.88\times 10^{-8}$\tablefootmark{e}  & AM17 \\
HL Tau & Tau & II & 04 31 38.44 & $+$18 13 57.65 & 140 &  $3.14\pm 0.03  \times 10^{-8}$\tablefootmark{d}  &    $0.35\times 10^{-8}$ &  AM17\\ 
HN Tau   & Tau & II & 04 33 39.35  & $+$17 51 52.37 & 140   &   $3.68\pm 0.55 \times 10^{-9}$\tablefootmark{b}    & $0.49\times 10^{-8}$  & AM17\\
IRAS04385+2550\tablefootmark{f} & Tau & II & 04 41 38.82 & $+$25 56 26.75 & 140  & $4.72\pm 0.74 \times 10^{-9}$\tablefootmark{b}  &   $7.76\times 10^{-9}$\tablefootmark{e} & AM17 \\
RW Aur   & Tau & II & 05 07 49.54  & $+$30 24 05.07 & 140  & $1.43\pm 0.28 \times 10^{-8}$\tablefootmark{b}  &   $(0.034-1.6)\times 10^{-6}$\tablefootmark{c} & AM17 \\
SU Aur   & Tau & II & 04 55 59.38  & $+$30 34 01.56 & 140   &   $8.03\pm 0.74 \times 10^{-9}$\tablefootmark{b}     & $18.03\times 10^{-8}$ & AM17\\
T Tau    & Tau & II & 04 21 59.43  & $+$19 32 06.37 & 140   & $1.12\pm 0.02 \times 10^{-6}$\tablefootmark{b}  & $2.36\times 10^{-7}$ & AM17 \\ 
UY Aur   & Tau & II & 04 51 47.38  & $+$30 47 13.50 & 140  & $2.24\pm  0.09 \times 10^{-8}$\tablefootmark{b}     & $6.74\times 10^{-8}$  & AM17\\
UZ Tau      & Tau & II & 04 32 42.89 & $+$25 52 32.60 & 140 & $2.76\pm 0.86 \times 10^{-9}$\tablefootmark{d}  & $0.45\times 10^{-8}$ & AM17 \\
V773 Tau & Tau & II & 04 14 12.92  & $+$28 12 12.45 & 140   &  $5.57\pm 0.74 \times 10^{-9}$\tablefootmark{b}     & $11.95\times 10^{-8}$ & AM17\\
XZ Tau   & Tau & II & 04 31 39.48  & $+$18 13 55.70 & 140   &   $5.90\pm 0.36  \times 10^{-8}$\tablefootmark{b}    & $0.71\times 10^{-8}$  & AM17\\
\hline\hline 
\end{tabular}
\tablefoot{  
\tablefoottext{a}{\small{Coordinates from \citet{howard_2013}.}} 
\tablefoottext{b}{\small{Calculated via Eq.\,\ref{equ:HM89_formula} from the given line [O\,I]$_{63}$ fluxes stated in Table C.3 in AM17.}} 
\tablefoottext{c}{\small{Accretion rates from \citet{podio_2012}.}} 
\tablefoottext{d}{\small{Calculated via Eq.\,\ref{equ:HM89_formula} from the given line [O\,I]$_{63}$ fluxes stated in Table\,1 in \citet{aresu_2014}.}}
\tablefoottext{e}{\small{\citet{white_2004}.}}
\tablefoottext{f}{\small{Other name: Haro 6-33.}} 
\tablefoottext{g}{\small{Accretion rates are calculated from the U-band excess \citep[see description in][]{alonso_martinez_2017}. }} 
}
\end{table*}

\end{document}